\documentclass[aps,rmp,twocolumn,nofootinbib,floatfix]{revtex4}

\usepackage{amsmath}
\usepackage{amsfonts}
\usepackage{amssymb}
\usepackage[latin1]{inputenc}        
\usepackage{graphicx} 
\usepackage{marvosym} 
\usepackage{fancyhdr}      
\usepackage{color}
\usepackage{verbatim}   
\usepackage{subfigure}  
\usepackage{pdfpages}

\newcommand{\bs}{\boldsymbol}
\newcommand{\beq}{\begin{equation}}
\newcommand{\eeq}{\end{equation}}
\newcommand{\bpm}{\begin{pmatrix}}
\newcommand{\epm}{\end{pmatrix}}
\newcommand{\bea}{\begin{eqnarray}}
\newcommand{\eea}{\end{eqnarray}}

\newcommand{\oL}{\omega_{\rm L}}
\newcommand{\oA}{\omega_{\rm A}}
\newcommand{\rhov}{\rho_{\rm v}}

\begin{document}
\title{Colloquium: Artificial gauge potentials for neutral atoms}
\author{Jean Dalibard}
\email{jean.dalibard@lkb.ens.fr}
\affiliation{Laboratoire Kastler Brossel, CNRS, UPMC, Ecole normale sup\'erieure, 24 rue Lhomond, 75005, Paris, France}
\author{Fabrice Gerbier}
\email{fabrice.gerbier@lkb.ens.fr}
\affiliation{Laboratoire Kastler Brossel, CNRS, UPMC, Ecole normale sup\'erieure, 24 rue Lhomond, 75005, Paris, France }
\author{Gediminas Juzeli\=unas}
\email{Gediminas.Juzeliunas@tfai.vu.lt}
\affiliation{Institute of Theoretical Physics and Astronomy, Vilnius
University, A. Go\v{s}tauto 12, Vilnius 01108, Lithuania}
\author{Patrik \"Ohberg}
\email{P.Ohberg@hw.ac.uk}
\affiliation{SUPA, Department of Physics, Heriot-Watt University, Edinburgh, EH14 4AS, United Kingdom}
\date{\today}
\begin{abstract}
When a neutral atom moves in a properly designed laser field, its center-of-mass motion may mimic the dynamics of a charged particle in a magnetic field, with the emergence of a Lorentz-like force.  In this Colloquium we present the physical principles at the basis of this artificial (synthetic) magnetism and relate the corresponding  Aharonov--Bohm phase to the Berry's phase that emerges when the atom follows adiabatically one of the dressed states of the atom-laser interaction. We also discuss some manifestations of artificial magnetism for a cold quantum gas, in particular in terms of vortex nucleation. We then generalise our analysis to the simulation of non-Abelian gauge potentials and present some striking consequences, such as the emergence of an effective spin-orbit coupling. We address both the case of bulk gases and discrete systems, where atoms are trapped in an optical lattice.
\end{abstract}
%
\maketitle
\tableofcontents

\section*{Introduction}

In 1982 Feynman introduced the concept of a quantum emulator,  as a possibility to circumvent the difficulty of simulating quantum physics with classical computers \cite{Feynman:1982}. His idea, based on the universality of quantum mechanics, was to use one controllable device to simulate other systems of interest. Nowadays Feynman's intuition is being implemented in various setups and among them, cold gases of neutral atoms play a central role \cite{Bloch:2008,Buluta:2009}. These gases constitute remarkably flexible playgrounds. They can be formed of bosons, fermions, or mixtures of both. Their environment can be controlled using the potential created by laser light, with harmonic, periodic, quasi-periodic or disordered energy landscapes. Interactions between particles can be adjusted using scattering resonances.  At first sight the only missing ingredient for implementing Feynman's idea with dilute atomic gases is the equivalent of orbital magnetism, which would allow one to simulate phenomena such as Quantum Hall effect. Here we discuss a general path that has recently emerged to fill this missing item, by using atom-light interaction to generate artificial gauge potentials acting on neutral matter.  The major part of this Colloquium addresses the case of a gauge field which has the U(1) Abelian symmetry, like electromagnetism, but we also discuss more elaborate configurations leading to non-Abelian gauge potentials.    

From the quantum mechanics point of view, the orbital magnetism of a particle with charge $e$ can be viewed as a consequence of the Aharonov--Bohm phase $\gamma$ acquired by the particle when it travels along a closed contour ${\cal C}$ \cite{Aharonov-Bohm59PR}. This phase has a geometric origin and does not depend on the duration needed for completing the trajectory. It reads $\gamma=2\pi \Phi/\Phi_0$, where $\Phi$ is the flux of the magnetic field through the contour ${\cal C}$ and $\Phi_0=h/e$ is the flux quantum. Therefore the quest for artificial magnetism amounts to realize situations where a neutral particle acquires a geometrical phase when it follows the contour ${\cal C}$. Similarly the generation of a non-Abelian field can be achieved for particles with an internal degree of freedom. After completion of the trajectory along ${\cal C}$, the initial internal state of the particle  $|\psi_i\rangle$ is changed into $U|\psi_i\rangle$, where $U$ is a unitary operator acting in the internal Hilbert space of the particle and depending only on the geometry of the contour. Note that through all this Colloquium the artificial gauge potentials that appear are not dynamical variables, i.e.  they are not influenced by the motion of the atoms.

One of the simplest examples of geometric phases, the so-called Berry's phase,  plays a central role in this paper. It emerges for example when a neutral particle with magnetic moment  $\bs \mu$ moves in a (real) non-homogenous 
magnetic field $\bs B_0(\bs r)$ \cite{Berry:1984}. If the particle is prepared at a point $\bs r_0$ in one of the local eigenstates $|m(\bs r_0)\rangle$ of the Hamiltonian $-\bs \mu\cdot \bs B_0(\bs r)$ and moves slowly enough, it follows adiabatically the local eigenstate $|m(\bs r_t)\rangle$. Once the trajectory along ${\cal C}$ is completed, the particle  returns to the internal state $|m(\bs r_0)\rangle$,  up to a phase factor containing a geometric component.

Berry's phase is also present in atom-light interactions \cite{Dum:1996}. The role of the magnetic states $|m(\bs r)\rangle$ is now played by the dressed states, \emph{i.e.} the eigenstates  of the atom-light coupling. The dressed states can vary on a short spatial scale (typically the wavelength of light) and the artificial gauge fields can be quite intense, in the sense that the geometric phase is large compared to $2\pi$ for a contour encircling a gas of realistic size. If the gas is superfluid, it will thus contain several vortices at equilibrium.  The situation considered here should not be confused with phase-imprinting methods, which correspond to imposing a dynamical rather than a geometrical phase to the atomic wavefunction [see \textcite{Andersen:2006} and refs. in].  In the latter case one acts on the atoms with a time-dependent  atom-light coupling, and the spatial phase profile of the light beam is dynamically transferred to the atom cloud. Here by contrast we look for a time-independent Hamiltonian and require that its ground state has a non-zero vorticity. 

The use of geometrical phases is not the only way to reach a Hamiltonian with an artificial gauge field for neutral particles. Rotating the system at angular frequency $\Omega$ around a given axis (say $z$) also leads to artificial magnetism in the rotating frame, with $B_z\propto\Omega$.  This has been widely used in the context of quantum gases [see \textcite{Cooper:2008} and \textcite{Fetter:2008} for reviews] and it is well suited when the confining potential is rotationally invariant around the $z$ axis. The Hamiltonian is then time-independent in the rotating frame, and the standard formalism of equilibrium statistical physics is applicable. However if the confining potential has a non-zero anisotropy in the lab frame, the Hamiltonian is time-dependent in any frame and the situation is difficult to handle from a theoretical point of view. By contrast, methods based on geometrical phases do not impose any constraint on the symmetry properties of the initial Hamiltonian and the artificial gauge fields are produced in the laboratory frame. Together with the possibility to extend the scheme to non-Abelian gauge potentials, this represents a significant advantage. Of course the use of laser beams also comes with some drawbacks that we will meet in this Colloquium, such as heating of the atoms because of residual spontaneous emission. 

This Colloquium is organized as follows. In Sec.~\ref{sec:two-level} we present a {toy model}, where the atomic transition that is coupled to the laser light is represented by a simple two-level system. It allows us to identify the important elements for the emergence of artificial gauge potentials: gradients of the phase of the light beam and of its intensity or detuning with respect to atomic resonance. This model is relevant for some atomic species like Ytterbium atoms that possess an electronic excited state with an extremely long lifetime, but it cannot be used as such for the the alkali-metal atoms, which are the most commonly used species in experiments.  Indeed the large rate of spontaneous emission processes and the ensuing random recoils of the atoms cause in this case a prohibitive heating of the gas. In Sec.~\ref{sec:Three-level-system} we thus turn to schemes where the ground state manifold is degenerate,  first introduced by \textcite{Dum:1996},  and we show that a gauge potential with a non-vanishing curl can then be generated  even if the  population of the excited states is negligible \cite{Juzeliunas:2004}.  We also discuss the recent implementation of an artificial magnetic field in a Bose--Einstein condensate of rubidium atoms by \textcite{Lin2009b}, which led to the observation of quantized vortices. Section \ref{sec:Non-Abelian} is devoted to the preparation of non-Abelian gauge fields and to the discussion of some of their physical consequences: negative refraction and reflection, implementation of the Klein paradox and the non-Abelian Aharonov--Bohm effect. In Sec.~\ref{sec:opticallattices} we study the production of artificial gauge potentials in optical lattices. Starting from the proposal by \textcite{Jaksch:2003}, we show that one can reach the strong magnetic field regime,  where the  phase $\gamma$ per lattice cell can take any value between 0 and $2\pi$. We also briefly discuss the implementation of non-Abelian schemes in a lattice. 

Throughout this Colloquium, we mostly address single atom physics, with some incursions to many-body physics at the mean-field level when dealing with the generation of quantized vortices in a superfluid. The application of a gauge field on a quantum gas can also lead to the emergence of strongly correlated many-body states. Their detailed discussion is outside the scope of this Colloquium, and we simply indicate a few lines of research in this direction in the final Outlook section. 


\section[Toy model]{Toy model: Two-level atom in a light beam}
\label{sec:two-level}

In order to present the essential ingredients of the physics of geometrical gauge fields, we start this Colloqium by discussing the simplest scheme for which artificial magnetism can occur. We consider a single quantum particle with a two-level internal structure, and we show how the adiabatic following of one particular internal state can provide the desired gauge fields. This will allow us to present with a minimum algebra the important physical concepts, which will subsequently be generalized to more complex schemes.  

We denote $\{|g\rangle,|e\rangle\}$ a basis of the 2-dimensional Hilbert space associated with the internal degree of freedom of the particle.  Later on these states will represent electronically ground and excited states of an atom, respectively. We assume that the particle evolves in space-dependent external fields that couple $|g\rangle$ and $|e\rangle$. At the present stage we do not specify the physical origin of these fields. In practice these can be the fields of optical lasers, microwave fields and/or static electric or magnetic fields acting on the electric or magnetic dipole moment of the particle. The general form for the Hamiltonian of the particle of mass $M$  is 
\begin{equation}
H=\left(\frac{P^2}{2M}+V\right)\,\hat 1+U \,,
\label{eq:hamiltonian1}
\end{equation}
where $\bs P=-i\hbar \bs \nabla$ is the momentum operator and $\hat 1$ is the identity operator in the internal Hilbert space. The coupling operator $U$ can be written in the matrix form 
\begin{equation}
U= \frac{\hbar \Omega}{2} \begin{pmatrix} 
\cos\theta & e^{-i\phi}\sin\theta \\
 e^{i\phi}\sin\theta& -\cos\theta
\end{pmatrix}\,.
\label{eq:hamiltonian2}
\end{equation} 
The particle dynamics is determined by four real quantities that may all depend on the position vector $\bs r$: The potential $V$ acts on the particle in a way that is independent of its internal state, whereas the \emph{generalized Rabi frequency} $\Omega$ characterizes the strength of the coupling that lifts the degeneracy between $|g\rangle$ and $|e\rangle$. The two remaining quantities are the \emph{mixing angle} $\theta$ and the \emph{phase angle} $\phi$. For a two-level atom in a monochromatic laser field [see  Eq.~(\ref{eq:matU}) below], $\Omega \cos\theta$ stands for the laser detuning from the atomic resonance, $\Omega \sin\theta$ is the magnitude of the atom-laser coupling and $\phi$ is the laser phase. 

In this section we first describe the atomic dynamics when the internal state of the particle follows adiabatically one of the eigenstates of $U$, and we give the expression of the geometrical gauge  potentials that appear in this case. Then we present a possible implementation of the Hamiltonian  (\ref{eq:hamiltonian1})-(\ref{eq:hamiltonian2}) with an alkaline-earth atom irradiated by a quasi-resonant laser beam, and we discuss the physical origin of the gauge potentials.  We also study under which condition the strength and the spatial extent of the geometric magnetic field are sufficient to induce a large circulation of the atomic phase. This can then allow for the nucleation of a lattice of quantized vortices if one applies this scheme to a collection of identical atoms forming a superfluid. 

\subsection{Adiabatic following of  a dressed state }
\label{subsec:adiabfollowing}

At a point $\bs r$  the eigenstates of the matrix $U$ are
 \beq
|\chi_1\rangle=\bpm \cos(\theta/2) \\ e^{i\phi}\sin(\theta/2) \epm
\, , \quad
|\chi_2\rangle=\bpm -e^{-i\phi}\sin(\theta/2) \\ \cos(\theta/2) \epm\, ,
 \label{eq:eigenstates}
 \eeq
with eigenvalues $\hbar\Omega/2$ and $-\hbar\Omega/2$, respectively. We will call them \emph{dressed states}, anticipating on the following discussions where they will correspond to the local eigenstates of the Hamiltonian describing the coupling between an atom and a light field. Since  the states $\{|\chi_j\rangle\}$  form a normalized, orthogonal basis, the quantity $i\langle \chi_j|\bs \nabla\chi_j\rangle$ is a real number and the relation $\langle \bs \nabla \chi_2 |\chi_1\rangle=-\langle \chi_2|\bs \nabla \chi_1\rangle$ holds, where  we set $|  \bs \nabla\chi_j\rangle\equiv\bs \nabla\left( |\chi_j\rangle\right)$.

Using the $\{|\chi_j\rangle\}$ basis for the internal Hilbert space, we can write the full state-vector of the particle as 
 \beq
|\Psi(\bs r,t)\rangle=\sum_{j=1,2}\psi_j(\bs r,t)|\chi_j(\bs r)\rangle\,.
 \eeq 
Suppose now that at initial time the particle is prepared in one particular internal dressed state, say $|\chi_1\rangle$. If the velocity distribution of the particle involves only sufficiently small components, the internal state of the particle  will remain proportional to $|\chi_1\rangle$ for all times. This is equivalent to the Born--Oppenheimer approximation in molecular physics: The position $\bs r$ of the particle and its internal degree of freedom play the role of the nuclear coordinates and of the electron dynamics, respectively. 

We now derive the equation of motion for $\psi_1$ in the case where $\psi_2$ remains negligible at all time. 
We consider first the action of the momentum operator $\bs P$ on the full state-vector $|\Psi\rangle$. 
Employing $\bs \nabla \left[ \psi_j\,|\chi_j\rangle \right]= \left[\bs \nabla \psi_j\right]\,|\chi_j\rangle + \psi_j|\bs \nabla \chi_j\rangle$ and the completeness relationship, one has  
  \beq
\bs P |\Psi\rangle=\sum_{j,l=1}^2 \left[ \left(\delta_{j,l}\bs P - \bs A_{jl}\right)\psi_l\right] |\chi_j\rangle\,,
 \eeq
 with $\bs A_{jl}(\bs r)=i\hbar \langle \chi_j | \bs \nabla\chi_l\rangle $. Assuming that $\psi_2=0$, we project the Schr\"odinger equation  $i \hbar |\dot\Psi \rangle=H|\Psi \rangle$ onto the internal dressed state $|\chi_1\rangle$, where $H$ is the full Hamiltonian (\ref{eq:hamiltonian1}). This leads to a closed equation for the probability amplitude $\psi_1$ to find the atom in the first dressed state \cite{Mead:1979,Jackiv88CAMP,Berry:1989,Moody:1989,Mead:1992}:
 \beq
i\hbar \frac{\partial \psi_1}{\partial t}= \left[
 \frac{({\bs P}-\bs A)^2}{2M} + V
 +\frac{\hbar \Omega}{2}+W\right]
\psi_1\,.
 \label{eq:motion}
 \eeq
 In addition to the terms $V$ and $\hbar \Omega/2$ that were already explicit in the initial Hamiltonian (\ref{eq:hamiltonian1}), two geometric potentials $\bs A$ and $W$ emerge in the adiabatic elimination of the state $|\chi_2\rangle$, due to the position dependence of the internal dressed states. The first one is the vector potential  
\begin{equation}
\bs A(\bs r)=i\hbar \langle \chi_1|  \bs \nabla\chi_1\rangle 
= \frac{\hbar}{2}\left(\cos \theta-1\right)\bs \nabla \phi\,.
\label{eq:vect}
\end{equation}
 The effective magnetic field associated with $\bs A$ is
\begin{equation}
\bs B(\bs r)= \bs \nabla \times \bs A= \frac{\hbar}{2}\, \bs \nabla (\cos\theta)   \times \bs \nabla
 \phi\,.
 \label{eq:chpmag}
\end{equation}  
When this artificial magnetic field is nonzero, the vector potential $\bs A$ cannot be eliminated from (\ref{eq:motion}) by a gauge transformation. The particle acquires an effective charge (that we set equal to 1 by convention) and its motion exhibits the usual features associated with orbital magnetism. A nonzero value of $\bs B$ occurs only if both the mixing angle $\theta$ and the phase angle $\phi$ vary in space with non collinear gradients. The second geometrical potential appearing in (\ref{eq:motion}) is the positive scalar term
\begin{equation}
W (\bs r)=\frac{\hbar^2}{2M} |\langle \chi_2| \bs \nabla
\chi_1\rangle|^2= \frac{\hbar^2}{8M}\left[
 (\bs \nabla\theta)^2+\sin^2 \theta (\bs \nabla \phi)^2
 \right] .
\label{eq:scalar}
\end{equation}
The first experimental evidence for the scalar potential in Quantum Optics was given by \textcite{Dutta:1999}. As such, the geometric scalar potential
is not a very useful tool in the sense that there exist many other ways to create a scalar potential on atoms, using for example the  AC Stark shift created by a far-detuned laser beam \cite{Grimm:2000}. However in the following we will estimate the value of the scalar potential for some relevant configurations, because it must be taken into account to determine the equilibrium shape of the cloud. 

The contributions of the vector and scalar potentials in Eq.~(\ref{eq:motion}) can be recovered in a systematic expansion in terms of the dimensionless adiabatic parameter, defined as the ratio between the short and long time scales \cite{Weigert:1993,Littlejohn:1993a}. The term $-\bs P\cdot \bs A/M$ appears at first order of the expansion, whereas $\bs A^2/2M$ and the scalar potential $W$ appear at second order. An extra term that is quadratic  with respect to $\bs P$ also shows up at second order, with a contribution that can reach that of the scalar potential when $P^2/M \sim \hbar \Omega$. We will not address this extra term in this review, because (i) we are mostly interested in the physics arising from the leading term of the expansion $-\bs P\cdot \bs A/M$ and (ii) we restrict to situations where the atomic kinetic energy is much smaller than $\hbar \Omega$.

The reason for which $\bs A$, $\bs B$ and $W$ are called `geometrical' is clear from expressions (\ref{eq:vect}), (\ref{eq:chpmag}) and (\ref{eq:scalar}), which depend only on the spatial variation of the angles $\theta$ and $\phi$, i.e. of the geometry of the coupling between $|g\rangle$ and $|e\rangle$, but not on the strength $\Omega$ of this coupling. 
Note that if we consider the adiabatic following of $|\chi_2\rangle$ instead of $|\chi_1\rangle$, the equation of motion for $\psi_2$ contains the same scalar potential $W$ and the opposite vector potential $-\bs A$. 

\subsection{Practical implementation with alkaline-earth atoms}
\label{subsec:alkalineearth}

We now discuss how the above model can be implemented in quantum optics, in order to create artificial orbital magnetism on a gas of cold neutral atoms  \cite{Dum:1996,Visser:1998}.  The simplest scheme consists in shining a single laser beam on an atom. We restrict the internal atomic  dynamics to a two-level transition of frequency $\oA$, between the ground state $|g\rangle$ and an electronically excited state $|e\rangle$. The laser light of frequency $\oL$ is supposed to be close to resonance with this transition. We suppose that the rate of spontaneous emission of photons from the excited state $|e\rangle$ is negligible on the relevant time scale, which is a realistic assumption if the experiment is performed using the intercombination line of alkali-earth atoms (calcium, strontium) or  Ytterbium atoms. Indeed for these atomic species, the radiative lifetime of the excited state $|e\rangle$ involved in the intercombination line  [also used in optical atomic clocks \cite{Ye:2008}]  is several seconds or even tens of seconds, large compared to the typical duration of cold atoms experiments. 

\begin{figure}[t]
\begin{center}
\includegraphics[width=6cm]{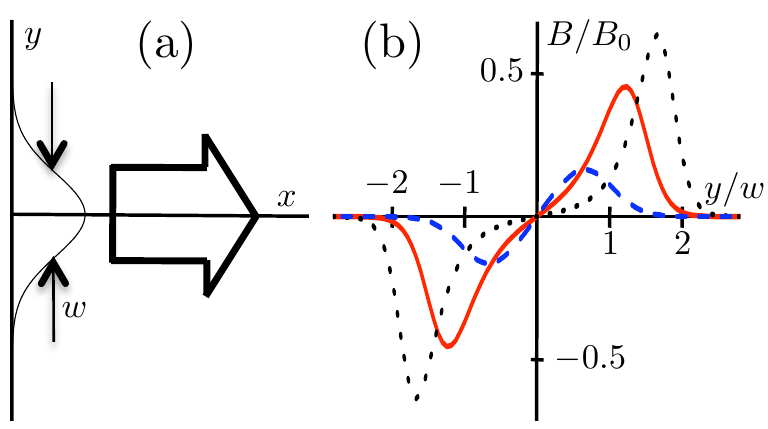}
\caption{(Color online) (a) A running wave propagating along the $x$ axis, with a gaussian profile along the $y$ axis (waist $w$) is used to create a geometrical gauge field on a two-level atom whose resonance frequency is close to the laser frequency. (b) Variation of the amplitude of the artificial magnetic field $B$ (in units of $B_0=\hbar k/w$) as a function of $y/w$;  $\kappa^{(0)}/\Delta=1$ (blue), 5 (red), 20 (black). }
\label{fig:Bgeom}
\end{center}
\end{figure}

We suppose that the atomic center-of-mass motion is restricted to the $xy$ plane using a suitable trapping potential that freezes the $z$ degree of freedom. The mode of the laser beam is a Gaussian traveling wave of wavenumber $k$ and wavelength $\lambda=2\pi/k$ propagating along the $x$ axis, with a waist $w$ in the $y$ direction (Fig.~\ref{fig:Bgeom}a). The states $|g\rangle$ and $|e\rangle$ stand for eigenstates of the internal atomic Hamiltonian in the absence of coupling with the radiation field. Using the rotating wave approximation, the coupling matrix $U$ can be written in the present case \cite{cohe92}
\begin{equation}
U=\frac{\hbar}{2}\bpm
\Delta & \kappa^*\\ \kappa  & -\Delta
\epm \,,
\label{eq:matU}
\end{equation}
where the detuning $\Delta=\oL-\oA$ and where the Rabi frequency $\kappa$ characterizes the strength of the atom-laser coupling. The spatially varying phase of the laser is incorporated into $\kappa$ which is therefore complex. 

In the following we neglect the diffraction of the laser beam and set $\kappa(\bs r)=\tilde \kappa(y)\,e^{ikx}$, where $\tilde \kappa$ is real and positive. The angle $\phi$ featured in Eq. (\ref{eq:hamiltonian2}) is then simply the running phase $kx$ of the propagating laser beam and $\bs \nabla \phi=k \,\bs e_x$, where $\bs e_x$ is the unit vector in the $x$ direction. The mixing angle $\theta$ is given by $\tan\theta=\tilde \kappa/\Delta$ and  two options are available to provide a non-zero $\bs \nabla \theta$. One can either use a gradient of the Rabi frequency $\tilde \kappa$ produced by a spatial variation in the laser intensity, or a gradient of the detuning $\Delta$. In this subsection we discuss these two configurations, which we refer to as ``grad-$\kappa$'' and ``grad-$\Delta$'' configurations, respectively.

\paragraph{Grad-$\kappa$ configuration.} A possible scheme leading to this configuration is shown in Fig.~\ref{fig:Bgeom}, where we take advantage of the transverse Gaussian profile of the laser beam. The Rabi frequency is
$\tilde \kappa (y)=\kappa^{(0)}\,e^{-y^2/w^2}$, 
which leads to an effective magnetic field $\bs B$ that is parallel to the $z$ axis, with the amplitude obtained  using Eq.  (\ref{eq:chpmag}):
\begin{equation}
B=B_0\, \frac{\Delta\tilde \kappa^2}{\Omega^3}\,\frac{y}{w}\,,
\label{eq:Btwolevel}
\end{equation}
with $\Omega^2=\Delta^2+\tilde \kappa^2$. Here we set $B_0=\hbar k/w$, which gives the typical scale for the effective magnetic field appearing in this case. The characteristic length scale associated to orbital magnetism is the so-called \emph{magnetic length} $\ell_B=(h/B)^{1/2}$, which gives the size of the elementary quantum cyclotron orbit. We obtain in this context $\ell_B \sim (w\lambda)^{1/2}$ (hence $\ell_B>\lambda$) for $B\sim B_0$. 

The variation of $B$ with $y/w$ is plotted in Fig.~\ref{fig:Bgeom}b for various values of the ratio $\kappa^{(0)}/\Delta$. When $\kappa^{(0)}/\Delta\gg 1$ the maximal value of  $B$ is obtained approximately at the point where $\tilde \kappa= \Delta$, i.e. $y_{\rm max}\approx w\, [\log(\kappa^{(0)}/\Delta)]^{1/2}\gg w$ and $B_{\rm max}\approx B_0 y_{\rm max}/(2\sqrt 2 w)$.  At first sight the limiting regime $\kappa^{(0)}/\Delta\to \infty$ is interesting since it leads to $B_{\rm max}\to \infty$. However the interval  $\Delta y$ over which $B$ takes significant values tends to zero as $1/B_{\rm max}$, and eventually becomes smaller than the magnetic length $\ell_B\propto 1/ B_{\rm max}^{1/2}$, making this regime $\kappa^{(0)}\gg \Delta$ not useful in practice. More generally we find that
\begin{equation}
\int_0^{+\infty} B\,dy=\frac{\hbar k}{2}\left[
1-\frac{1}{\sqrt{1+(\kappa^{(0)}/\Delta)^2}}
\right] < \frac{\hbar k}{2}\,.
\label{eq:inequality}
\end{equation}
Consider for example the case $\kappa^{(0)}/\Delta=5$. The maximum value $B_{\rm max}\approx 0.45\,B_0$ is obtained for $|y|/w\approx 1.2$, and $B>B_0/4$ over an interval of width $\sim w/2$.  

The calculation of the scalar potential leads to
\begin{equation}
W(y) = \frac{E_{\rm R}}{4}
\frac{\tilde \kappa^2(y)}{\Delta^2+\tilde\kappa^2(y)}\left[
1+\frac{4y^2}{w^4 k^2}\frac{\Delta^2}{\Delta^2+\tilde \kappa^2(y)}
\right]\,,
\label{eq:Wtwolevel}
\end{equation}
where we define the recoil energy $E_{\rm R}=\hbar^2k^2/2M$  as the kinetic energy of an atom initially at rest when it absorbs or emits a single photon. Realistic values of the waist $w$ are such that $kw\gg 1$. When $\kappa^{(0)}\lesssim \mbox{a few }\Delta$, the main contribution to the scalar potential (\ref{eq:scalar})  is the $(\bs \nabla \phi)^2$ term, which corresponds to the first term in the bracket of Eq. (\ref{eq:Wtwolevel}). 
The scalar potential then creates a potential bump that is maximal in $y=0$, with a height scaling like $E_{\rm R}$. When  $\kappa^{(0)}/ \Delta \gg 1$, the contribution proportional to $(\bs \nabla \theta)^2$ to the scalar potential becomes dominant [second term in the bracket of Eq. (\ref{eq:Wtwolevel})]. The corresponding force is now very large around the point $y_{\rm max}$ and may create strong distortions of the trapped atom cloud.

\paragraph{Grad-$\Delta$ configuration.} We suppose now that the laser waist $w$ is very large, so that the spatial mode of the laser is well approximated by a plane wave and the Rabi frequency $\tilde{\kappa}$ can be taken uniform. We also assume that a gradient of detuning is present along the $y$ axis, so that $\Delta = \Delta' (y-y_{0})$. For alkaline-earth or Ytterbium atoms, it can be created using an additional off-resonant laser, which produces a differential light shift for the $g$ and $e$ states. The effective magnetic field $\bs B$ is parallel to the $z$ axis, with an amplitude given by 
\begin{equation}
B=B_0\, {\cal L}^{3/2}(y)\, , \qquad
{\cal L}(y)=\frac{1}{1+(y-y_{0})^2/\ell_\kappa^2}\, ,
\label{eq:BtwolevelDet}
\end{equation}
where we introduced the characteristic length $\ell_{\kappa}=\tilde{\kappa}/ \vert \Delta' \vert$ and set $B_0=\hbar k/(2\ell_\kappa)$. The scalar potential is
\begin{equation}
\label{eq:WtwolevelDet}
W(y)=\frac{E_{\rm R}}{4}\left( 
{\cal L}(y) + \frac{1}{k^2\ell_\kappa^2}{\cal L}^2(y) \right)
\,.
\end{equation} 
The magnetic field can be made arbitrarily large by taking $\Delta'\to \infty$, but as for the grad-$\kappa$ case this is not relevant in practice. Indeed the field then takes a large value only over a domain whose size $\ell_\kappa \propto 1/\Delta'$ becomes smaller than the magnetic length $\ell_B=(\ell_\kappa \lambda/2)^{1/2}$. 

\subsection{Validity of the adiabatic approximation}
\label{subsec:validity}

We now discuss the validity of the adiabatic approximation \cite{Messiah_adiab} underlying the projection mechanism generating the gauge potentials. For simplicity, we adopt a semi-classical point of view in which we treat classically the atomic center-of-mass motion, but still use quantum mechanics to describe its internal dynamics. The atom  is supposed to be initially at rest  in the internal state $|\chi_1\rangle$, and then accelerated in order to reach the velocity $\bs v$. The population $\Pi_{2}$ of the state $|\chi_2\rangle$ at the end of the acceleration process is not strictly zero.  For small $\vert {\bs v}\vert$ it is given by  $\Pi_{2}\approx \left| {\bs v\cdot \langle \chi_2 | \bs \nabla \chi_1\rangle/\Omega}\right |^2$, where the motional coupling is $\langle \chi_2 | \bs \nabla \chi_1\rangle = \left[ \bs \nabla \theta -i \sin\theta\, \bs \nabla \phi \right]e^{-i\phi}/2$ \cite{Cheneau:2008}.

We first address the grad-$\kappa$ implementation discussed above, for which $\bs \nabla \theta =\Delta \,\bs \nabla \tilde \kappa\,/\,(\Delta^2+\tilde \kappa^2)$, and we restrict to the practical case where $\kappa^{(0)}\lesssim \mbox{a few }\Delta$. Since the gradient of the laser intensity always occurs with the spatial scale $w\gg k^{-1}$, we find that $|\bs\nabla \theta|\ll k$. The motional coupling is dominated by the contribution of $\bs \nabla \phi$ and reaches the maximum value $| \langle \chi_2 | \bs \nabla \chi_1\rangle \approx k \vert \sin(\theta) \vert = k \vert \kappa^{(0)}/\Omega \vert$. Using the fact that the atomic velocity changes by an amount equal to the recoil velocity $v_{\rm R}=\hbar k/M$ during each photon absorption or emission process, a necessary condition for the adiabatic approximation to hold ($\Pi_{2}\ll 1$)  is thus
\begin{equation}
\hbar \Omega \gg \sqrt{\hbar\kappa^{(0)}E_{\rm R}}.
\label{eq:validity}
\end{equation}
In most situations of practical interest, $\kappa^{(0)}\sim \Delta$ and Eq.~(\ref{eq:validity}) reduces to the intuitive condition 
\begin{equation}
\hbar\kappa^{(0)}\gg E_{\rm R},
\label{eq:validity_intuitive}
\end{equation}
stating that the coupling strength should be much larger than the recoil energy. Of course the validity condition for the adiabatic approximation is more stringent if the atomic velocity is notably larger than $v_{\rm R}$. 

We now address the case of the grad-$\Delta$ scheme. Far away from any resonance point where $\Delta$ vanishes, the adiabaticity condition (\ref{eq:validity}) still holds. Near a resonance point, the validity condition for the adiabatic approximation may be more stringent, since the gradient of the mixing angle $|\bs \nabla \theta|\sim 1/\ell_{\kappa}$ can be large. More precisely suppose that on each side of the resonance point, $|\Delta|$ is large compared to the coupling $\tilde\kappa$. The dressed states then nearly coincide with the bare states, with either \{$|\chi_1\rangle=|e\rangle$, $|\chi_2\rangle=|g\rangle$\} or \{$|\chi_1\rangle=|g\rangle$, $|\chi_2\rangle=|e\rangle$\}. The switch between these two configurations occurs within a region of size $\sim l_{\kappa}$ around the resonance point. Taking again the atomic velocity on the order of the recoil velocity $v_{\rm R}$, the adiabatic condition reads
\begin{equation}
\hbar\tilde\kappa \frac{k \ell_{\kappa}}{\left[1+\left( k \ell_{\kappa}\right)^2\right]^{1/2}} \gg E_{\rm R} \,.
\label{eq:adiab_gradDelta}
\end{equation}
When $k \ell_{\kappa}\gg 1$, this reduces to Eq.~(\ref{eq:validity}) in $y=y_0$, i.e.  $\hbar \tilde \kappa \gg E_{\rm R}$. In the opposite limit $k \ell_{\kappa}\ll 1$, the peak value of $B$ is very large, but it is reached only in a small region of width $\ell_{\kappa}\ll k^{-1}$ and the adiabaticity condition $\tilde{\kappa}\ell_{\kappa} \gg v_{\rm R} $ is more demanding. This is thus not a convenient configuration for a continuum implementation, but it becomes relevant for the lattice case (Sec.~\ref{sec:opticallattices}). 

\subsection{Physical interpretation of the geometric potentials}
\label{subsec:physicalinterpretation}

We now turn to the physical interpretation of the geometrical gauge fields. We focus our discussion on the two-level system of Sec.~\ref{subsec:adiabfollowing} and Sec.~\ref{subsec:alkalineearth}, but the following physical images can be generalized to the schemes that will be analysed in the next sections.

The scalar potential $W$ can be interpreted as the kinetic energy associated with the fast micromotion of the particle. This was first explained for a classical continuous internal degree of freedom by \textcite{Aharonov:1992}. Here we outline the reasoning of  \textcite{Cheneau:2008} that addresses the case of a quantized internal degree of freedom. Consider a particle prepared in the internal state $|\chi_1\rangle$, with a center-of-mass state that consists in a wave packet localized around a given point $\bs r$ with an extension small compared to the scale of variation of $\theta$ and $\phi$. We introduce the force operator $\bs F=-\bs \nabla U$ and note that the eigenstates $|\chi_j\rangle$ of the coupling $U$ are not eigenstates of $\bs F$. The force acting on the particle thus exhibits quantum fluctuations, $\langle \bs F^2\rangle \neq \langle \bs F\rangle^2$, which are characterized in the Heisenberg picture by the symmetrized correlation function 
\begin{eqnarray}\nonumber
C(\tau)&=&\frac{1}{2}\langle \bs{\delta F}(0)\cdot\bs{\delta F}(\tau)+\bs{\delta F}(\tau)\cdot\bs{\delta F} (0)\rangle\,,
\label{eq:corfunction}\\
&=& \hbar^2\Omega^2 \left| \langle \chi_2|\bs \nabla \chi_1\rangle \right| ^2 \,\cos(\Omega \tau)\,.
\end{eqnarray}
with $\bs{\delta F}(\tau)=\bs F(\tau)-\langle \bs F\rangle$ . We now recall that in classical mechanics, a particle submitted to a rapidly oscillating force ${ \bs{\delta F}}$  undergoes a micromotion with the average kinetic energy 
\begin{equation}
E_{\rm K}=\int \frac{\tilde C(\omega)}{2M\omega^2}\,d\omega\,,
\label{eq:micromotion}
\end{equation}
where $\tilde C(\omega)$ is the Fourier transform of $C(\tau)$. Inserting the value (\ref{eq:corfunction}) of $C(\tau)$ into (\ref{eq:micromotion}), we find that $E_{\rm K}$ coincides with the scalar geometric potential given in (\ref{eq:scalar}). 

The vector potential $\bs A$ given in (\ref{eq:vect}) is related to the Berry's phase $\gamma$ that appears when a quantum system -- here, the two-state system associated with the internal degree of freedom of the particle -- is slowly transported round a contour ${\cal C}$, while remaining in an eigenstate $|\chi(\bs r)\rangle$ of its Hamiltonian \cite{Berry:1984}:
\begin{equation}
\gamma({\cal C})=i\oint \langle \chi|\bs \nabla \chi\rangle \cdot d\bs r=\frac{1}{\hbar}\oint \bs A\cdot d\bs r
 \,,
\label{eq:phaseBerry}
\end{equation}
where the second equality holds for the two-level system considered above when it is prepared in the state $|\chi\rangle$.  

When $\bs B=\bs \nabla \times \bs A\neq 0$, the Lorentz force $\bs F=\bs v\times \bs B$ acts on the atom when it moves with velocity $\bs v$. For the scheme discussed in Sec.~\ref{subsec:alkalineearth}, the momentum change imparted by the Lorentz force has a simple physical interpretation. Suppose that the atom moves along the $y$-axis, with a trajectory starting from $y_1\gg w$ at time $t_1$, and ending in $y_2=0$ at time $t_2$. Since $\bs B$ is directed along the $z$ direction and $\bs v$ along the $y$ direction, the average momentum change $\langle \bs{\Delta P}\rangle$ is directed along   $x$:
\begin{equation}
\langle \Delta P_x\rangle = -\int_{t_1}^{t_2} v_y(t) B\,dt = \int_0^{y_1}B\,dy \,.
\label{eq:DeltaPx{}}
\end{equation}
Using (\ref{eq:inequality}) and $y_1\gg w$, we find that $\Delta P_x\approx \hbar k/2$ in the case where $\kappa^{(0)}\gg\Delta$. This result has a simple physical interpretation. When the atom is located in $y_1\gg w$, the occupied dressed state is $\approx |g\rangle$. When the atom arrives at $y_2=0$ where the atom-laser coupling is maximal, the dressed state is $ \approx \left( |g\rangle+ |e\rangle e^{ikx} \right)/\sqrt 2$ in the limit $\kappa^{(0)}\gg\Delta$. Hence a measurement of the atomic momentum $\Delta P_x$ can give the results $0$ or $\hbar k$, both with probability $1/2$, hence $\langle \Delta P_x\rangle =\hbar k/2$. A more detailed discussion of the physical origin of the Lorentz force for other atomic trajectories is given by \textcite{Cheneau:2008}.

\subsection{Achieving states with a nonzero circulation}
\label{subsec:circulation}

After this discussion of some simple schemes that generate artificial gauge potentials, we now turn to the effect of this potential on the external (center-of-mass) degree of freedom of the atom. One of the main motivations for the generation of artificial magnetic fields is  indeed to create some extended regions where the orbital magnetism is sufficient to favor states with a nonzero circulation. For instance, if a superfluid is placed in such a region, its ground state will exhibit a vortex lattice. We now explore under which condition this can occur for the simple schemes outlined above. 

When a charged superfluid (with $e=1$ here by convention) is placed in a magnetic field, the vortex density is  $\rhov=B/(2\pi \hbar)$ or in other words $\rho_v=\ell_B^{-2}$, wher $\ell_B$ is the magnetic length \cite{tink96}. If the magnetic field obtained from the geometric potential $\bs A$ keeps a value $\sim B$ on a disk of radius $r$, one therefore expects that $N_{\rm v}\approx \pi r^2 \rhov=r^2B/(2\hbar)$ vortices will be present in steady state in a superfluid filling this disk. We now wish to determine whether one can reach a situation with $N_{\rm v}\gg 1$, which is equivalent to requiring that the phase $\gamma({\cal C})$ defined in (\ref{eq:phaseBerry}) is large compared to $2\pi$.

Consider again the grad-$\kappa$ scheme represented in Fig.~\ref{fig:Bgeom}, where we choose for example $\kappa^{(0)}=5\,\Delta$. One gets in this way a fictitious magnetic field that is approximately uniform with a value $\sim B_0/4$ in the band parallel to the $x$ axis, centered on $y=1.2\,\,w$, with a width $\ell_y\approx w/2$. The length of this band along $x$ is limited only by the diffraction of the laser beam, which occurs on a distance $\gg w$, if the waist $w$ is chosen much larger than the laser wavelength $\lambda=2\pi/k$.  In order to study the physics of vortex lattices in this geometry, the requirement is thus simply to fit several vortex rows in the width $\ell_y$. Since the distance between two vortex rows is $\approx \rhov^{-1/2}$, this requirement can be written
\begin{equation}
N_{\rm vortex\ rows}\approx \frac{1}{4}\sqrt{w/\lambda}\gg 1\,.
\label{eq:Nvortexrows}
\end{equation} 
Clearly this method is well adapted to the study of vortex lattices only if $w \gg \lambda$. The choice of a small waist (of the order of a few $\lambda$ only) is not appropriate, because the scaling $B_0\propto 1/w$, which would tend to favor small waists, is compensated by the other scaling $\ell_y \sim w$ over which the field keeps a significant value. A similar argument can be made for the grad-$\Delta$ scheme with $\ell_y=\ell_\kappa$.

So far we have restricted our discussion to the case of a single laser travelling wave and the spatial scale of variation for the mixing angle $\theta$ is thus the beam waist $w$. It is interesting to consider also the case where several traveling waves irradiate the atom at different angles, so that interference phenomena can introduce a much shorter length scale for $\theta$, typically $\lambda/(2\pi)$. For simplicity we restrict to the case of two waves and we choose the corresponding wavevectors equal to  $\bs k_\pm=k (\bs e_x \pm \bs e_y)/\sqrt 2$. The resulting light field still has a spatially varying phase $\phi=kx/\sqrt 2$, and it presents an interference pattern along the $y$ direction with a spatial period $\lambda/\sqrt 2$. Hence $|\bs \nabla \phi|\sim |\bs \nabla \theta |\sim k/\sqrt 2$, and we find using (\ref{eq:chpmag}) that the maximal modulus of the artificial magnetic field is $|B|\sim 0.1\,\hbar k^2\kappa^{(0)}/\Delta$.  This field is directed along the $z$ axis, and is a periodic function of $y$ with changes of sign every $\lambda/(2\sqrt 2)$. The same reasoning as above shows that one can marginally localize one quantum of circulation in each disk of area $k^{-2}$ over which the field is approximately uniform. In order to obtain a circulation $\gg 2 \pi$, one needs to rectify this spatially alternating field. Practical solutions will be detailed in Sec.~\ref{sec:opticallattices} devoted to artificial gauge fields in optical lattices.


\section[Multi-level systems]{Gauge potentials for multi-level systems\label{sec:Three-level-system}}
\label{sec:three-level}

In the model discussed in Sec.~\ref{sec:two-level}, the internal state of the atom is at any place a linear combination of ground and excited states, and each of these two states must have a relatively large weight in order to obtain a non negligible artificial gauge potential. Therefore this configuration can be used only if the excited electronic state has a very long radiative lifetime, as is the case for alkaline-earth species. In order to address a larger class of atoms (including the more widely used alkali atoms), we now turn to schemes that take advantage of the (quasi-) degeneracy of the electronic ground level. Denoting $\{|g_j\rangle,\;j=1,\ldots,N\}$ a basis set of the ground state manifold, we look for configurations where some dressed states are linear combinations of the $|g_j\rangle$ states, with a negligible contribution of the excited state manifolds, $|\chi\rangle\approx\sum_j \alpha_j |g_j\rangle$. As we will see, this can be obtained by taking benefit of a so-called \emph{dark state} \cite{Arimondo:1996}, or by choosing a laser frequency that is strongly detuned with respect to the atomic resonance lines.   
If the atom is prepared in such a dressed state and moves sufficiently slowly to follow it adiabatically, geometrical gauge potentials show up as in  Sec.~\ref{subsec:alkalineearth} \cite{Dum:1996}. Since we use laser beams to provide the relevant stimulated Raman couplings between the states $|g_j\rangle$, the $\alpha_{j}$ coefficients can vary significantly on a short length scale, typically an optical wavelength. One can thus produce geometrical fields with comparable amplitudes  to those found in Sec.~\ref{sec:two-level}, while avoiding the strong heating that would be caused by spontaneous emission processes. 

In this section we first consider the dark state case, which occurs for a $\Lambda$ level scheme, where two sublevels of the electronic ground  state $|g_1\rangle$ and $|g_2\rangle$ are coupled to a single excited state $|e\rangle$ by  two laser beams. The dark state  is an eigenstate of the atom-laser coupling that is a linear combination of $|g_1\rangle$ and $|g_2\rangle$ with a strictly zero contribution of the excited state. We then discuss two possible practical implementations  of this dark state scheme, first using laser beams carrying orbital angular momentum, and then using counterpropagating Gaussian beams with a spatial shift of their axis. Finally we describe  an alternative scheme involving a position-dependent detuning. This scheme that is not relying on dark states but on a large detuning, has led to the first experimental observation by  \textcite{Lin2009b} of a geometric magnetic field in the context of cold atom physics. 

\subsection{Artificial magnetic field in a $\Lambda$ scheme}
\label{subsec:Lambda_scheme}

\begin{figure}[t]
\centerline{\includegraphics{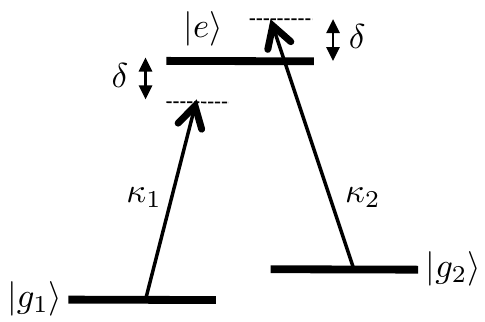}}
\caption{Atomic $\Lambda$-level structure providing a dark state that depends parametrically on the  Rabi frequencies $\kappa_{1}$ and $\kappa_{2}$.}
\label{fig:Lambda_scheme}
\end{figure}

We consider the $\Lambda$-type atomic level structure represented in Fig.~\ref{fig:Lambda_scheme}, where two laser beams couple the atomic states $|g_1\rangle$ and $|g_2\rangle$ to the third one $|e\rangle$.  The lasers are tuned symmetric with respect to the average of the frequencies of the $g_1-e$ and $g_2-e$ transitions. The full atomic Hamiltonian is given in Eq.~(\ref{eq:hamiltonian1}), and the coupling operator between the light and the atom written in the \{$|g_1\rangle$, $|e\rangle$, $|g_2\rangle$\} basis reads using the rotating wave approximation 
 \begin{equation}
U=\frac{\hbar}{2}\left(\begin{array}{ccc}
-2\delta & \kappa_{1}^{*} & 0\\
\kappa_{1} &0 & \kappa_{2}\\
0 & \kappa_{2}^{*} &2 \delta\end{array}\right)\,.
 \label{eq:H-matrix}
 \end{equation}
Here $\kappa_{1,2}$ are the complex, space-dependent Rabi frequencies, which include the spatially varying phases of the laser beams as in Sec.~\ref{subsec:alkalineearth}. The frequency $2\delta$ is the detuning of the two-photon excitation with respect to the Raman resonance between $g_1$ and $g_2$.  

Suppose that the two-photon (Raman) excitation is resonant ( $\delta=0$). In this case the coupling matrix $U$ possesses an eigenstate with zero energy called \emph{dark} (or \emph{uncoupled}). This state contains no contribution from the excited state $|e\rangle$ and reads:
\begin{equation}
|D\rangle=(\kappa_{2}|g_1\rangle-\kappa_{1}|g_2\rangle)/\kappa\,,
\label{eq:D-state}
\end{equation}
where $\kappa=\left(|\kappa_{1}|^{2}+|\kappa_{2}|^{2}\right)^{1/2}$.  The two other eigenstates of $U$ have the eigenenergies $\pm \hbar \kappa/2$, and read $|\pm\rangle= (|B\rangle \pm |e\rangle)/\sqrt{2}$, where  $|B\rangle$ is the \emph{bright} (\emph{coupled}) state
\begin{equation}
|B\rangle=(\kappa_{1}^{*}|g_1\rangle+\kappa_{2}^{*}|g_2\rangle)/\kappa\,.
\label{eq:brightstate_Lambda}
\end{equation}
Dark states are frequently encountered in quantum optics applications such as sub-recoil cooling \cite{Aspect:1988}, electromagnetically induced transparency \cite{Arimondo:1996,Harris97Physics-Today,Lukin:2003,Fleischhauer05RMP} and Stimulated Raman Adiabatic Passage \cite{Bergmann:1998,Vitanov01AAMOP,Shapiro07RMP}. These applications rely on the robustness of the state $|D\rangle$ with respect to the decoherence caused by spontaneous emission. 

Like in Sec.~\ref{sec:two-level}, the full atomic state-vector can be cast into the eigenstates of the operator $U$ as
\begin{equation}
|\Psi(\bs{r})\rangle=\sum_{X=D,\pm}\psi_{X}(\bs r)|X(\bs {r})\rangle\,,\label{eq:State-vector-Lambda}\end{equation}
 where the wave-functions $\psi_{D}(\bs r)$ and $\psi_{\pm}(\bs r)$ describe the translational motion of an atom in the internal states $|D(\bs{r})\rangle$ and $|\pm(\bs{r})\rangle$, respectively. The adiabatic approximation assumes that the atom stays in the dark state, so that one can write approximately $|\Psi(\bs r)\rangle\approx\psi_{D}(\bs r)|D(\bs r)\rangle$. Projecting the Schr\"odinger equation onto the dark state and neglecting the couplings to the two other
internal states $|\pm\rangle$, we arrive at an equation of motion for the dark state wave function $\psi_{D}(\bs r)$ similar to (\ref{eq:motion}):
\begin{equation}
i\hbar\frac{\partial \psi_{D}}{\partial t}=\left[\frac{(\bs P-\bs A)^2}{2M}+V+W\right]\psi_{D}\,,\label{schrod_gauge}
\end{equation}
 where $\bs A=i\hbar\langle D|\bs \nabla D\rangle$ and $W=\hbar^{2}\left|\langle B|\bs \nabla D\rangle\right|^{2}/(2M)$ are the effective vector and scalar potentials emerging due to the spatial dependence of the dark state. 
 
 We therefore recover a situation similar to that of  Sec.~\ref{subsec:adiabfollowing}, provided we replace $|\chi_1\rangle$ by $|D\rangle$ and  $|\chi_2\rangle$ by $|B\rangle$. Comparing Eq.~(\ref{eq:eigenstates}) and Eq.~(\ref{eq:D-state}), we set:
 \begin{equation}
\sqrt \zeta\equiv\frac{|\kappa_1|}{|\kappa_2|}=-\tan\frac{\theta}{2}\,,\qquad \phi_1-\phi_2=\phi\,,
\label{eq:zeta}
\end{equation}
where the $\phi_{j}$ are the phases of the Rabi frequencies $\kappa_{j}=\tilde \kappa_{j}\,e^{i\phi_{j}}$ ($j=1,2$). The artificial magnetic field  $\bs{B}=\bs \nabla\times\bs{A}$ given in (\ref{eq:chpmag}) can be expressed in terms of $\zeta$ and $\phi$:
\begin{equation} 
 \bs B=\hbar\frac{\nabla \phi \times\nabla\zeta}{(1+\zeta)^{2}}\,.
 \label{eq:Bfield_Lambda}
\end{equation}
This effective magnetic field $\bs{B}$ is non-zero only if the gradients of the intensity ratio $\zeta$ and the relative phase $\phi$ are both non-zero and not parallel to each other. We discuss in the next subsections some practical implementations of this $\Lambda$ scheme, using either light beams with orbital angular momentum or counterpropagating Gaussian beams with an axis offset.

\begin{figure}[t]
\centerline{\includegraphics{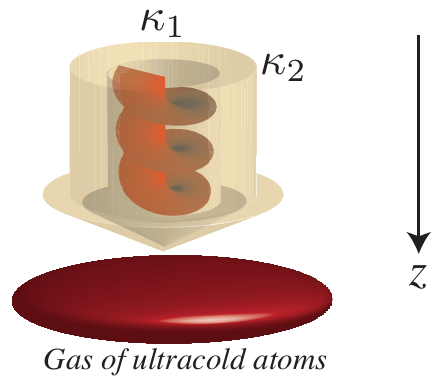}}
\caption{(Color online) Two copropagating Gaussian beams, one of them being prepared in a Laguerre--Gauss mode with a non-zero orbital angular momentum, drive the two transitions of an atom with a $\Lambda$ level scheme. For an atom prepared in the dark state $|D\rangle$ (Eq.~\ref{eq:D-state}), the nontrivial phase and intensity ratios between the Rabi frequencies $\kappa_{1}$ and $\kappa_{2}$ produce an artificial magnetic field parallel to the beam propagation axis. }
\label{fig:orbital}
\end{figure}

\subsection{Using light beams with orbital angular momentum }
\label{subsec:orbital_angular_momentum}

We consider here the situation where the atoms located in the plane $z=0$ are irradiated by two laser beams propagating along the $z$ axis (Fig. \ref{fig:orbital}). The beams are prepared in Laguerre--Gauss modes, and they carry the orbital angular momenta $\hbar\ell_{1}$ and $\hbar\ell_{2}$ per photon. This scheme was first proposed by \textcite{Juzeliunas:2004} [see also \cite{Juzeliunas:2005,Juzeliunas:2005b,Zhang:2005}]. The complex Rabi frequencies can be written $\kappa_{j}(\bs r) = \tilde\kappa_{j}(\rho)\;e^{i\ell_{j}\varphi}$ ($j=1,2$),  where $\varphi$ is the azimuthal angle around the $z$ axis and $\rho$ is the radial coordinate in the $x-y$ plane. The effective magnetic field (\ref{eq:Bfield_Lambda}) is directed along the $z$ direction and its amplitude reads 
 \begin{equation}
B(\rho)=\frac{\hbar\ell}{\rho}\frac{\partial_{\rho}\zeta}{(1+\zeta)^{2}}
 \end{equation}
where $\ell=\ell_{1}-\ell_{2}$ is the relative winding number of the two beams. 

For concreteness we consider beams with equal waists $w$ and we choose $\ell_{1}=\ell$, $\ell_{2}=0$ so that
\begin{equation}
 \tilde \kappa_{1}(\rho) =\kappa^{(0)} \rho^{\ell}e^{-\rho^{2}/w^{2}}\,,\quad
 \tilde \kappa_{2}(\rho) =\kappa^{(0)}\rho_c^{\ell}e^{-\rho^{2}/w^{2}}\,,
\label{eq:kappa-vortex}
\end{equation}
where  $\rho_{c}$ is the radius at which the two beams have equal intensities. The winding number $\ell$ determines the shape of the gauge potentials. (i) For $\ell>1$, the effective magnetic field is zero at the center of the beams and is maximum at $\rho=\rho_{c}\,\left[(\ell-1)/(\ell+1) \right]^{1/2\ell}$. (ii) For $\ell=1$, the effective magnetic field is maximum at the origin with the value $B(0)=2\hbar/\rho_{c}^2$. Its amplitude is inversely proportional to $\rho_{c}$, and can thus be controlled by changing the intensity ratio of the laser beams. Typically $\rho_c$ has the same order of magnitude as the waist $w$. Since $w\gg k^{-1}$, the magnetic field generated in this way is notably smaller than the value $B_0\approx \hbar k/w$ found in Sec.~\ref{subsec:alkalineearth}, where we took advantage of the rapid spatial variation of the phase $kx$ of a plane running wave.

The effective magnetic flux (in units of the Planck constant $h$) through a circle ${\cal C}$ of radius $r_{0}$ reads [see Eq.~(\ref{eq:phaseBerry})]
 \begin{equation}
\frac{\gamma({\cal C})}{2\pi}=\frac{1}{h}\oint\bs{A}\cdot\mathrm{d}\bs{l}=\ell\,\frac{\zeta_{0}}{1+\zeta_{0}}\,,\label{flux}\end{equation}
 where $\zeta_{0}$ is the intensity ratio at the radius $\rho=r_{0}$. As explained in Sec.~\ref{subsec:circulation}, this flux gives the maximal number of vortices that can be observed if a superfluid gas is placed in this laser configuration. Its maximal value is $\ell$ and it is approximately reached for $\zeta_{0}\gg 1$, i.e. for radii $\rho_0$ such that the intensity of the beam 1 (with orbital angular momentum) largely exceeds that of the beam 2 (with no angular momentum). In practice the winding number $\ell$ can reach a few tens \cite{Hadzibabic:2010}; this configuration is therefore better suited to generate small vortex patterns rather than large vortex arrays. 

\begin{figure}[t]
\centerline{\includegraphics{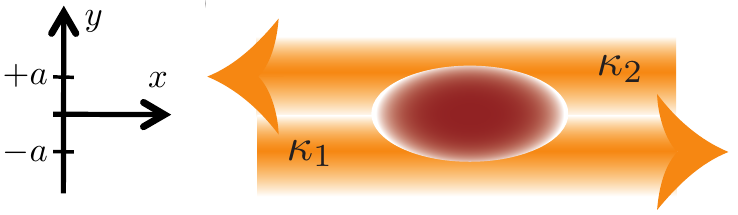}}
\caption{(Color online)  Two counterpropagating beams with an offset $2a$ between their propagation axes drive the two transitions of an atom with a $\Lambda$ level scheme. When the atom prepared in the dark state $|D\rangle$, the laser beams induce an artificial magnetic field perpendicular to the figure plane.}
\label{fig:shifted_beams}
\end{figure}

\subsection{Using spatially shifted laser beams}
\label{subsec:spatial_shift}

We now turn to the scheme represented in Fig.~\ref{fig:shifted_beams}, which constitutes a direct generalisation of the configuration studied in section~\ref{subsec:alkalineearth} for a two-level atom.  It uses two Gaussian beams that are counterpropagating along the $x$ axis, so that the phase difference between the two beams provides the necessary gradient of the phase angle $\phi$ entering in Eq.~(\ref{eq:Bfield_Lambda}). The gradient of the intensity ratio $\zeta$ is obtained by a spatial shift $\pm a$ of each beam axis along the $y$ direction. This configuration was first proposed by \textcite{Juzeliunas:2006}, and a modified version that is more flexible in terms of choice of laser polarizations  was later suggested by \textcite{Gunter:2009}. It offers the possibility to reach large magnetic field values over an extended region of space, while taking advantage of the dark state configuration to minimize the spontaneous emission rate.  This configuration with spatially shifted laser beams can also be used to generate the spin-Hall effect \cite{Zhu:2006PRA} and the Stern--Gerlach effect for chiral molecules \cite{Li-Brud-Sun07PRL}.
 
To simplify the mathematical treatment we assume that both beams have equal waists $w$ and equal central Rabi frequency $\kappa^{(0)}$, and we neglect their diffraction along the $x$ axis. Furthermore we use the fact that the frequencies of the transitions $g_1-e$ and $g_2-e$ are very close in practice and we set $k_1=k_2=k$. The complex Rabi frequencies are then given by
\begin{equation}
\kappa_j(\bs r) = \kappa^{(0)}\,e^{\pm ikx}\,e^{-(y\pm a)^2/w^2}\,,
\label{eq:kappaj_offset}
\end{equation}
where the $+$ (resp. $-$) sign stands for $j=1$ (resp $j=2$). The relative phase between the two beams is $\phi=2kx$, and the intensity ratio reads $\zeta=\exp(8ya/w^2)$. The effective magnetic field is oriented along the $z$ direction and its amplitude is obtained from Eq.~(\ref{eq:Bfield_Lambda}):
 \begin{equation}
 B(y)=\frac{4\hbar k a}{w^2}\,\frac{1}{\cosh^2(4ya/w^2)}\,.
 \end{equation}
The offset $a$ of the beams along the $y$ axis is \emph{a priori} arbitrary. In practice, one should not take $a\gg w$, in order to keep a significant laser intensity along the line $y=0$ where $B$ is maximum, and ensure that the atoms follow adiabatically the dark state $|D\rangle$ at this location. Taking as a typical value $a=w/2$ we find $B(0)=2\hbar k/w$, which is comparable to the result of  Sec.~\ref{subsec:alkalineearth}. Therefore the conclusions of Sec.~\ref{subsec:circulation} concerning the possibility to generate large vortex arrays remain valid for this configuration. Like the geometric magnetic field $B(y)$, the scalar potential $W(y)$ is maximum along the line $y=0$. An extra trapping potential $V$ is therefore needed to prevent the atoms from flying away from this area. 

\subsection{Gauge potentials involving a gradient of detuning}
\label{subsection:detuninggradient}

We explained in Sec.~\ref{subsec:alkalineearth} that the necessary gradient of the mixing angle $\theta$ can be provided by a gradient of the detuning of the laser frequency, as well as by a gradient of the laser intensity. The same distinction applies to the $\Lambda$ scheme and more generally to schemes involving multiple atomic levels. The first observation of a geometrical magnetic field by \textcite{Lin2009b} was actually based on a gradient of detuning for optical Raman transitions occurring between various ground sublevels.  In this section we present the main features of this experiment and connect it with the already discussed configurations.

\begin{figure}
\centerline{\includegraphics{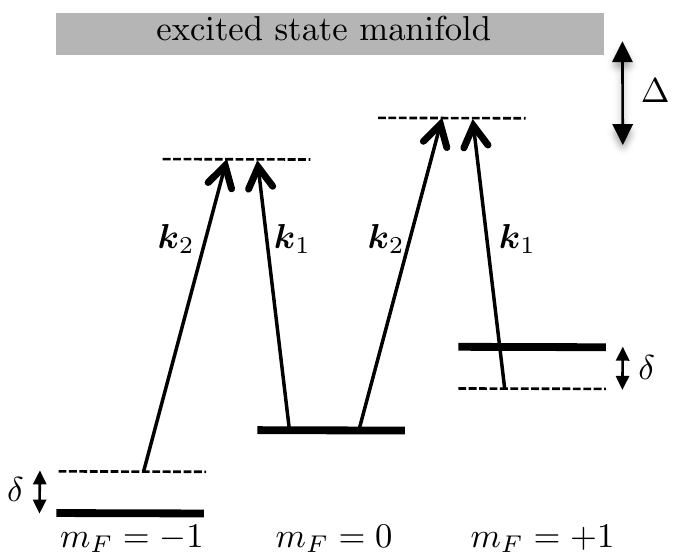}}
\caption{Atomic scheme used in \textcite{Lin2009b}.  $^{87}$Rb atoms with a spin $F=1$ ground state are placed in an external magnetic field that displaces the sublevels $|m=\pm 1\rangle$ (dotted lines) with respect to $|m=0\rangle$. The atoms are irradiated by two laser beams of wave vectors $\bs k_1$ and $\bs k_2$ that induce a resonant Raman coupling $\Delta m_F=\pm 1$ between the Zeeman sublevels. An additional spatial gradient of the external magnetic field induces an additional displacement of the $|m=\pm 1\rangle$ sublevels (full line). It thus creates a spatially varying two-photon detuning $\delta$ which allows one to generate an artificial gauge potential. Note that the frequency scale is not respected in the drawing (in practice $\Delta \gg \delta$).}
\label{fig:multilevel} 
\end{figure}

The experiment by \textcite{Lin2009b} has been performed with $^{87}$Rb atoms in their $F=1$ hyperfine level. In the process all three Zeeman sublevels $|m_F\rangle$ with $m_F=0,\pm 1$ acquire a significant population.  The atoms are irradiated by two laser beams with wave vectors $\bs k_1$ and $\bs k_2$ that create a quasi-resonant Raman coupling between Zeeman sublevels with $\Delta m_F=\pm 1$   (see Fig. \ref{fig:multilevel} and its caption for details).  The coupling occurs via the absorption of a photon in one beam and the stimulated emission of a photon in the other beam, accompanied by a change of the atom momentum of $\pm \hbar \bs k_{\rm d}$, where the difference of wave vectors $\bs k_{\rm d}= \bs k_1-\bs k_2=k_{d}\bs e_{x}$ is chosen along $x$.  An important ingredient is the application of a real magnetic field in addition to the laser beams. The Zeeman shift created by the magnetic field lifts the initial degeneracy of the sublevels $m_{F}=0,\pm1$. It gives a control knob  on the two-photon detuning $\delta$. In the initial experiment by \textcite{Lin:2009} a spatially uniform real magnetic field was applied, and it led to a spatially uniform vector potential,  corresponding to a zero geometric magnetic field. Subsequently \textcite{Lin2009b} used a non-homogeneous real magnetic field, making the two-photon detuning $\delta$ position-dependent. In the following we assume the linear form $\delta=\delta' (y-y_{0})$, where $\delta'>0$ is the uniform gradient of detuning. 

This experiment can first be analyzed using the adiabatic framework used so far. We assume that the coupling lasers can be considered as plane waves on the scale of the atomic cloud, so that the problem is translationally invariant along $x$. Therefore the two-photon coupling $\kappa$ induced by the laser pair can be written as $\kappa=\kappa^{(0)}e^{i\phi}$, with $\phi= \bs k_{\rm d}\cdot \bs r= k_{\rm d} x$ being the phase of the Raman coupling. The single-photon detuning $\Delta$ with respect to the excited state manifold involved in the transition is supposed to be very large compared to the Rabi frequencies. One can then perform an adiabatic elimination of the excited states to obtain an effective Hamiltonian acting on the ground state manifold. Keeping only the terms relevant for the subsequent discussion, this coupling written in the basis $\{|m_F=-1\rangle,\,|m_F=0\rangle,\,|m_F=+1\rangle\}$ has the form of Eq.~(\ref{eq:H-matrix}) with $\kappa_2=\kappa_{1}^{*}=\kappa$. 

We focus on the eigenstate of $U$ associated to the lowest eigenvalue $-\hbar [\delta^2+(\kappa^{(0)})^2 / 2]^{1/2}$,  
\begin{equation}
 |\chi\rangle=e^{i\phi}\cos^2\frac{\theta}{2}|-1\rangle-\frac{\sin\theta}{\sqrt 2}|0\rangle+e^{-i\phi}\sin^2\frac{\theta}{2}|+1\rangle
\label{eq:dark_state_Spielman}
\end{equation}
where we have set $\tan \theta=\kappa^{(0)}/\sqrt 2\,\delta$. The vector potential is now given by $\bs A=i\hbar \langle \chi|\bs \nabla \chi\rangle=-\hbar\bs k_{\rm d} \cos \theta$. The artificial magnetic field can be written in a form that is reminiscent of the result for the grad-$\Delta$ case of Sec.~\ref{subsec:alkalineearth} [Eq.~(\ref{eq:BtwolevelDet})] (subject to the replacement of $\Delta$ by $\delta$):
\begin{equation}
\bs B=\bs e_{z}\;B_0 \,{\cal L}^{3/2}(y-y_0) ,
\label{eq:Bspielman}
\end{equation}
where the characteristic length entering in the Lorentzian function ${\cal L}$ is $\ell_{\kappa}=\kappa^{(0)}/\sqrt 2\,\delta' $, and where the peak value of the artificial magnetic field is $B_0=\hbar k_{\rm d}/\ell_\kappa$. The prediction (\ref{eq:Bspielman}) is in good agreement with the experimental results of  \textcite{Lin2009b} ($\ell_{\kappa}\approx 40\,\lambda$ in the experiment). The scalar potential also takes a form that is similar to the one found for a two-level atom in the grad-$\Delta$ case: $W(y)\approx\hbar^2 k_{\rm d}^2/(4M)\;{\cal L}(y-y_0)$, where we assumed $k_d\ell_\kappa \gg 1$. For an external trapping potential centered in $y_0$, $W(y)$ weakens the confinement along the $y$ direction by creating in the vicinity of $y_0$  the anti-trapping potential $-M\omega_W^2(y-y_0)^2/2$, with $\omega_W=\hbar k_{\rm d}/\sqrt{2}M\ell_\kappa$ ($\omega_W/2\pi \simeq 30$\,Hz for the parameters of \textcite{Lin2009b}).

Alternatively, the scheme described above can be analyzed using an original framework introduced by \textcite{Spielman:2009}. This framework is usable in the particular case where the only $x$-dependence of the atom-laser coupling is the phase term $e^{\pm i k_{\rm d} x}$, and it has then a larger range of validity than the standard adiabatic approximation. In a first step one ``freezes'' the $y$ motion and diagonalizes exactly the $x$-dependent Hamiltonian $H_x=P_x^2/(2M)+U$ , where $U$ is the $3\times 3$ matrix given in Eq.~(\ref{eq:H-matrix}) with $\kappa_2=\kappa_1^*=\kappa^{(0)}e^{i k_{\rm d} x}$. The eigenstates are 3-component spinors that can be labelled by their momentum $\hbar K_x$. For each $K_x$, this diagonalization yields three dispersion relations $E_{n}(K_{x})$ ($n=1,2,3$), each with a minimum at a momentum $K_{n,\rm min}$ depending on the two-photon detuning $\delta$ and Rabi frequency $\kappa^{(0)}$.  At low energy and large coupling strength one obtains $E_n \approx \hbar^2(K_{x}-K_{n, \rm min})^2/2M^\ast$, where $M^\ast$ is  an effective mass. In a second step, one considers the motion along $y$ and takes into account the spatial variation of the detuning $\delta=\delta' (y-y_{0})$. For the lowest energy state $n=1$, this motion is described in an approximate manner by the projected Hamiltonian $H=P_y^2/(2M)+ E_1$, which leads to the identification of an effective vector potential $\bs A=\hbar K_{1,\rm min} \bs e_{x}$. The spatial dependence of $\delta$, hence of $K_{1,\rm min}$, along the $y$ axis corresponds to an effective magnetic field $\bs {B}=\hbar(\partial K_{1,\rm min} / \partial y)\, \bs e_{z}$, described here in the Landau gauge. The exact diagonalization and the adiabatic approaches agree when the latter is valid (see Sec.~\ref{subsec:validity}), with $M^\ast \approx M$ in particular. For Rabi frequencies smaller than those required for the adiabatic approximation [see Eq. (\ref{eq:validity_intuitive})], novel features appear in the exact diagonalization approach such as the possibility to simulate a spin-orbit coupling between two Bose--Einstein condensates in different dressed atomic states, experimentally observed by \textcite{Lin:2011}.
 
The discussion of Sec.~\ref{subsec:circulation} about the observable vortex pattern applies directly to the present scheme. The  number of vortex rows that can fit along the $y$ axis is again given by $\sim \ell_{\kappa}\sqrt{\rhov}\sim \sqrt{\ell_{\kappa}/\lambda}$. The limit of large vortex arrays can thus in principle be reached with this configuration, provided one uses a relatively small gradient of detuning. \textcite{Lin2009b} generated about $10$ vortices in a rubidium Bose--Einstein condensate. These vortices did not order into a regular lattice, presumably because of the heating of the cloud and atom losses ($1/e$ lifetime of 1.4~s) caused by the residual photon scattering. 

One could think that increasing the detuning of the laser coupling would solve the photon scattering problem. Unfortunately the situation is not so favorable, at least for alkali-metal species. Indeed, by contrast to the scalar light shift that scales as $U_{\rm scal.}\propto I / \Delta$ where $I$ is the light intensity, the two-photon Raman coupling $\kappa$ is proportional to the vector part of the light shift $U_{\rm vec.}\propto I \Delta_{\rm FS}/\Delta^2$. Here $\Delta_{\rm FS}$ denotes the fine structure splitting, which is only a few percents of $\oA$ for alkalis (2\% for rubidium atoms, 5\% for cesium atoms) and the above scaling for $\kappa$ is valid for $\Delta_{\rm FS}\lesssim \Delta$. Therefore in the limit of large detunings the Raman coupling $\kappa$ decreases with $\Delta$ as fast as the photon scattering rate $\propto I \Gamma/\Delta^2$. In other words, the usual trick that consists in increasing both $\Delta$ and $I$ to enhanced the role of the scalar light shift with respect to spontaneous emission processes is not applicable in this context. 


\section{Non-Abelian gauge potentials}

\label{sec:Non-Abelian}

The idea of non-Abelian geometric gauge potentials goes back to the work by \textcite{Wilczek:1984}, who considered the generalization of the adiabatic theorem to the case where the Hamiltonian of the system of interest possesses a group of eigenstates that remain degenerate (or quasi-degenerate) and well-isolated from other levels in the course of the time evolution. This analysis was followed by applications to many areas including {\it inter alia} molecular and condensed matter physics \cite{Bohm:2003,Xiao10RMP}.  In particular, the possibility to generate non-Abelian magnetic monopoles was demonstrated in the rotational dynamics of diatomic molecules \cite{Moody86PRL,Zygelman:1990,Bohm92JMP} and  in the nuclear quadrupole resonance \cite{Zee88PRA}.

\begin{figure}
\centerline{\includegraphics{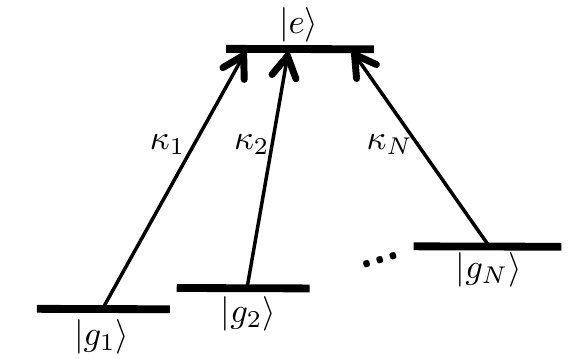}}
\caption{Multipod configuration. An atomic state $|e\rangle$ is coupled to
$N$ different atomic states $|g_j\rangle$ ($j=1,...,N$) by $N$ resonant
laser fields.}
\label{fig:figure-Npod} 
\end{figure}

In this section we are interested in the non-Abelian dynamics of cold atoms in light fields, which can emerge when a $(N+1)$-state atomic system with $N\geq 3$ (see Fig. \ref{fig:figure-Npod}) is excited by a suitable configuration of laser beams.  The first study in this direction was performed  in the context of Stimulated Raman Adiabatic Passage (STIRAP) by \textcite{Unanyan98OC,Unanyan99PRA}, who considered the behaviour of a four-state atomic system (tripod configuration) when it is driven by three successive laser pulses. Subsequently \textcite{Osterloh:2005} and \textcite{Ruseckas:2005} transposed  the concept of gauge potentials to the case of a continuous atom-laser excitation  relevant for the present review and identified situations where the non-Abelian gauge potentials emerge. 

We start this section by providing a general formulation of the adiabatic motion of atoms when some internal atomic states remain degenerate in the presence of the atom-light coupling. We show how non-Abelian gauge potentials appear and discuss the structure of these potentials for some typical laser configurations. We also present some physical phenomena that are associated with non-Abelian gauge potentials, such as the generation of magnetic monopoles, Rashba spin-orbit coupling, and non-Abelian Aharonov--Bohm effect. 

\subsection{Emergence of non-Abelian gauge potentials}
\label{subsec:emergence_non_abelian}

In this section we consider atoms with $N+1$ internal levels, which are coupled to a light field. We assume that we can cast the atom-laser interaction at any given point $\bs r$ as a time-independent $(N+1)\times (N+1)$ matrix $U(\bs r)$ using the rotating wave approximation. Our starting point is then  still the full atomic Hamiltonian of Eq.~(\ref{eq:hamiltonian1}). For a fixed position $\bs{r}$,  $U(\bs{r})$ can be diagonalized to give a set of $N+1$ dressed states $\left|\chi_{n}(\bs{r})\right\rangle $, with eigenvalues $\varepsilon_{n}(\bs{r})$ ($n=1,\ldots,N+1$). The key feature that leads to non-Abelian gauge potentials is that some dressed states form a degenerate (or nearly degenerate) manifold, at any point $\bs r$ in space. More specifically  we shall assume that the first $q$ atomic dressed states form a degenerate subspace ${\cal E}_q$, and these levels are well separated from the remaining ones.

The full quantum state of the atom describing both internal and motional degrees of freedom can be written $|\Psi\rangle=\sum_{n=1}^{N+1}\psi_{n}(\bs{r})\left|\chi_{n}(\bs{r})\right\rangle$ where each $\psi_{n}$ is the wave-function for the centre of mass motion of the atom in the internal state $|\chi_n\rangle$. We are interested here in the dynamics of the atom when it is initially prepared in the subspace ${\cal E}_q$. Neglecting transitions to states outside ${\cal E}_q$, we can project the full Schr\"odinger equation onto ${\cal E}_q$ and we arrive at a closed Schr\"odinger equation for the 
reduced column vector $\tilde{\Psi}=\left(\psi_{1},\dots,\psi_{q}\right)^{\top}$
\begin{equation}
i\hbar\frac{\partial \tilde{\Psi}}{\partial t}=\left[\frac{(\bs P-\bs{A})^{2}}{2M}+(V+\varepsilon)\hat 1_q+W \right]\tilde{\Psi}\,, \label{eq:SE-reduced-non-Abel}\end{equation}
where $\hat 1_q$ is the identity matrix in ${\cal E}_q$ and $\varepsilon(\bs{r})$ is a diagonal matrix of eigenenergies $\varepsilon_{n}(\bs{r})$  ($n=1,\ldots,q$). This equation is very reminiscent of Eq.~(\ref{eq:motion}); However $\bs{A}$ and $W$ are now $q\times q$ matrices with the elements 
\begin{eqnarray}
\bs{A}_{n,m} & = & i\hbar\langle\chi_{n}(\bs{r})|\nabla\chi_{m}(\bs{r})\rangle,\label{eq:A-nm}\\
W_{n,m} & = & 
\frac{1}{2M}\sum_{l=q+1}^{N+1}\bs{A}_{n,l}\cdot\bs{A}_{l,m}\,,\label{eq:fi-nm}
\end{eqnarray}
with $n,m\in(1,\dots,q)$. The effective vector potential $\bs{A}$ is called the \emph{Mead-Berry connection}.

The effective magnetic field (or \emph{curvature}) $\bs{B}$ associated to $\bs A$ is: 
 \begin{eqnarray}
B_{i}=\frac{1}{2}\epsilon_{ikl}\, F_{kl},\quad F_{kl}=\partial_{k}A_{l}-\partial_{l}A_{k}-\frac{i}{\hbar}[A_{k},A_{l}].\label{eq:B-non-Abel}
 \end{eqnarray}
Note that the term $\frac{1}{2}\varepsilon_{ikl}[A_{k},A_{l}]=(\bs{A}\times\bs{A})_{i}$ does not vanish in general, because the vector components of $\bs{A}$ do not necessarily commute. Therefore the magnetic field  $\bs{B}$ can be non-zero even if the vector potential $\bs{A}$ is uniform in space. 
This property is specific of non-Abelian dynamics and occurs only if $q\geq 2$, whereas for $q=1$ we recover  simply the Abelian dynamics discussed in the two preceding sections.

\subsection{The multipod scheme}
\label{subsec:N_1podscheme}

A generic way to obtain a degenerate subspace in atom-laser interaction is to realize the situation depicted in Fig.~\ref{fig:figure-Npod}, where one single level labelled $|e\rangle$ is coupled to  $N$  levels labelled $|g_j\rangle$ ($j=1,\ldots,N$), with complex Rabi frequencies $\kappa_j$ \cite{Juzeliunas:2010}. We will discuss at the end of this subsection how this scheme can be implemented in practice for $N=3$ (tripod) and $N=4$ (quadrupod). The atom-light coupling operator is
 \begin{equation}
U=\sum_{j=1}^{N}\frac{\hbar\kappa_{j}(\bs{r})}{2}\,|e\rangle\langle g_j|+\mathrm{H.c.}\,,\label{eq:H-0}\end{equation}
which can be conveniently rewritten as
 \begin{equation}
U=\frac{\hbar\kappa(\bs{r})}{2}\left(|e\rangle\langle B(\bs{r})|+|B(\bs{r})\rangle\langle e|\right)\,,\label{eq:H-0-alternative}\end{equation}
where $|B\rangle=\sum_{j=1}^{N}\kappa_{j}^{*}|g_j\rangle/\kappa$ is the bright (coupled) state generalizing (\ref{eq:brightstate_Lambda}) and $\kappa$ is the total Rabi frequency, $\kappa^{2}=\sum_{j=1}^{N}|\kappa_{j}|^{2}$. 

The diagonalization of $U$ is straightforward. First the coupling between $|B\rangle$ and $|e\rangle$ gives rise to the two eigenstates $|\pm\rangle=\left(|e\rangle\pm|B\rangle\right)/\sqrt{2}$ with energies $\pm\hbar\kappa/2$. Then the remaining orthogonal subspace of dimension $N-1$ corresponds to degenerate dark states that are all eigenstates of $U$ with energy $\varepsilon=0$. This provides the degenerate subspace ${\cal E}_q$ with $q=N-1$ introduced in 
Sec.~\ref{subsec:emergence_non_abelian}, which is required for the emergence of non-Abelian dynamics.  In the following we denote $|D_{n}\rangle$ ($n=1,\ldots,N-1$) a normalized orthogonal basis of ${\cal E}_{N-1}$. 

\begin{figure}
\centerline{ \includegraphics[width=1\columnwidth]{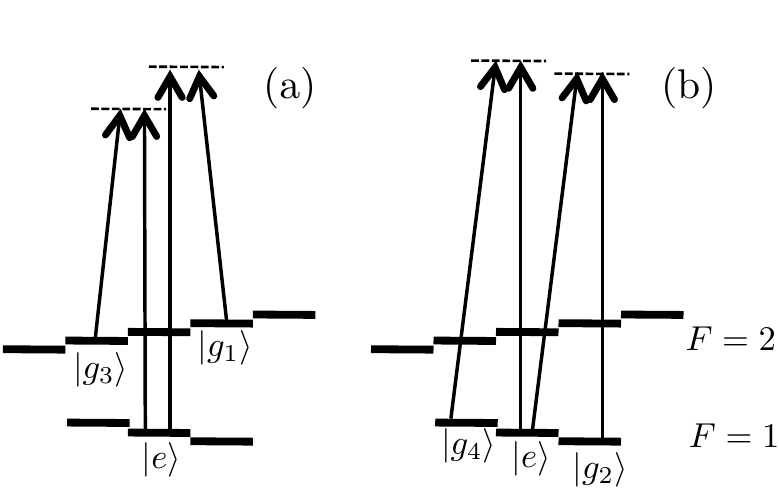}}
\caption{Implementation of the multipod scheme for $N=4$ for alkali-metal atoms with two hyperfine levels of angular momentum $F=1$ and $F=2$. The couplings involved in this scheme correspond to stimulated Raman transitions between hyperfine states of the ground levels. We choose $|e\rangle\equiv|F=1,m_{F}=0\rangle$. A small magnetic field lifts the degeneracy within the levels $F=1$ and $F=2$ and allows one to address selectively the various Raman transitions, using pairs of laser beams with properly chosen polarization and frequencies. (a) Two pairs of laser beams induce the transitions $|e\rangle\to|g_1\rangle\equiv|F=2,m_{F}=1\rangle$ and $|e\rangle\to|g_3\rangle\equiv|F=2,m_{F}=-1\rangle$. (b) A pair of laser beams induce the transitions $|e\rangle\to|g_2\rangle\equiv|F=1,m_{F}=1\rangle$ and $|e\rangle\to|g_4\rangle\equiv|F=1,m_{F}=-1\rangle$. }
\label{fig:exp_scheme} 
\end{figure}

Let us briefly discuss how such a multipod scheme can be implemented in quantum optics for the two cases $N=3$ and $N=4$. The scheme of Fig.~\ref{fig:figure-Npod} for $N=3$ is realized in a straightforward way by considering  an atomic transition between a ground electronic state with angular momentum $J_g=1$, and an excited electronic state with angular momentum $J_e=0$. Such a transition occurs for alkali-metal species such as $^{23}$Na or $^{87}$Rb, as well as for helium atoms prepared in metastable electronic spin triplet states ($^3S_1 - ^3\hskip-1.5mm P_0$ transition).  The scheme $N=4$ is a bit more subtle to achieve and we briefly outline the proposal detailed by \textcite{Juzeliunas:2010}. The idea is to use the two hyperfine ground levels of an alkali-metal atom like $^{87}$Rb, with angular momenta equal to $F=1$ and $F=2$, respectively (see Fig.~\ref{fig:exp_scheme}). For the state $|e\rangle$ we choose one particular ground state $|e\rangle\equiv|F=1,m=0\rangle$. The four other states playing the role of the states $|g_j\rangle$ are the Zeeman sublevels $|F=1,m_F=\pm 1\rangle$ and $|F=2,m_F=\pm 1\rangle$. The couplings between $|e\rangle$ and the levels $|g_j\rangle$ are induced by resonant two-photon Raman transitions like in Sec.~\ref{subsection:detuninggradient}. The decoherence and heating due to photon scattering can be minimized in this scheme by chosing a large single photon detuning (that is, of the order of the fine structure splitting $\Delta_{\rm FS}$), as explained in Sec. \ref{subsection:detuninggradient}.

\subsection{Generating a magnetic monopole}
\label{sub:monopole}

One of the interesting properties of non-Abelian gauge potentials is the possibility to generate magnetic monopoles, as first pointed out by \textcite{Moody86PRL}. In this section we present a possible way to implement such a monopole using atom-light interaction. We concentrate on the physical aspects of the problem and we refere the reader to the original publication by \textcite{Ruseckas:2005} for technical details.

We will consider here the tripod configuration obtained using $N=3$ ground levels in the formalism of Sec.~\ref{subsec:N_1podscheme}. It is convenient to parametrize the Rabi frequencies $\kappa_{j}$ with angle and phase variables according to 
\begin{eqnarray}
\kappa_{1} &=& \kappa\,\sin\alpha\,\cos\beta\,\mathrm{e}^{i\phi_{1}}
\nonumber\\
\kappa_{2}&=&\kappa\,\sin\alpha\,\sin\beta\,\mathrm{e}^{i\phi_{2}}
\label{eq:kappa_nonAbelian} \\
\kappa_{3}&=&\kappa\,\cos\alpha\,\mathrm{e}^{i\phi_{3}} \nonumber
\end{eqnarray}
in which case we can choose the following expression for the dark states in the basis $\{|g_1\rangle,|g_2\rangle,|g_3\rangle\}$ 
\begin{equation}
|D_{1}\rangle = \left(
 \begin{array}{c}
 \  \ \ \sin\beta \,\mathrm{e}^{i\phi_{31}} \\
 -\cos\beta\,\mathrm{e}^{i\phi_{32}}\\
  0 
 \end{array}
 \right)
 \quad
|D_{2}\rangle =\begin{pmatrix} \cos\alpha\,\cos\beta\,\mathrm{e}^{i\phi_{31}}\\
\cos\alpha\,\sin\beta\,\mathrm{e}^{i\phi_{32}}\\
 -\sin\alpha
  \end{pmatrix}
 \label{eq:D2}
 \end{equation}
with $\phi_{ij}=\phi_{i}-\phi_{j}$. The vector potential then reads   
\begin{eqnarray}
\bs{A}_{11} & = & \hbar\left(\cos^{2}\beta\;\bs\nabla \phi_{23}+\sin^{2}\beta\;\bs\nabla \phi_{13}\right)\nonumber \\
\bs{A}_{12} & = & \hbar\cos\alpha\left(\frac{1}{2}\sin(2\beta)\;\bs\nabla \phi_{12}-i\bs\nabla\beta\right) \,\label{eq:A-special}\\
\bs{A}_{22} & = & \hbar\cos^{2}\alpha\left(\cos^{2}\beta\;\bs\nabla \phi_{13}+\sin^{2}\beta\;\bs\nabla \phi_{23}\right).\nonumber 
\end{eqnarray}
 
To generate a magnetic monopole, we consider the laser setup formed with two of the laser beams propagating along the $z$ axis, and the third one along the $x$ axis. More precisely the first two beams are prepared in the Laguerre--Gauss modes with orbital angular momentum $\ell=\pm 1$, and the third beam is  prepared in the first order Hermite--Gauss mode so that  
\begin{equation}
\kappa_{1,2}=\kappa_{0}\frac{\rho}{R}\,\mathrm{e}^{i(kz\mp \varphi)},\quad\kappa_{3}=\kappa_{0}\frac{z}{R}\,\mathrm{e}^{ikx}.
 \label{eq:kappa-monopole}
 \end{equation}
Here $\rho$ is the distance from the $z$ axis and $\varphi$ the azimuthal angle around this axis. We suppose that the beam waists are large compared to the other physical scales of the problem and we therefore omit the Gaussian profiles in Eq. (\ref{eq:kappa-monopole}). The calculation of the vector potential using the formalism of Sec.~\ref{subsec:N_1podscheme} gives 
\begin{eqnarray}
 \bs{A}&=&-\frac{\hbar}{r\tan\vartheta}\,\bs{{e}}_{\varphi}\;\hat \sigma_x  \nonumber  \\
 &+&\frac{\hbar k}{2}({\bs{e}}_{z}-{\bs{e}}_{x})
 \left[(1+\cos^{2}\vartheta)\hat 1+(1-\cos^{2}\vartheta)\hat \sigma_z\right],\nonumber 
\end{eqnarray}
where $r$, $\vartheta$ and $\varphi$ are the spherical coordinates, $\hat \sigma_j$ are the Pauli matrices and $\hat 1$ is the  $2\times 2$ unity matrix. The first term in the vector potential proportional to $\hat \sigma_{x}$ represents the field of a magnetic monopole of unit strength at the origin ($r=0$):
\begin{equation}
\bs{B}=\frac{\hbar}{r^{2}}\,{\bs{e}}_{r}\,\hat \sigma_x+\cdots\,.
\end{equation}
The dots indicate non-monopole field contributions\footnote{A pure monopole would emerge if the Rabi frequencies $\kappa_j$ could be taken proportional to the corresponding three Cartesian coordinates  $\kappa_1=Ax$,  $\kappa_2=Ay$ and  $\kappa_3=Az$. This is however not possible in practice, as this spatial dependence  is not consistent with the Maxwell equations for the driving laser field.} proportional to the Pauli matrices and to the unity matrix.    
Note also that the total intensity of the laser fields (\ref{eq:kappa-monopole}) vanishes at the origin representing a singular point. Thus one should apply an additional potential that will expel the atoms from this region, in order to avoid non-adiabatic transitions in the vicinity of $r=0$.

This configuration was further analyzed by \textcite{Pietila09PRL}, who studied the behavior of an interacting Bose--Einstein condensate in this monopole configuration. They showed numerically that the existence of the monopole gives rise to a pseudospin texture with a topological charge that cancels the monopole charge.

\subsection{Generating a spin-orbit coupling}
\label{subsec:SO_coupling}

Non-Abelian light-induced gauge potentials can be used for generating a spin-orbit coupling for cold atoms, simulating the one appearing for electrons in condensed matter. The electron's spin degree of freedom plays a key role in the emerging area of semiconductor spintronics \cite{Fert:2008,Grunberg:2008,Zutic04RMP}. A first scheme for a semiconductor device is the spin field-effect Datta-Das transistor (DDT). It was proposed 20 years ago \cite{dattadas} and implemented recently \cite{Koo:2009Science}. Atomic and polaritonic analogs of the electron spin transistor have also been suggested \cite{Vaisnav08PRL-DDT,Johne09arxiv-Polariton-DDT}. An important ingredient of the DDT is the spin-orbit coupling of the Rashba \cite{Rashba60,Winkler03Review} or Dresselhaus \cite{Dresselhaus55PR,Schliemann03PRL-DDT-Balanced} types. This Rashba-Dresselhaus (RD) coupling scheme is described by a vector potential which can be made proportional to the spin-$1/2$ operator of a particle within a plane \cite{Cserti2006PRB}.

We explain in this section how an effective spin-orbit coupling can be generated for atoms for the cases of an effective spin $s=1/2$ and $s=1$. We start with the general formalism of a $N$-pod level scheme in the case where the coupling lasers are plane waves of equal amplitude propagating in the $xy$ plane. More precisely we assume that the wave-vectors $\bs k_j$ form a regular polygon
 \begin{equation}
\bs{k}_{j}=k\left(-\bs{e}_{x}\cos \alpha_j+\bs{e}_{y}\sin\alpha_j\right)\,,\quad
 \alpha_{j}=2\pi j/N\,.
 \label{eq:Omega_j-plane-wave}
 \end{equation}
We set  $\kappa_{j}=\kappa^{(0)}\, e^{i\bs{k}_{j}\cdot\bs{r}}/\sqrt N$ and use the set of  orthogonal normalized dark states 
 \begin{equation}
|D_{n}\rangle=\frac{1}{\sqrt{N}}\sum_{j=1}^{N}|g_j\rangle\, e^{i\alpha_{j}n-i\bs{k}_{j}\cdot\bs{r}}\,,\quad n=1,\ldots,N-1 \,.
  \label{eq:D-n} 
\end{equation}
 Substituting $|D_{n}\rangle$ into Eqs.~(\ref{eq:A-nm})-(\ref{eq:fi-nm}),
one arrives at constant (yet non-Abelian) scalar and vector potentials
\begin{eqnarray}
W_{n,m} & = & \frac{\hbar^{2}k^{2}}{4M}(\delta_{m,1}\delta_{n,1}+\delta_{m,N-1}\delta_{n,N-1})\,,\label{eq:Phi-general-result}\\
\bs{A}_{n,m} & = & -\frac{\hbar k}{\sqrt 2}\sum_{\pm}\bs{e}_{\pm}\delta_{n,m\pm1}\,,\label{eq:A-nm-poligon}\end{eqnarray}
with $\bs e_\pm=(\bs e_x\pm i\bs e_y)/\sqrt 2$. For a constant external potential $V$, a basis of stationary solutions of the Schr\"odinger equation (\ref{eq:SE-reduced-non-Abel}) are the plane waves $\tilde \Psi_{\bs{K}}(\bs{r},t)=\Phi_{\bs{K}}e^{i(\bs{K}\cdot\bs{r}-\Omega_{\bs{K}}t)}$ with the amplitude $\Phi_{\bs{K}}$ obeying the eigenvalue equation $H_{\bs{K}}\Phi_{\bs{K}}=\hbar\omega_{\bs{K}}\Phi_{\bs{K}}$ where $H_{\bs{K}}$ is the $\bs K$-dependent $(N-1)\times(N-1)$ matrix:
\begin{equation}
 H_{\bs{K}}=\frac{\left(\hbar \bs K\,\hat 1-\bs{A}\right)^{2}}{2M}+W+V\hat{1}\,.
 \label{eq:H-k}
 \end{equation}

In the tripod setup ($N=3$) the wave vectors $\bs{k}_{j}$ form an equilateral triangle and $W$, $\bs A$ and $H_{\bs K}$ are $2\times 2$ matrices. The scalar potential is proportional to the unit matrix, $W=\hbar^{2}k^{2}/(4M)\,\hat 1$, whereas the vector potential $\bs{A}=- k\boldsymbol{\hat S}_{\bot}$ is proportional to the spin $1/2$ operator $\bs{\hat S}_{\bot}=\hat S_x \bs e_x+\hat S_y\bs e_y$ in the $xy$ plane ($\bs {\hat S}= \hbar \bs{\hat \sigma}/2$, where the $\bs{\hat \sigma}$ are the Pauli matrices). This provides a spin-orbit coupling of the RD type, characterized by two dispersion branches $\Omega_{\bs{K}}^{\pm}=\hbar\left(K\,\pm \,k/2\right)^{2}/2M$ for $V=-\hbar^{2}k^{2}/(4M)$. A number of other  arrangements of laser beams have been considered to produce the same RD spin-orbit coupling \cite{Jacob07APB,Juz08PRA,Oh09-tripod,Stanescu07PRL,Vaishnav08PRL,Larson09PRA}.

In the tetrapod setup ($N=4$) the choice of Eq.~(\ref{eq:Omega_j-plane-wave}) corresponds to two orthogonal pairs of counterpropagating laser fields. The vector potential $\bs{A}=-k \bs{\hat J_{\bot}}/\sqrt 2$ is proportional to the projection of a spin $1$ operator in the $xy$ plane $\bs{\hat J_\bot}=\hat J_{x}\bs{e}_{x}+\hat J_{y}\bs{e}_{y}$, whereas the scalar potential is proportional to the squared $z$ component of the spin operator, $W=\hat J_{z}^{2}\, k^{2}/(4M)\,.$ The eigenfrequencies are now
 \begin{equation}
 \hbar\Omega_{K}^{\beta}=\frac{\hbar^{2}}{2M}\Bigl(K^{2}+\sqrt 2\,Kk\beta+k^{2}\Bigr)+V\,,\quad\beta=0
 \,, \pm 1,
  \label{eq:omega-k-beta-quatro-pod}
 \end{equation}
with $K=\vert {\bs K}\vert$. For $\beta=\pm1$ the dispersion curves are analogous to those of the spin-$1/2$ RD model. The additional dispersion curve with $\beta=0$ corresponds to a parabola centered at $K=0$.

A spectacular consequence of spin-orbit RD coupling is the Zitterbewegung, a phenonenon which was analyzed for cold atoms \cite{Vaishnav08PRL,Merkl08EPL,Song2009PRA,Larson2010PRA}, electrons in solids \cite{Schlieman06PRB,Cserti2006PRB,Rusin2009PRA} and trapped ions \cite{Lamata2007PRL,Rusin2010PRD}. It was recently observed for the latter systems \cite{Gerritsma2010Nature}. Another manifestation of the RD coupling is the negative refraction and reflection that occurs when a matter wave is incident on a potential step. This problem was investigated for spin-$1/2$ atoms \cite{Juz08PRL} and electrons \cite{Winkler09PRB,Dargys10SM}.  For small incident wavenumbers, $K\ll k$, the transmission probability is close to unity for normal incidence. This nearly complete transmission is a manifestation of the Klein paradox appearing also for electron tunneling in graphene \cite{Katsnelson06NP,Castro-Neto:2009RMP}. The transmission probability decreases when the angle of incidence increases, and the transmitted matter wave experiences negative refraction, again similar to electrons in graphene \cite{Cheianov07Science}. Particles with a spin larger than $1/2$ have additional internal degrees of freedom, which modifies the continuity conditions at the potential step. For the negative refraction phenomenon at non-normal incidence, \textcite{Juzeliunas:2010} have shown that the amplitude of the refracted beam is significantly increased in comparison with the spin $1/2$ case.

Spin-orbit coupling with its multicomponent dispersion curves also offers interesting scenarios in the presence of collisional interactions \cite{Stanescu07PRL,Stanescu08PRA,Wang:2010,Yip:2010}. In two dimensions with a symmetric RD gauge potential of the form $\hbar k(\sigma_x{\bs e}_x+\sigma_y{\bs e}_y)$, the dispersion curve has a minimum at a constant non-zero radius in momentum space. This corresponds to a massive degeneracy of the single-particle ground level, since all plane waves with momentum $\bs K$ such that $|\bs K|=k$ are possible ground states. For a Bose gas,  this precludes in principle  the formation of a condensate, and atomic interactions will lead to the formation of a strongly correlated ground state. However in the presence of an asymmetry in the RD coupling, the massive degeneracy is lifted and the minimum of the dispersion occurs for two opposite values of the atomic momenta  \cite{Stanescu07PRL,Stanescu08PRA}. Weakly-interacting bosons with such RD coupling will then behave essentially as a two-component system at low temperature.

We conclude by signalling a possible drawback of the tripod scheme:  The two dark state states forming the effective spin-1/2 system are not the lowest single-particle energy states, hence collisional deexcitation can transfer the atoms out of  the dark state manifold down to the ground dressed state. To overcome this difficulty, a scheme involving $N$ ground or metastable internal states cyclically coupled by laser fields was recently proposed  \cite{Camplbell:2011}.  By properly setting the direction and phases of the laser fields, a pair of degenerate pseudospin states with lowest energy emerge. The states are subjected to the RD coupling and are immune to the collisional decay.

\subsection{Non-Abelian Aharonov--Bohm effect}
\label{subsec:Non_Abelian_AB}

The spin-orbit coupling that we presented in the preceding subsection can also be the source of quasi-relativistic dynamics, where the atomic centre of mass motion is governed by an effective Dirac equation in the limit of small momenta and strong gauge potentials. The quasi-relativistic dynamics of the atomic centre of mass is a remarkable effect which connects phenomena from high energy physics, graphene and photonic crystals with atomic physics and quantum optics. It relies on the two-  or three-component nature of the effective atomic internal structure, but it is not explicitly linked to the non-Abelian nature of the coupling. 

To observe a non-Abelian effect one should study spin dynamics. In our case it will be a pseudo spin, \emph{i.e.} an effective multi-component system created by the interaction between incident laser beams and the atoms. 
Previously we have shown that the gauge potential affecting the pseudo spin can be for instance proportional to the Pauli spin matrices in the $x$ and $y$-direction such that the matrices corresponding to the two directions do not commute. This is indeed also the requirement for having a non-Abelian system. 
A simple illustration of a non-Abelian situation is the so called non-Abelian Aharonov--Bohm experiment \cite{Jacob07APB}. Here we envisage a particle with an internal structure, the pseudo spin, which is allowed to move along two different paths from $A$ to $B$ along straight lines in $x$ and $y$-direction. Let us suppose the path consists of two equidistant trips of length $L$ along the $x$ and $y$ directions. The question is, if we start at $A$, with a given orientation of the pseudo spin, what is the spin at the final point $B$? If the spin is subject to a non-Abelian field, the result will depend on which path the particle took. This is easily seen by neglecting any external dynamics and only considering the effects from the gauge potential, which is of the form ${\bs A}={A_x\bs e_x}+{ A_y\bs e_y}$. The final state based on the path going first along $x$-direction and then $y$-direction will be given by $\exp[i{A_y} L]\exp[i{A_x} L]\tilde\Psi_{\rm in}$ where $\tilde \Psi_{\rm in}$ is the initial pseudo spin. If  we take the path along the $y$-direction first and then in the $x$-direction, we will have the final state  $\exp[i{A_x} L]\exp[i{A_y} L]\Psi_{\rm in}$. These two states are not necessarily the same, because $A_x$ and $A_y$ do not commute if the system is non-Abelian. For instance, when $A_{x}\propto\hat{\sigma}_{y}$ and $A_{y}\propto\hat{\sigma}_{x}$, these operations are equivalent to spin rotations around $x$ and $y$, which do not yield the same final state in general. Interestingly this situation is similar to the scattering of protons onto a non-Abelian flux line where the protons are anticipated to be converted into neutrons \cite{Horvathy:1986}.


\section{Gauge potentials in optical lattices}

\label{sec:opticallattices}
Optical lattices have recently emerged as a major tool for the field of quantum gases [see \textcite{Lewenstein:2007} and \textcite{Bloch:2008} for recent reviews]. They correspond to a periodic array of trapping sites connected to each other by quantum tunneling. Such a potential landscape is created using the interference pattern of several off-resonant lasers. In addition to the obvious analogy with the effective periodic potential exerted by the ionic matrix on electrons in a real solid, optical lattices allow one to bring quantum gases into the strongly correlated regime, where interactions dominate the behavior of the system. The entrance in the correlated regime is obtained by increasing the lattice depth and it follows from two effects, the increase of the on-site interaction energy due to the stronger confinement of the atoms near the lattice sites, and the reduction of the tunneling probability from site to site which decreases the kinetic energy. 

In this section we first review some basic elements on the band structure in a periodic potential (\S~\ref{subsec:lattices_reminder}) and the effects of a magnetic field on the single particle spectrum (\S~\ref{subsec:lattices_Harper}). We then discuss the concept of \emph{laser assisted tunnelling}, which allows one to control the phase of the tunneling matrix elements and realize artificial gauge potentials (\S~
\ref{subsec:lattices_flux_cell} and \ref{subsec:lattices_rectification}). Very recently  another class of lattices with orbital magnetism, the so-called \emph{flux lattices}, has been proposed by \textcite{Cooper:2011}. We briefly relate this proposal as well as those based on laser assisted tunnelling to the dressed state approach discussed in the previous section in \S~\ref{subsec:lattice_connection_DS}. Finally we give a short account on the possibility to generate non-Abelian fields in lattices in \S~\ref{subsec:lattice_non_Abelian}.

Let us mention briefly some alternative methods that do not use laser coupling to realize effective magnetic fields for cold atoms in optical lattices. One is simply to rotate the lattice, which was realized experimentally in \textcite{Tung:2006} and \textcite{Williams:2010} for lattices with large lattice spacing, {\it i.e.} outside the Hubbard regime.  Another method relies on a temporal modulation of the lattice potential with $x-y$ \cite{Sorensen:2005} or $x^2-y^2$ \cite{Lim:2008a} symmetries (see also \textcite{Kolovsky:2011EPL}). One can also control the sign of the tunnelling matrix element by modulating of the lattice potential [\textcite{Eckardt:2005}, \textcite{Lignier:2007}], which can provide artificial magnetism for some specific lattice geometries. Finally on can obtain the desired gauge potential through interactions with a bath of atoms \cite{Klein:2009a}. 

\subsection{Reminder on band structure}
\label{subsec:lattices_reminder}

Consider  a particle moving in the two-dimensional square lattice potential with period $d$ and depth $V_0$
 \begin{eqnarray}
V_{\rm lat}(x,y)=V_{0}\left[\sin^2\left({\pi x}/{d}\right) + \sin^2\left( {\pi y}/{d}\right)\right].
\end{eqnarray}
The energy eigenstates are the Bloch waves $\psi_{\eta,{\bs q}} ({\bs r})=e^{i{\bs q}\cdot {\bs r}}\,u_{\eta,{\bs q}} ({\bs r})$, where $\eta=0,1,\ldots$ denotes the band index and the quasi momentum ${\bs q}$ takes values in the first Brillouin Zone (BZ)  [$-\pi/d,\pi/d$[$\times$[$-\pi/d,\pi/d$[ [see for example \textcite{Ashcroft:1976}]. The function $u_{\eta,{\bs q}}$ is a periodic function of $\bs r$, with period $d$ along both directions of space. The energies $\epsilon_{\eta}({\bs q}$) associated with the Bloch waves form bands when $\bs q$ is varied inside the first Brillouin zone. 

We assume in the following that the lattice is in the tight-binding (TB) regime, where the energy width of each band is much smaller than the gaps between two consecutive bands. More specifically, we consider particles confined to the lowest Bloch band $\eta=0$ as it is usually the target regime for experiments on correlated quantum gases, and we subsequently drop the band index. In this limit, it is useful to transform to the orthogonal basis of Wannier functions 
\begin{equation}
w_{n,m}({\bs r})=\mathcal{N}\int_{\rm BZ} e^{-i {\bs q}\cdot{\bs r}_{n,m}}\, \psi_{{\bs q}}(\bs r)\;d^2q
\label{eq:Wannier}
\end{equation}
where $\cal{N}$ is a normalization factor and ${\bs r}_{n,m}=d(n {\bs e}_{x}+m {\bs e}_{y})$, with $n,m$ being integers.  In the TB regime, the Wannier function $w_{n,m}$ is localized near the lattice site at position ${\bs r}_{n,m}$ and the initial Hamiltonian $P^2/(2M)+V_{\rm lat}$ can be replaced by
\begin{equation}
H_{\rm TB}=-J \sum_{\rm n.n.} \hat{a}_{n,m}^\dagger \hat{a}_{n',m'} +{\rm h.c.},
\label{eq:TBfree}
\end{equation}
where $\hat{a}_{n,m}$ is an annihilation operator for a particle in the state $w_{n,m}$. The sum runs over nearest neighbors (n.n.) only, and $J$ is the tunelling energy characterizing the hopping between neighboring sites. Hopping to more distant sites is neglected in this approximation. The single-particle dispersion is in the TB approximation $\epsilon({\bs q})=-2J [\cos(q_{x}d) + \cos(q_{y}d)]$, corresponding to an energy width of $8J$ for the band.

\subsection{Harper equation and Hofstadter butterfly}
\label{subsec:lattices_Harper}

Suppose now that the square lattice considered above is placed in a uniform magnetic field $\bs B=B\,\bs e_z$, associated to the vector potential $\bs A=(-By,0,0)$ (Landau gauge). For a particle carrying a charge $e$, the Hamiltonian (\ref{eq:TBfree}) can be written 
\begin{eqnarray}\nonumber \label{eq:HAB}
H&=&-J \sum_{n,m,\pm} e^{\pm i \phi_{n,m}}\hat{a}_{n \pm 1,m}^\dagger \hat{a}_{n,m}\\
\label{eq:Harper} & & -J \sum_{n,m,\pm} \hat{a}_{n,m \pm 1}^\dagger \hat{a}_{n,m}+{\rm h.c.},
\end{eqnarray}
with
\begin{equation}\label{eq:AB}
\phi_{n,m}=\frac{e}{\hbar} \int_{{\bs r}_{n,m}}^{{\bs r}_{n+1,m}} {\bs A}\cdot d{\bs s},
\end{equation}
where the integral is taken along a straight line joining neighboring sites. The appearance of the phase factors $e^{\pm i  \phi_{n,m}}$ in Eq.~(\ref{eq:HAB}) can be understood in terms of the Aharonov--Bohm phase accumulated along a straight line joining two adjacent sites of the lattice.  With the choice of the Landau gauge,
these phase factors are 
\begin{equation} \label{eq:phiLandau}
{\bs r}_{n,m} \rightarrow {\bs r}_{n+1,m}:\quad \phi_{n,m} =  2\pi\alpha m,
\end{equation}
where $\alpha=\Phi/\Phi_0$, $\Phi=Bd^2$ is the magnetic flux through a unit cell and $\Phi_0=h/e$ is the flux quantum. A different gauge choice would lead to different phase factors $\phi_{n,m}$, but the phase accumulated on a closed circuit around an elementary cell
\begin{equation}
\gamma=\sum_{\square} \phi_{n,m} = {e B d^2}/{\hbar} = 2\pi\alpha 
\label{eq:sumplaquette}
\end{equation}
is gauge invariant (the symbol $\square$ indicates that the sum runs over an elementary cell of the lattice). 

The reduction of the full Schr\"odinger equation to Eq.~(\ref{eq:HAB}) involves another approximation, the so-called {\it Peierls's substitution} [see \textcite{Luttinger:1951} for the discussion of this approximation in the TB limit].  It  is valid if the Landau energy $\hbar \omega_c$, where $\omega_{c}=B/M$ is the cyclotron frequency, remains much smaller than the energy gap between the bands $n=0$ and $n=1$. Outside of the TB regime, its validity has to be carefully examined \cite{Nenciu:1991}. In the context of our Colloquium, these considerations are not relevant as we will discuss methods to simulate the Hamiltonian (\ref{eq:HAB}) directly.

\begin{figure}[t]
\centerline{\includegraphics{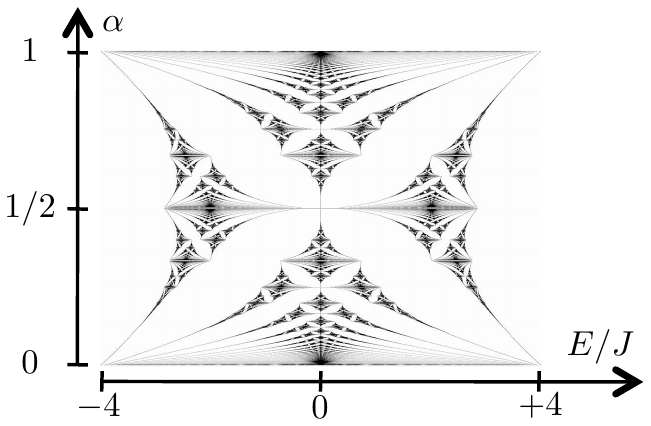}}
\caption{Hofstadter butterfly: single particle spectrum of the Hamiltonian of Eq. (\ref{eq:Harper}) in units of the tunneling amplitude $J$, for $\alpha=\phi/\phi_0$ varying between 0 and 1.}
\label{fig:hofstadter}
\end{figure}

The single particle eigenstates of the Hamiltonian (\ref{eq:HAB}) can be searched in the form $\sum_{m,n}C_m e^{iq_x nd}\hat{a}_{n,m}^\dagger|\mbox{vac}\rangle$. The corresponding eigenvalue equation for the coefficients $C_m$  is known as the Harper equation  \cite{Harper:1955}.  Its solutions have been extensively studied in connection with the quantum Hall effect \cite{Thouless:1982a,Kohmoto:1989a}. The energy spectrum, represented in Fig.~\ref{fig:hofstadter}, is known as the \emph{Hofstadter butterfly} \cite{Hofstadter:1976}, and it exhibits a remarkable self-similarity. This spectrum has a much more complex structure than its counterpart for a continuous system,  the latter being composed of uniformly spaced, infinitely degenerate Landau levels. This structure can be qualitatively understood on a relatively simple ground. For a rational value of $\alpha$, that is $\alpha=p/q$ with $p,q$ integers, the Hamiltonian is still periodic but with a larger period $qd$ along the $y$ axis. One can then divide the system into ``macro-cells'' of size $q d$ and look for new eigenstates [the so-called {\it magnetic Bloch functions} \cite{Blount:1966}] appropriate for the ``macro-lattice''. Because the macro-cell contains $q$ original lattice sites, each energy level is $q-$fold degenerate and  the fundamental band splits into $q$ subbands separated by small energy gaps (with the exception of $\alpha=1/2$, where the two bands touch at the edges of the reduced Brillouin zone). The Landau levels structure is recovered in the weak flux limit where $\alpha=p/q \ll 1$. We note finally the special case of $\alpha=1/2$, corresponding to a real tunneling matrix element alternating in sign from one column to the next. In this case, the excitation spectrum  features two `Dirac points' for $(k_x,k_y)=(\pm\pi/d_x, \pi/2d_y)$ around which the dispersion relation is linear \cite{Hatsugai:1990,Lim:2008a,Hou:2009a}. The behavior of an ultracold Fermi gas in such a situation is expected to be similar to that observed in graphene \cite{Castro-Neto:2009RMP}.

\subsection{Simulating a magnetic flux through each lattice cell} 
\label{subsec:lattices_flux_cell}

We now discuss how the Hamiltonian (\ref{eq:Harper}) can be realized for cold atoms in an optical lattice. We  proceed in two steps. First we introduce the notion of laser assisted tunneling and show how it can be used to obtain a non-zero flux through each lattice cell. However laser assisted tunneling in its simplest version provides a flux that alternates in sign between neighboring cells, and thus does not simulate a uniform magnetic field. In a second step (Sec. \ref{subsec:lattices_rectification}), we will show how one can rectify this magnetic  flux and reach the Hamiltonian (\ref{eq:Harper}). 

\begin{figure}[t]
\centering{\includegraphics[width=80mm]{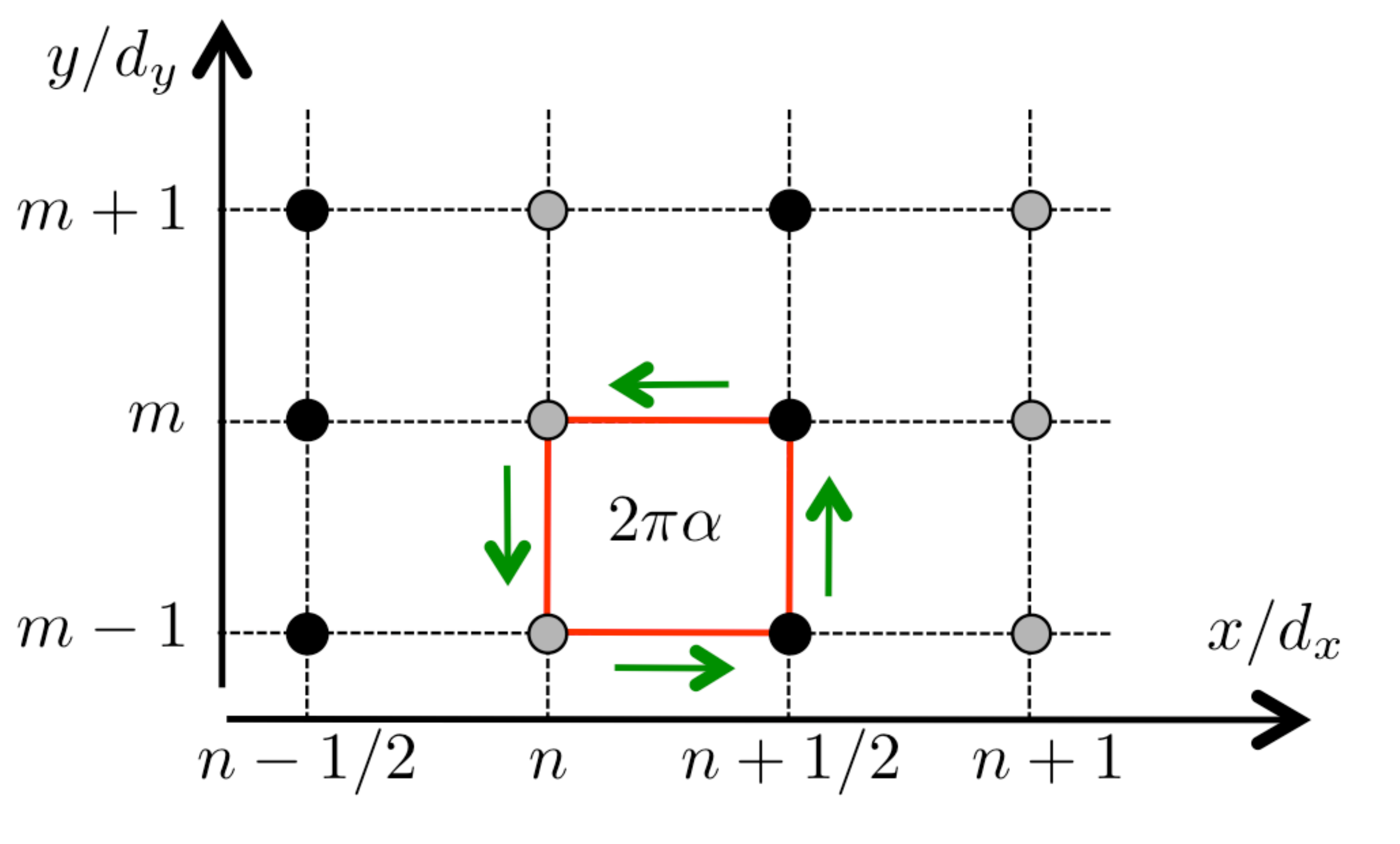}}
\caption{(Color online)  Sketch of the state-dependent 2D lattice potential. Grey (black) dots mark lattice sites for atoms in the ground $g$ (excited $e$) state, respectively.  We highlight a trajectory around an elementary cell of the coupled lattice in which the particle acquires the Aharonov--Bohm phase $2\pi \alpha$. The $g \to g$ and $e\to e$ transitions occur thanks to standard tunneling. The $g\to e$ and $e\to g$ transitions correspond to resonant laser assisted tunneling. In a $g\to e$ transition the atom absorbs a photon and jumps from a state localized around  $x=nd_x$ to a state localized around $x=(n\pm 1/2)d_x$ (and vice versa for a  $e\to g$ transition).}
\label{fig:2Dlattice}
\end{figure}

The notion of laser assisted tunneling has been introduced in this context by \textcite{Ruostekoski:2002} and \textcite{Jaksch:2003}. The first ingredient is to design a  \emph{state-dependent lattice}. Consider atoms with two internal states $g$ and $e$, trapped in spatially separated sublattices (see Fig.~\ref{fig:2Dlattice}). We focus on the situation where each sublattice has rectangular symmetry with lattice spacings $d_{x}$  and $d_{y}$. The $e$ sublattice is deduced from the $g$ sublattice by a translation ${d_{x}}/{2}$ along the $x$ axis. We label 
\begin{eqnarray}
{\bs r}_{2n,m}^{(g)}&=&n d_x {\bs e}_{x}+m d_y {\bs e}_{y}\,, \\ 
{\bs r}_{2n+1,m}^{(e)} &= & \left(n+\;{1}/{2} \right)d_x {\bs e}_{x}+m d_y{\bs e}_{y}\,,
\label{eq:lattices_sites}
\end{eqnarray}
the positions of the trapping sites on each sublattice, with the corresponding Wannier functions $w_{2n,m}^{(g)}$ and $w_{2n+1,m}^{(e)}$. Tunneling energies along $x$ and $y$ within a given sublattice ($g$ or $e$) are denoted by $J_{x}$ and $J_{y} $, respectively.

In practice, there are two options to generate such a state-dependent lattice. The first one, suitable for alkali atoms, selects $g$ and $e$ as two Zeeman or hyperfine states in the electronic ground state manifold (electronically excited states are too short lived for this application). In this case, the leading terms in the optical lattice potential is the sum of scalar and vector terms, the latter behaving as an effective magnetic field \cite{Dupontroc:1973a}. The states $g$ and $e$ are chosen such that they have opposite magnetic moments, and the detuning from resonance of the light beams forming the $x$ lattice is adjusted to that the polarizability of $g$ and $e$ are opposite \cite{Mandel:2003a}. This provides the desired state dependent lattice potential along the $x$ axis. Note however that the required detuning lies relatively close to the resonance lines, so that heating due to photon scattering can cause serious practical problems. 

The second option is to use atomic species already considered in Sec.~\ref{subsec:alkalineearth} with a long-lived excited state \cite{Gerbier:2010,Yi:2008}. Practical examples are alkaline-earth atoms or Ytterbium atoms, where the $^{3}P_{0}$ internal state has a typical lifetime over ten seconds. One then chooses the electronic (spin singlet) ground state for $g$ and the $^{3}P_{0}$ excited state for $e$. For such atoms, both states couple to different electronic states, and it is possible to find a \emph{magic wavelength} (used for the $y$ lattice) such that the polarizabilites are identical for both states, and an \emph{anti-magic wavelength} (used for the $x$ lattice) such that they are opposite. Both are far detuned from any resonance, so that heating by spontaneous emission is not an issue in this case.

In a state-dependent lattice, tunneling from a given site on the $g$ sublattice to a neighboring one on the $e$ sublattice can be driven by a resonant light field that couples $g$ and $e$. We assume this coupling laser is a plane running wave with wavevector ${\bs k}$. The corresponding complex tunneling matrix element is
 \begin{equation}\label{eq:Jeff}
g,{\bs r}_{2n,m}^{(g)} \rightarrow e,{\bs r}_{2n\pm 1,m}^{(e)}: J_{\rm eff} = \hbar \kappa^{(0)} \,\mathcal{O}\,  e^{i {\bs k}\cdot {\bs r}_{2n,m}^{(g)}},
\end{equation}
where $\kappa^{(0)}$ characterizes the strength of the  atom-laser coupling and the dimensionless number  
\begin{eqnarray}\label{eq:overlap}
\mathcal{O} & = & \int \left( w_{2n+1,m}^{(e)}({\bs r})\right)^\ast w_{2n,m}^{(g)}({\bs r}) \,e^{i {\bs k}\cdot {\bs r}}  \;d^{2}{r},
\end{eqnarray}
is the overlap integral between neighboring Wannier functions \cite{Jaksch:2003} (note that $w_{2n,m}^{(g)}$ and $w_{2n+1,m}^{(e)}$ are not orthogonal since they belong to different sublattices). Eq.~(\ref{eq:Jeff}) displays the essential phase factor that allows to reproduce the Aharonov--Bohm phase (\ref{eq:AB}). By choosing a laser  propagating in the $y-z$ plane, and under the assumption that state independent tunneling along $x$ is negligible ($J_{x }\ll \vert J_{\rm eff}\vert$), one obtains 
\begin{eqnarray}
\nonumber
H&=&-\vert J_{\rm eff}\vert \sum_{n,m,\pm} e^{ i \phi_{m}}\hat{b}_{2n \pm 1,m}^\dagger \hat{a}_{2n,m} +{\rm h.c.}\\
\label{eq:Halter} & -& J_{y} \sum_{n,m,\pm} \hat{a}_{2n,m \pm 1}^\dagger \hat{a}_{2n,m} + \hat{b}_{2n+1,m \pm 1}^\dagger \hat{b}_{2n+1,m} 
\end{eqnarray}
where $\hat a_{2n,m}^\dagger$ (resp. $\hat b_{2n+1,m}^\dagger$) creates an atom in the internal state $g$ (resp. $e$)  in the state $ w_{2n,m}^{(g)}$ (resp. $w_{2n+1,m}^{(e)}$). Here $e^{ i \phi_{m}}$ is the complex phase of $J_{\rm eff}$, characterizing the phase acquired by an atom when it tunnels from a site of the sublattice $g$ to a site of the sublattice $e$:
\begin{equation}
g,{\bs r}_{2n,m}^{(g)} \rightarrow e,{\bs r}_{2n\pm 1,m}^{(e)}: \quad \phi_{2n,m}=2\pi \alpha m
\label{eq:lattices_transitions}
\end{equation}
where $\alpha=k_y d_y/(2\pi)$ can be adjusted between 0 and 1 by changing the angle of the wave vector $\bs k$ of the coupling laser with the $y$ axis.

Although the Hamiltonian (\ref{eq:Halter}) contains complex hopping amplitudes, it does {\it not } yet coincide with the Hamiltonian ({\ref{eq:HAB}}) that we wish to simulate. In the initial Hamiltonian, the phase picked up across each $x$-link has the same sign for a given link direction in real space: $2\pi \alpha m$ in the $+x$ direction (and thus $- 2\pi \alpha m$ in the $-x$ direction since the Hamiltonian is Hermitian). Here, the phase is tied to the direction in internal space: $2\pi \alpha m$ for transitions to $g$ to $e$, irrespective of the fact that they occur in the $+x$ or $-x$ directions. 

In total, we achieve with  (\ref{eq:Halter}) a situation where the flux through a cell changes sign from one cell  to the next one along the $x$ axis. We anticipated this difficulty in Sec.\ref{subsec:circulation}, where we noted that realizing artificial gauge potentials with variations on the scale of the optical wavelength $\lambda_{L}$ would naturally result in an oscillating (or {\it staggered}) effective magnetic field. Although this configuration can lead to interesting situations \cite{Wang:2006a,Hou:2009a,Moller:2010}, it does not realize our goal of producing a uniform magnetic field, except for the specific case $\alpha=1/2$, since phase changes of $\pm \pi$ are equivalent. 

\subsection{Rectification of the magnetic field in the lattice}
\label{subsec:lattices_rectification}

The need for rectification of the magnetic field that we found in the preceding subsection can be formulated on more general grounds. We are looking for an effective Hamiltonian that must break time-reversal symmetry. For our two-level system, the time reversal operator is  $\mathcal{T}=\mathcal{K}_{\rm s}\mathcal{K}_{\rm c}$ \cite{Messiah_T_sym}, where $\mathcal{K}_{\rm s}=e^{i\pi\hat \sigma_{y}/2}= i \hat \sigma_{y}$ is such that $\mathcal{K}_{\rm s}|g\rangle=|e\rangle$, $\mathcal{K}_{\rm s}|e\rangle=-|g\rangle$, and $\mathcal{K}_{\rm c}$ is the conjugation operator. Now, the action of  $\mathcal{T}$ on the configuration studied in the previous section results in the same configuration, except for a mere translation by an amount $d_x/2$ along the $x$ axis. Hence, in order to produce a uniform magnetic field, one must add another ingredient that is not symmetric under the ``sublattice translation'' operation. 

\textcite{Jaksch:2003} proposed to realize this by adding a linear potential gradient along $x$, $V_{\rm grad}(n)=\zeta n$, so that the laser frequency for the $(g; 2n,m) \rightarrow (e; 2n+1,m)$ transition becomes shifted by $+\zeta$ while that for the $ (g; 2n+2,m)\rightarrow (e; 2n+1,m) $ transition is shifted by $-\zeta$. By choosing $\zeta \gg \vert J_{\rm eff}\vert$, the two transition frequencies become non-degenerate and must be adressed by two different lasers. One then chooses the laser resonant with $(g; 2n,m) \rightarrow (e; 2n+1,m)$ to propagate along $+{\bs e}_{y}$, and the laser resonant with $(e; 2n+1,m) \rightarrow (g; 2n+2,m)$ to propagate along $-{\bs e}_{y}$. According to Eq.~(\ref{eq:Jeff}), the sign of the phase factor changes on every second column, which results in a rectification of the staggered field to a uniform one (provided off-resonant terms are negligible compared to resonant ones). 
Practically, such a potential gradient can be implemented by a real electric field affecting $g$ and $e$ equally, or by the AC Stark shift exerted by an off-resonant laser. The large electric fields or optical powers involved, combined with the fact that the  tilting potential must be linear over the whole cloud to ensure a uniform transition freqency, make this option challenging from an experimental point of view. 

A possibly more practical configuration  is based on a superlattice potential with period $2d_{x}$ along $x$ on the top of the main lattice, which can perform the same role as the potential gradient of the previous paragraph \cite{Gerbier:2010}. One needs three (instead of two) different coupling laser frequencies in this case, but this is not a significant increase in difficulty since the frequency differences are in the tens of kHz range, and can thus be easily generated using frequency modulators. Another proposal by \textcite{Mueller:2004} realizes the same goal by using three internal states in three different sublattices. In that case, the distinction between neighboring transitions is automatic, and there is no need for an additional external potential. Unfortunately this trapping configuration seems difficult to implement in practice for the commonly used atomic species. 

\subsection{Connection with dressed state approach}
\label{subsec:lattice_connection_DS}

It is interesting to establish a link between the optical lattice case and the schemes discussed in sections \ref{sec:two-level} and \ref{sec:Three-level-system}, where we assumed that the atoms were following adiabatically one internal dressed state, defined as one of the local eigenstates of the atom-laser coupling matrix. We first discuss the case of lattices with laser assisted tunneling that we have considered in the first part of the present section. We then briefly present the concept of \emph{flux lattices} recently introduced by \textcite{Cooper:2011}.

\paragraph*{Lattices with laser assisted tunnelling.} 
The dressed states can  of course also be defined in the lattice case. For simplicity we take a uniform Rabi frequency $\kappa^{(0)}$ for the coupling laser and a spatially varying detuning $\Delta({\bs r})=\Delta_0 +(V_e(\bs r)-V_{g}(\bs r))/\hbar$, where $V_{g}$ and $V_{e}$ denote the potential energies for the $g$ and $e$ sublattices, respectively. We denote $V_0$ the characteristic amplitude of oscillation of $V_e$ and $V_g$, and we assume that $\hbar \kappa^{(0)}\ll V_0$ to ensure that the coupling laser does not lead to a strong deformation of the state-dependent lattice potential. As explained in Sec.~\ref{subsec:validity}, this corresponds to a situation such that the mixing angle $\theta$ varies rapidly around the points where the resonance condition $\Delta(\bs r)=0$ occurs.  We use as a typical value $|\bs \nabla \Delta|\approx k V_0/\hbar$, where $k$ is a typical optical wave number. From the discussion of Sec.~\ref{subsec:validity}, we obtain the validity condition for the adiabatic approximation 
$\kappa^{(0)}\gg \sqrt{V_{0}E_{\rm R}}/\hbar$. In the TB limit the right hand-side is approximately the  frequency gap $\omega_{\rm gap}$ between the ground ($\eta=0$) and the first excited ($\eta=1$) band. With the adiabaticity condition now written as $\kappa^{(0)}\gg \omega_{\rm gap}$, we immediately conclude that the adiabatic following of a given dressed state is intrinsically incompatible with the single-band model used this section. In other words the laser-assisted tunneling scheme considered in this section is a ``diabatic process" [\textcite{Smith69PR}] with respect to the dressed state basis, and it is therefore qualitatively different from the adiabatic scheme considered in sections I and II to generate artificial gauge fields in a bulk system. 

\paragraph*{Flux lattices \cite{Cooper:2011}.}
A flux lattice is based on a purely periodic 2D pattern of interfering laser fields. Atoms are modelled as two-level systems like in section \ref{sec:two-level}, with a coupling matrix $U$ in Eq.~(\ref{eq:hamiltonian2}) whose coefficients $\Omega$, $\theta$ and $\phi$ are periodic functions of $x$ and $y$. The atom-laser coupling is chosen large enough so that one approaches the regime where the atom follows adiabatically one dressed state (Eq.~(\ref{eq:validity_intuitive})), which is thus opposite to the case of laser assisted tunnelling discussed above.  Because the vector potential $\bs A$ calculated from Eq.~(\ref{eq:vect}) is also a periodic function of $x$ and $y$, one could think that the flux of the magnetic field across a unit cell of the lattice, which is equal to the circulation of $\bs A$ on the contour of the cell, is always zero. However this does not hold if the laser field is chosen such that $\bs A$ has  singularities inside the cell. Such singularities occur at locations where $\sin \theta=0$, so that $\phi$ is ill-defined. Contrary to $\bs A$, the magnetic field $\bs B$ in these points, calculated from Eq.~(\ref{eq:chpmag}), is non-singular and \textcite{Cooper:2011} showed that there exist configurations where $B_z$ keeps a constant sign over the cell,\footnote{More precisely the total field consists of a regular background component with non-zero average, plus a periodic array of gauge-dependent Dirac strings -- located in points where $\cos \theta=-1$ for the gauge choice of Eq.\, (\ref{eq:eigenstates}) -- and carrying a flux of opposite sign as compared to the background. The regular and singular components together give a zero net flux over an elementary cell, as expected for a periodic Hamiltonian. However the Dirac strings are non-measurable and only the background component has an observable effect.}  with a magnitude $\sim \hbar k^2$ . In particular the Chern number of the lowest allowed energy band \cite{Thouless:1982a} can be non-zero. This scheme thus constitutes an attractive alternative to the ones based on laser assisted tunnelling.

\subsection{Non Abelian gauge fields in a lattice}
\label{subsec:lattice_non_Abelian}

Optical lattices are also well suited for the generation of artificial non-Abelian magnetic fields, as first proposed by \textcite{Osterloh:2005}, and we conclude this section by a brief description of a possible implementation. As in Sec.~\ref{sec:Non-Abelian}, the basic idea to generate non-Abelian potentials is to use several internal atomic levels that are coupled by various laser beams. Here we consider an atomic species with  $2N$ quasi-degenerate sublevels,  trapped either in sublattice $g$ (sublevels $ g_{i}$, $i=1,\cdots, N$) or in sublattice $e$ (sublevels $e_j$, $j=1,\cdots, N$). Instead of a simple phase, laser-induced couplings must induce a rotation in internal space, generated by a $N\times N$ matrix ${\bs M}$ with non-commuting components $M_x,M_y,M_z$. Consider for instance a species with spin $1/2$ in both the ground and excited manifolds (the spin $1/2$ plays the role of a fictitious ``color'' charge here), which moves in an optical lattice that is now state-dependent both along $x$ and $y$. Laser-induced tunneling is used along both axes, with the additional possibility of changing the spin index $m_{z}=\pm1/2$. For instance, the lasers inducing tunneling along $y$ can be chosen to flip the spin ($m_{z}\rightarrow m'_{z}=-m_z$), so that $M_{y}\propto \alpha \hat{\sigma}_{x} + \beta \hat{\sigma}_{y}$, where $\alpha, \beta$ are numerical coefficients depending on the laser phases. The lasers inducing tunneling along $x$ are then chosen to preserve the spin ($m_{z}\rightarrow m'_{z}=m_z$), so that $M_{x}\propto \gamma {\hat 1} + \delta  \hat{\sigma}_{z}$. Provided that $\delta\neq 0$, {\it i.e.} the $x$-lasers apply a different phase conditionally on the internal ``color'', this realizes a non-Abelian potential.  The physical effects that one expects from this non-Abelian settings are similar to those discussed in Section \ref{sec:Non-Abelian}, in particular the emergence of a Rashba-like spin-orbit coupling \cite{Dudarev04PRL,Osterloh:2005,Goldman:2007b,Satija:2008,Goldman:2009a,Goldman:2009b,Goldman:book}.


\section{Outlook}

We have presented in this Colloquium the physical principles that lead to the generation of artificial gauge potentials on neutral atoms, using their coupling to laser fields. We have considered the cases of bulk systems and discrete lattices, and we have shown that both Abelian and non Abelian gauge potentials are accessible, provided one chooses a  suitable atomic level structure and proper incident light fields. We have also explored several physical consequences of these gauge fields, such as the nucleation of quantized vortices in a superfluid, the generation of a spin-orbit coupling \emph{via} a non-Abelian gauge field, or the possibility to address the strong magnetic field limit in a 2D lattice with the corresponding Hofstadter butterfly energy spectrum. The variety of the items in this (incomplete) list shows the richness of the situations that can be realized. 

These artificial fields also constitute novel tools for characterizing the properties of an assembly of atoms. \textcite{Cooper:2010} suggested to use a small artificial magnetic field to probe the superfluidity of a gas. The field  is generated with a laser scheme close to the one of Sec.~\ref{subsec:orbital_angular_momentum}, and it is used to simulate a \emph{rotating bucket} experiment and measure the reduced moment of inertia of a fluid when it acquires a superfluid component. In presence of the artificial field the normal component will stay at rest in the lab frame, whereas the superfluid component will rotate. A key feature of this proposal is that one can use spectroscopic methods to access the respective populations of the superfluid and normal states, by measuring the populations of the various ground states $g_j$ involved in the process.

In order to keep this Colloquium within reasonable limits, we intentionally restricted our analysis to single particle or mean-field physics.  However it is clear that these gauge fields can lead to very interesting phenomena when combined with the richness of strongly correlated states that emerge in many-body physics. For the Abelian case in the continuum limit, the situation is similar to the case of rotating gases, and paths to Quantum Hall physics that have been identified for Bose and Fermi gases remain of order [for a review see \textcite{Cooper:2008}]. 
In the context of the Bose--Hubbard model in presence of a uniform magnetic field, \textcite{Moller:2009} have recently proposed an approach based on the composite fermion theory to establish the existence of strongly correlated phases that have no equivalent in the continuum limit (see also \textcite{Polak:2009}). Another line of research with optical lattices deals with the combination of artificial gauge fields and nearest-neighbor interactions, as recently explored by \textcite{Ruostekoski:2009}.  

To end on a somehow futuristic tone, we note that with the generation of non-Abelian gauge fields on an atomic gas, one will have at hand matter with intriguing topological characteristics. A possible application is the simulation of topological insulators with neutral atoms [see e.g. \textcite{Wu:2008}, \textcite{Stanescu:2010} and \textcite{Goldman:2010a}]. Another intriguing perspective could be topological quantum computing \cite{Sarma:2006}. In this respect the artificial gauge fields provide a different scenario compared to the anticipated emergent non-Abelian excitations in for instance fractional quantum Hall systems. With artificially created gauge potentials the pseudo-spin is effectively turned into a non-Abelian anyon, since two pseudo-spins which swap places will be represented by different final states depending on wether they were interchanged clockwise or counter clockwise. Such effects may provide an atomic building block for fault tolerant topological quantum computing.
 
\begin{acknowledgments} 
This work is partially supported by IFRAF, ANR (BOFL project), the European Union (MIDAS and NAME-QUAM STREP projects), EGIDE, Research Council of Lithuania (TAP-44/2010 and TAP-17/2010 projects), and DARPA (OLE project).  We are grateful to A.~Dargys, A.~Fetter, M.~Fleischhauer, K.\, J.~G\"unter, M.~Lewenstein, W.\,D.~Phillips,  J.~Ruseckas, I.\,B.~Spielman and A.~Zee for several enlightening discussions.
\end{acknowledgments}


\bibliographystyle{apsrmp}

\begin{thebibliography}{141}
\expandafter\ifx\csname natexlab\endcsname\relax\def\natexlab#1{#1}\fi
\expandafter\ifx\csname bibnamefont\endcsname\relax
  \def\bibnamefont#1{#1}\fi
\expandafter\ifx\csname bibfnamefont\endcsname\relax
  \def\bibfnamefont#1{#1}\fi
\expandafter\ifx\csname citenamefont\endcsname\relax
  \def\citenamefont#1{#1}\fi
\expandafter\ifx\csname url\endcsname\relax
  \def\url#1{\texttt{#1}}\fi
\expandafter\ifx\csname urlprefix\endcsname\relax\def\urlprefix{URL }\fi
\providecommand{\bibinfo}[2]{#2}
\providecommand{\eprint}[2][]{\url{#2}}

\bibitem[{\citenamefont{Aharonov and Bohm}(1959)}]{Aharonov-Bohm59PR}
\bibinfo{author}{\bibnamefont{Aharonov}, \bibfnamefont{Y.}}, and
  \bibinfo{author}{\bibfnamefont{D.}~\bibnamefont{Bohm}}, \bibinfo{year}{1959},
  \bibinfo{journal}{Phys. Rev.} \textbf{\bibinfo{volume}{115}},
  \bibinfo{pages}{485}.

\bibitem[{\citenamefont{Aharonov and Stern}(1992)}]{Aharonov:1992}
\bibinfo{author}{\bibnamefont{Aharonov}, \bibfnamefont{Y.}}, and
  \bibinfo{author}{\bibfnamefont{A.}~\bibnamefont{Stern}},
  \bibinfo{year}{1992}, \bibinfo{journal}{Phys. Rev. Lett.}
  \textbf{\bibinfo{volume}{69}}(\bibinfo{number}{25}), \bibinfo{pages}{3593}.

\bibitem[{\citenamefont{Andersen} \emph{et~al.}(2006)\citenamefont{Andersen,
  Ryu, Clad\'e, Natarajan, Vaziri, Helmerson, and Phillips}}]{Andersen:2006}
\bibinfo{author}{\bibnamefont{Andersen}, \bibfnamefont{M.~F.}},
  \bibinfo{author}{\bibfnamefont{C.}~\bibnamefont{Ryu}},
  \bibinfo{author}{\bibfnamefont{P.}~\bibnamefont{Clad\'e}},
  \bibinfo{author}{\bibfnamefont{V.}~\bibnamefont{Natarajan}},
  \bibinfo{author}{\bibfnamefont{A.}~\bibnamefont{Vaziri}},
  \bibinfo{author}{\bibfnamefont{K.}~\bibnamefont{Helmerson}}, and
  \bibinfo{author}{\bibfnamefont{W.~D.} \bibnamefont{Phillips}},
  \bibinfo{year}{2006}, \bibinfo{journal}{Phys. Rev. Lett.}
  \textbf{\bibinfo{volume}{97}}, \bibinfo{pages}{170406}.

\bibitem[{\citenamefont{Arimondo}(1996)}]{Arimondo:1996}
\bibinfo{author}{\bibnamefont{Arimondo}, \bibfnamefont{E.}},
  \bibinfo{year}{1996}, in \emph{\bibinfo{booktitle}{Progress in Optics, vol.
  35}}, edited by \bibinfo{editor}{\bibfnamefont{E.}~\bibnamefont{Wolf}}
  (\bibinfo{publisher}{Elsevier}), p. \bibinfo{pages}{259}.

\bibitem[{\citenamefont{Ashcroft and Mermin}(1976)}]{Ashcroft:1976}
\bibinfo{author}{\bibnamefont{Ashcroft}, \bibfnamefont{N.~W.}}, and
  \bibinfo{author}{\bibfnamefont{N.~D.} \bibnamefont{Mermin}},
  \bibinfo{year}{1976}, \emph{\bibinfo{title}{Solid State Physics}}
  (\bibinfo{publisher}{Holt, Rinehardt and Winston}, \bibinfo{address}{New
  York}).

\bibitem[{\citenamefont{Aspect} \emph{et~al.}(1988)\citenamefont{Aspect,
  Arimondo, Kaiser, Vansteenkiste, and Cohen-Tannoudji}}]{Aspect:1988}
\bibinfo{author}{\bibnamefont{Aspect}, \bibfnamefont{A.}},
  \bibinfo{author}{\bibfnamefont{E.}~\bibnamefont{Arimondo}},
  \bibinfo{author}{\bibfnamefont{R.}~\bibnamefont{Kaiser}},
  \bibinfo{author}{\bibfnamefont{N.}~\bibnamefont{Vansteenkiste}}, and
  \bibinfo{author}{\bibfnamefont{C.}~\bibnamefont{Cohen-Tannoudji}},
  \bibinfo{year}{1988}, \bibinfo{journal}{Phys. Rev. Lett.}
  \textbf{\bibinfo{volume}{61}}, \bibinfo{pages}{826}.

\bibitem[{\citenamefont{Bergmann} \emph{et~al.}(1998)\citenamefont{Bergmann,
  Theuer, and Shore}}]{Bergmann:1998}
\bibinfo{author}{\bibnamefont{Bergmann}, \bibfnamefont{K.}},
  \bibinfo{author}{\bibfnamefont{H.}~\bibnamefont{Theuer}}, and
  \bibinfo{author}{\bibfnamefont{B.~W.} \bibnamefont{Shore}},
  \bibinfo{year}{1998}, \bibinfo{journal}{Rev. Mod. Phys.}
  \textbf{\bibinfo{volume}{70}}(\bibinfo{number}{3}), \bibinfo{pages}{1003}.

\bibitem[{\citenamefont{Berry}(1984)}]{Berry:1984}
\bibinfo{author}{\bibnamefont{Berry}, \bibfnamefont{M.~V.}},
  \bibinfo{year}{1984}, \bibinfo{journal}{Proc. Roy. Soc. London A}
  \textbf{\bibinfo{volume}{392}}, \bibinfo{pages}{45}.

\bibitem[{\citenamefont{Berry}(1989)}]{Berry:1989}
\bibinfo{author}{\bibnamefont{Berry}, \bibfnamefont{M.~V.}},
  \bibinfo{year}{1989}, in \emph{\bibinfo{booktitle}{Geometric {P}hases in
  {P}hysics}}, edited by
  \bibinfo{editor}{\bibfnamefont{A.}~\bibnamefont{Shapere}} and
  \bibinfo{editor}{\bibfnamefont{F.}~\bibnamefont{Wilczek}}
  (\bibinfo{publisher}{World Scientific}, \bibinfo{address}{Singapore}), pp.
  \bibinfo{pages}{7--28}.

\bibitem[{\citenamefont{Bloch} \emph{et~al.}(2008)\citenamefont{Bloch,
  Dalibard, and Zwerger}}]{Bloch:2008}
\bibinfo{author}{\bibnamefont{Bloch}, \bibfnamefont{I.}},
  \bibinfo{author}{\bibfnamefont{J.}~\bibnamefont{Dalibard}}, and
  \bibinfo{author}{\bibfnamefont{W.}~\bibnamefont{Zwerger}},
  \bibinfo{year}{2008}, \bibinfo{journal}{Rev. Mod. Phys}
  \textbf{\bibinfo{volume}{80}}(\bibinfo{number}{3}), \bibinfo{eid}{885}.

\bibitem[{\citenamefont{Blount}(1962)}]{Blount:1966}
\bibinfo{author}{\bibnamefont{Blount}, \bibfnamefont{E.~I.}},
  \bibinfo{year}{1962}, \bibinfo{journal}{Phys. Rev.}
  \textbf{\bibinfo{volume}{126}}(\bibinfo{number}{5}), \bibinfo{pages}{1636}.

\bibitem[{\citenamefont{Bohm} \emph{et~al.}(1992)\citenamefont{Bohm, Kendrick,
  Loewe, and Boya}}]{Bohm92JMP}
\bibinfo{author}{\bibnamefont{Bohm}, \bibfnamefont{A.}},
  \bibinfo{author}{\bibfnamefont{B.}~\bibnamefont{Kendrick}},
  \bibinfo{author}{\bibfnamefont{M.}~\bibnamefont{Loewe}}, and
  \bibinfo{author}{\bibfnamefont{L.}~\bibnamefont{Boya}}, \bibinfo{year}{1992},
  \bibinfo{journal}{J. Math. Phys.}
  \textbf{\bibinfo{volume}{33}}(\bibinfo{number}{3}), \bibinfo{pages}{977}.

\bibitem[{\citenamefont{Bohm} \emph{et~al.}(2003)\citenamefont{Bohm,
  Mostafazadeh, Koizumi, Niu, and Zwanziger}}]{Bohm:2003}
\bibinfo{author}{\bibnamefont{Bohm}, \bibfnamefont{A.}},
  \bibinfo{author}{\bibfnamefont{A.}~\bibnamefont{Mostafazadeh}},
  \bibinfo{author}{\bibfnamefont{H.}~\bibnamefont{Koizumi}},
  \bibinfo{author}{\bibfnamefont{Q.}~\bibnamefont{Niu}}, and
  \bibinfo{author}{\bibfnamefont{J.}~\bibnamefont{Zwanziger}},
  \bibinfo{year}{2003}, \emph{\bibinfo{title}{Geometric {P}hases in {Q}uantum
  {S}ystems}} (\bibinfo{publisher}{Springer}, \bibinfo{address}{Berlin,
  Heidelberg, New York}).

\bibitem[{\citenamefont{Buluta and Nori}(2009)}]{Buluta:2009}
\bibinfo{author}{\bibnamefont{Buluta}, \bibfnamefont{I.}}, and
  \bibinfo{author}{\bibfnamefont{F.}~\bibnamefont{Nori}}, \bibinfo{year}{2009},
  \bibinfo{journal}{Science} \textbf{\bibinfo{volume}{326}},
  \bibinfo{pages}{108}.

\bibitem[{\citenamefont{Campbell} \emph{et~al.}(2011)\citenamefont{Campbell,
  Juzeli\=unas, and Spielman}}]{Camplbell:2011}
\bibinfo{author}{\bibnamefont{Campbell}, \bibfnamefont{D.~L.}},
  \bibinfo{author}{\bibfnamefont{G.}~\bibnamefont{Juzeli\=unas}}, and
  \bibinfo{author}{\bibfnamefont{I.~B.} \bibnamefont{Spielman}},
  \bibinfo{year}{2011}, \bibinfo{journal}{arXiv:1102.3945} .

\bibitem[{\citenamefont{Cheianov} \emph{et~al.}(2007)\citenamefont{Cheianov,
  Fal'ko, and Altshuler}}]{Cheianov07Science}
\bibinfo{author}{\bibnamefont{Cheianov}, \bibfnamefont{V.~V.}},
  \bibinfo{author}{\bibfnamefont{V.}~\bibnamefont{Fal'ko}}, and
  \bibinfo{author}{\bibfnamefont{B.~L.} \bibnamefont{Altshuler}},
  \bibinfo{year}{2007}, \bibinfo{journal}{Science}
  \textbf{\bibinfo{volume}{315}}, \bibinfo{pages}{1252}.

\bibitem[{\citenamefont{Cheneau} \emph{et~al.}(2008)\citenamefont{Cheneau,
  Rath, Yefsah, G\"{u}nter, Juzeli\=unas, and Dalibard}}]{Cheneau:2008}
\bibinfo{author}{\bibnamefont{Cheneau}, \bibfnamefont{M.}},
  \bibinfo{author}{\bibfnamefont{S.~P.} \bibnamefont{Rath}},
  \bibinfo{author}{\bibfnamefont{T.}~\bibnamefont{Yefsah}},
  \bibinfo{author}{\bibfnamefont{K.~J.} \bibnamefont{G\"{u}nter}},
  \bibinfo{author}{\bibfnamefont{G.}~\bibnamefont{Juzeli\=unas}}, and
  \bibinfo{author}{\bibfnamefont{J.}~\bibnamefont{Dalibard}},
  \bibinfo{year}{2008}, \bibinfo{journal}{Europhys. Lett.}
  \textbf{\bibinfo{volume}{83}}(\bibinfo{number}{6}), \bibinfo{pages}{60001}.

\bibitem[{\citenamefont{Cohen-Tannoudji and
  Dupont-Roc}(1972)}]{Dupontroc:1973a}
\bibinfo{author}{\bibnamefont{Cohen-Tannoudji}, \bibfnamefont{C.}}, and
  \bibinfo{author}{\bibfnamefont{J.}~\bibnamefont{Dupont-Roc}},
  \bibinfo{year}{1972}, \bibinfo{journal}{Phys. Rev. A}
  \textbf{\bibinfo{volume}{5}}(\bibinfo{number}{2}), \bibinfo{pages}{968}.

\bibitem[{\citenamefont{Cohen-Tannoudji}
  \emph{et~al.}(1992)\citenamefont{Cohen-Tannoudji, Dupont-Roc, and
  Grynberg}}]{cohe92}
\bibinfo{author}{\bibnamefont{Cohen-Tannoudji}, \bibfnamefont{C.}},
  \bibinfo{author}{\bibfnamefont{J.}~\bibnamefont{Dupont-Roc}}, and
  \bibinfo{author}{\bibfnamefont{G.}~\bibnamefont{Grynberg}},
  \bibinfo{year}{1992}, \emph{\bibinfo{title}{Atom-Photon Interactions}}
  (\bibinfo{publisher}{Wiley}, \bibinfo{address}{New York}).

\bibitem[{\citenamefont{Cooper}(2008)}]{Cooper:2008}
\bibinfo{author}{\bibnamefont{Cooper}, \bibfnamefont{N.~R.}},
  \bibinfo{year}{2008}, \bibinfo{journal}{Advances in Physics}
  \textbf{\bibinfo{volume}{57}}(\bibinfo{number}{6}), \bibinfo{pages}{539}.

\bibitem[{\citenamefont{Cooper}(2011)}]{Cooper:2011}
\bibinfo{author}{\bibnamefont{Cooper}, \bibfnamefont{N.~R.}},
  \bibinfo{year}{2011}, \bibinfo{journal}{Phys. Rev. Lett.}
  \textbf{\bibinfo{volume}{106}}(\bibinfo{number}{17}),
  \bibinfo{pages}{175301}.

\bibitem[{\citenamefont{Cooper and Hadzibabic}(2010)}]{Cooper:2010}
\bibinfo{author}{\bibnamefont{Cooper}, \bibfnamefont{N.~R.}}, and
  \bibinfo{author}{\bibfnamefont{Z.}~\bibnamefont{Hadzibabic}},
  \bibinfo{year}{2010}, \bibinfo{journal}{Phys. Rev. Lett.}
  \textbf{\bibinfo{volume}{104}}(\bibinfo{number}{3}), \bibinfo{pages}{030401}.

\bibitem[{\citenamefont{Cserti and David}(2006)}]{Cserti2006PRB}
\bibinfo{author}{\bibnamefont{Cserti}, \bibfnamefont{J.}}, and
  \bibinfo{author}{\bibfnamefont{G.}~\bibnamefont{David}},
  \bibinfo{year}{2006}, \bibinfo{journal}{Phys. Rev. B}
  \textbf{\bibinfo{volume}{74}}, \bibinfo{pages}{172305}.

\bibitem[{\citenamefont{Dargys}(2010)}]{Dargys10SM}
\bibinfo{author}{\bibnamefont{Dargys}, \bibfnamefont{A.}},
  \bibinfo{year}{2010}, \bibinfo{journal}{Superlattice. Microstruct.}
  \textbf{\bibinfo{volume}{48}}, \bibinfo{pages}{221}.

\bibitem[{\citenamefont{Datta and Das}(1990)}]{dattadas}
\bibinfo{author}{\bibnamefont{Datta}, \bibfnamefont{S.}}, and
  \bibinfo{author}{\bibfnamefont{B.}~\bibnamefont{Das}}, \bibinfo{year}{1990},
  \bibinfo{journal}{Appl. Phys. Lett.} \textbf{\bibinfo{volume}{56}},
  \bibinfo{pages}{665}.

\bibitem[{\citenamefont{Dresselhaus}(1955)}]{Dresselhaus55PR}
\bibinfo{author}{\bibnamefont{Dresselhaus}, \bibfnamefont{G.}},
  \bibinfo{year}{1955}, \bibinfo{journal}{Phys. Rev.}
  \textbf{\bibinfo{volume}{100}}, \bibinfo{pages}{580}.

\bibitem[{\citenamefont{Dudarev} \emph{et~al.}(2004)\citenamefont{Dudarev,
  Diener, Carusotto, and Niu}}]{Dudarev04PRL}
\bibinfo{author}{\bibnamefont{Dudarev}, \bibfnamefont{A.~M.}},
  \bibinfo{author}{\bibfnamefont{R.~B.} \bibnamefont{Diener}},
  \bibinfo{author}{\bibfnamefont{I.}~\bibnamefont{Carusotto}}, and
  \bibinfo{author}{\bibfnamefont{Q.}~\bibnamefont{Niu}}, \bibinfo{year}{2004},
  \bibinfo{journal}{Phys. Rev. Lett.} \textbf{\bibinfo{volume}{92}},
  \bibinfo{pages}{153005}.

\bibitem[{\citenamefont{Dum and Olshanii}(1996)}]{Dum:1996}
\bibinfo{author}{\bibnamefont{Dum}, \bibfnamefont{R.}}, and
  \bibinfo{author}{\bibfnamefont{M.}~\bibnamefont{Olshanii}},
  \bibinfo{year}{1996}, \bibinfo{journal}{Phys. Rev. Lett.}
  \textbf{\bibinfo{volume}{76}}(\bibinfo{number}{11}), \bibinfo{pages}{1788}.

\bibitem[{\citenamefont{Dutta} \emph{et~al.}(1999)\citenamefont{Dutta, Teo, and
  Raithel}}]{Dutta:1999}
\bibinfo{author}{\bibnamefont{Dutta}, \bibfnamefont{S.~K.}},
  \bibinfo{author}{\bibfnamefont{B.~K.} \bibnamefont{Teo}}, and
  \bibinfo{author}{\bibfnamefont{G.}~\bibnamefont{Raithel}},
  \bibinfo{year}{1999}, \bibinfo{journal}{Phys. Rev. Lett.}
  \textbf{\bibinfo{volume}{83}}(\bibinfo{number}{10}), \bibinfo{pages}{1934}.

\bibitem[{\citenamefont{Eckardt} \emph{et~al.}(2005)\citenamefont{Eckardt,
  Weiss, and Holthaus}}]{Eckardt:2005}
\bibinfo{author}{\bibnamefont{Eckardt}, \bibfnamefont{A.}},
  \bibinfo{author}{\bibfnamefont{C.}~\bibnamefont{Weiss}}, and
  \bibinfo{author}{\bibfnamefont{M.}~\bibnamefont{Holthaus}},
  \bibinfo{year}{2005}, \bibinfo{journal}{Phys. Rev. Lett.}
  \textbf{\bibinfo{volume}{95}}(\bibinfo{number}{26}), \bibinfo{eid}{260404}.

\bibitem[{\citenamefont{Fert}(2008)}]{Fert:2008}
\bibinfo{author}{\bibnamefont{Fert}, \bibfnamefont{A.}}, \bibinfo{year}{2008},
  \bibinfo{journal}{Rev. Mod. Phys.}
  \textbf{\bibinfo{volume}{80}}(\bibinfo{number}{4}), \bibinfo{pages}{1517}.

\bibitem[{\citenamefont{Fetter}(2009)}]{Fetter:2008}
\bibinfo{author}{\bibnamefont{Fetter}, \bibfnamefont{A.~L.}},
  \bibinfo{year}{2009}, \bibinfo{journal}{Rev. Mod. Phys.}
  \textbf{\bibinfo{volume}{81}}(\bibinfo{number}{2}), \bibinfo{pages}{647}.

\bibitem[{\citenamefont{Feynman}(1982)}]{Feynman:1982}
\bibinfo{author}{\bibnamefont{Feynman}, \bibfnamefont{R.~P.}},
  \bibinfo{year}{1982}, \bibinfo{journal}{International Journal of Theoretical
  Physics} \textbf{\bibinfo{volume}{21}}, \bibinfo{pages}{467}.

\bibitem[{\citenamefont{Fleischhauer}
  \emph{et~al.}(2005)\citenamefont{Fleischhauer, Imamoglu, and
  Marangos}}]{Fleischhauer05RMP}
\bibinfo{author}{\bibnamefont{Fleischhauer}, \bibfnamefont{M.}},
  \bibinfo{author}{\bibfnamefont{A.}~\bibnamefont{Imamoglu}}, and
  \bibinfo{author}{\bibfnamefont{J.~P.} \bibnamefont{Marangos}},
  \bibinfo{year}{2005}, \bibinfo{journal}{Rev. Mod. Phys}
  \textbf{\bibinfo{volume}{77}}, \bibinfo{pages}{633}.

\bibitem[{\citenamefont{Gerbier and Dalibard}(2010)}]{Gerbier:2010}
\bibinfo{author}{\bibnamefont{Gerbier}, \bibfnamefont{F.}}, and
  \bibinfo{author}{\bibfnamefont{J.}~\bibnamefont{Dalibard}},
  \bibinfo{year}{2010}, \bibinfo{journal}{New Journal of Physics}
  \textbf{\bibinfo{volume}{12}}(\bibinfo{number}{3}), \bibinfo{pages}{033007}.

\bibitem[{\citenamefont{Gerritsma} \emph{et~al.}(2010)\citenamefont{Gerritsma,
  Kirchmair, Zahringer, Solano, Blatt, and Roos}}]{Gerritsma2010Nature}
\bibinfo{author}{\bibnamefont{Gerritsma}, \bibfnamefont{R.}},
  \bibinfo{author}{\bibfnamefont{G.}~\bibnamefont{Kirchmair}},
  \bibinfo{author}{\bibfnamefont{F.}~\bibnamefont{Zahringer}},
  \bibinfo{author}{\bibfnamefont{E.}~\bibnamefont{Solano}},
  \bibinfo{author}{\bibfnamefont{R.}~\bibnamefont{Blatt}}, and
  \bibinfo{author}{\bibfnamefont{C.~F.} \bibnamefont{Roos}},
  \bibinfo{year}{2010}, \bibinfo{journal}{Nature}
  \textbf{\bibinfo{volume}{463}}, \bibinfo{pages}{68}.

\bibitem[{\citenamefont{Goldman}(2007)}]{Goldman:2007b}
\bibinfo{author}{\bibnamefont{Goldman}, \bibfnamefont{N.}},
  \bibinfo{year}{2007}, \bibinfo{journal}{Europhys. Lett.}
  \textbf{\bibinfo{volume}{80}}(\bibinfo{number}{2}), \bibinfo{pages}{20001}.

\bibitem[{\citenamefont{Goldman}(2009)}]{Goldman:book}
\bibinfo{author}{\bibnamefont{Goldman}, \bibfnamefont{N.}},
  \bibinfo{year}{2009}, \emph{\bibinfo{title}{Quantum Transport in Lattices
  Subjected to External Gauge Fields}} (\bibinfo{publisher}{VDM Verlag}).

\bibitem[{\citenamefont{Goldman}
  \emph{et~al.}(2009{\natexlab{a}})\citenamefont{Goldman, Kubasiak, Bermudez,
  Gaspard, Lewenstein, and Martin-Delgado}}]{Goldman:2009b}
\bibinfo{author}{\bibnamefont{Goldman}, \bibfnamefont{N.}},
  \bibinfo{author}{\bibfnamefont{A.}~\bibnamefont{Kubasiak}},
  \bibinfo{author}{\bibfnamefont{A.}~\bibnamefont{Bermudez}},
  \bibinfo{author}{\bibfnamefont{P.}~\bibnamefont{Gaspard}},
  \bibinfo{author}{\bibfnamefont{M.}~\bibnamefont{Lewenstein}}, and
  \bibinfo{author}{\bibfnamefont{M.~A.} \bibnamefont{Martin-Delgado}},
  \bibinfo{year}{2009}{\natexlab{a}}, \bibinfo{journal}{Phys. Rev. Lett.}
  \textbf{\bibinfo{volume}{103}}(\bibinfo{number}{3}), \bibinfo{eid}{035301}
  (pages~\bibinfo{numpages}{4}).

\bibitem[{\citenamefont{Goldman}
  \emph{et~al.}(2009{\natexlab{b}})\citenamefont{Goldman, Kubasiak, Gaspard,
  and Lewenstein}}]{Goldman:2009a}
\bibinfo{author}{\bibnamefont{Goldman}, \bibfnamefont{N.}},
  \bibinfo{author}{\bibfnamefont{A.}~\bibnamefont{Kubasiak}},
  \bibinfo{author}{\bibfnamefont{P.}~\bibnamefont{Gaspard}}, and
  \bibinfo{author}{\bibfnamefont{M.}~\bibnamefont{Lewenstein}},
  \bibinfo{year}{2009}{\natexlab{b}}, \bibinfo{journal}{Phys. Rev. A}
  \textbf{\bibinfo{volume}{79}}(\bibinfo{number}{2}), \bibinfo{eid}{023624}
  (pages~\bibinfo{numpages}{11}).

\bibitem[{\citenamefont{Goldman} \emph{et~al.}(2010)\citenamefont{Goldman,
  Satija, Nikolic, Bermudez, Martin-Delgado, Lewenstein, and
  Spielman}}]{Goldman:2010a}
\bibinfo{author}{\bibnamefont{Goldman}, \bibfnamefont{N.}},
  \bibinfo{author}{\bibfnamefont{I.}~\bibnamefont{Satija}},
  \bibinfo{author}{\bibfnamefont{P.}~\bibnamefont{Nikolic}},
  \bibinfo{author}{\bibfnamefont{A.}~\bibnamefont{Bermudez}},
  \bibinfo{author}{\bibfnamefont{M.~A.} \bibnamefont{Martin-Delgado}},
  \bibinfo{author}{\bibfnamefont{M.}~\bibnamefont{Lewenstein}}, and
  \bibinfo{author}{\bibfnamefont{I.~B.} \bibnamefont{Spielman}},
  \bibinfo{year}{2010}, \bibinfo{journal}{Phys. Rev. Lett.}
  \textbf{\bibinfo{volume}{105}}(\bibinfo{number}{25}),
  \bibinfo{pages}{255302}.

\bibitem[{\citenamefont{Grimm} \emph{et~al.}(2000)\citenamefont{Grimm,
  Weidem{{\"u}}ller, and Ovchinnikov}}]{Grimm:2000}
\bibinfo{author}{\bibnamefont{Grimm}, \bibfnamefont{R.}},
  \bibinfo{author}{\bibfnamefont{M.}~\bibnamefont{Weidem{{\"u}}ller}}, and
  \bibinfo{author}{\bibfnamefont{Y.~B.} \bibnamefont{Ovchinnikov}},
  \bibinfo{year}{2000}, \bibinfo{journal}{Adv. At. Mol. Opt. Phys.}
  \textbf{\bibinfo{volume}{42}}, \bibinfo{pages}{95}.

\bibitem[{\citenamefont{Gr\"unberg}(2008)}]{Grunberg:2008}
\bibinfo{author}{\bibnamefont{Gr\"unberg}, \bibfnamefont{P.~A.}},
  \bibinfo{year}{2008}, \bibinfo{journal}{Rev. Mod. Phys.}
  \textbf{\bibinfo{volume}{80}}(\bibinfo{number}{4}), \bibinfo{pages}{1531}.

\bibitem[{\citenamefont{G\"{u}nter}
  \emph{et~al.}(2009)\citenamefont{G\"{u}nter, Cheneau, Yefsah, Rath, and
  Dalibard}}]{Gunter:2009}
\bibinfo{author}{\bibnamefont{G\"{u}nter}, \bibfnamefont{K.~J.}},
  \bibinfo{author}{\bibfnamefont{M.}~\bibnamefont{Cheneau}},
  \bibinfo{author}{\bibfnamefont{T.}~\bibnamefont{Yefsah}},
  \bibinfo{author}{\bibfnamefont{S.~P.} \bibnamefont{Rath}}, and
  \bibinfo{author}{\bibfnamefont{J.}~\bibnamefont{Dalibard}},
  \bibinfo{year}{2009}, \bibinfo{journal}{Phys. Rev. A}
  \textbf{\bibinfo{volume}{79}}, \bibinfo{pages}{011604}.

\bibitem[{\citenamefont{Hadzibabic}(2011)}]{Hadzibabic:2010}
\bibinfo{author}{\bibnamefont{Hadzibabic}, \bibfnamefont{Z.}},
  \bibinfo{year}{2011}, \bibinfo{title}{{P}rivate communication}.

\bibitem[{\citenamefont{Harper}(1955)}]{Harper:1955}
\bibinfo{author}{\bibnamefont{Harper}, \bibfnamefont{P.~G.}},
  \bibinfo{year}{1955}, \bibinfo{journal}{Proceedings of the Physical Society.
  Section A} \textbf{\bibinfo{volume}{68}}(\bibinfo{number}{10}),
  \bibinfo{pages}{874}.

\bibitem[{\citenamefont{Harris}(1997)}]{Harris97Physics-Today}
\bibinfo{author}{\bibnamefont{Harris}, \bibfnamefont{S.}},
  \bibinfo{year}{1997}, \bibinfo{journal}{Physics Today}
  \textbf{\bibinfo{volume}{50}}, \bibinfo{pages}{36}.

\bibitem[{\citenamefont{Hatsugai and Kohmoto}(1990)}]{Hatsugai:1990}
\bibinfo{author}{\bibnamefont{Hatsugai}, \bibfnamefont{Y.}}, and
  \bibinfo{author}{\bibfnamefont{M.}~\bibnamefont{Kohmoto}},
  \bibinfo{year}{1990}, \bibinfo{journal}{Phys. Rev. B}
  \textbf{\bibinfo{volume}{42}}(\bibinfo{number}{13}), \bibinfo{pages}{8282}.

\bibitem[{\citenamefont{Hofstadter}(1976)}]{Hofstadter:1976}
\bibinfo{author}{\bibnamefont{Hofstadter}, \bibfnamefont{D.~R.}},
  \bibinfo{year}{1976}, \bibinfo{journal}{Phys. Rev. B}
  \textbf{\bibinfo{volume}{14}}(\bibinfo{number}{6}), \bibinfo{pages}{2239}.

\bibitem[{\citenamefont{Horv\'athy}(1986)}]{Horvathy:1986}
\bibinfo{author}{\bibnamefont{Horv\'athy}, \bibfnamefont{P.~A.}},
  \bibinfo{year}{1986}, \bibinfo{journal}{Phys. Rev. D}
  \textbf{\bibinfo{volume}{33}}(\bibinfo{number}{2}), \bibinfo{pages}{407}.

\bibitem[{\citenamefont{Hou} \emph{et~al.}(2009)\citenamefont{Hou, Yang, and
  Liu}}]{Hou:2009a}
\bibinfo{author}{\bibnamefont{Hou}, \bibfnamefont{J.-M.}},
  \bibinfo{author}{\bibfnamefont{W.-X.} \bibnamefont{Yang}}, and
  \bibinfo{author}{\bibfnamefont{X.-J.} \bibnamefont{Liu}},
  \bibinfo{year}{2009}, \bibinfo{journal}{Phys. Rev. A}
  \textbf{\bibinfo{volume}{79}}(\bibinfo{number}{4}), \bibinfo{eid}{043621}.

\bibitem[{\citenamefont{Jackiw}(1988)}]{Jackiv88CAMP}
\bibinfo{author}{\bibnamefont{Jackiw}, \bibfnamefont{R.}},
  \bibinfo{year}{1988}, \bibinfo{journal}{Comments At. Mol. Phys.}
  \textbf{\bibinfo{volume}{21}}, \bibinfo{pages}{71}.

\bibitem[{\citenamefont{Jacob} \emph{et~al.}(2007)\citenamefont{Jacob,
  \"Ohberg, Juzeli\=unas, and Santos}}]{Jacob07APB}
\bibinfo{author}{\bibnamefont{Jacob}, \bibfnamefont{A.}},
  \bibinfo{author}{\bibfnamefont{P.}~\bibnamefont{\"Ohberg}},
  \bibinfo{author}{\bibfnamefont{G.}~\bibnamefont{Juzeli\=unas}}, and
  \bibinfo{author}{\bibfnamefont{L.}~\bibnamefont{Santos}},
  \bibinfo{year}{2007}, \bibinfo{journal}{Appl. Phys. B}
  \textbf{\bibinfo{volume}{89}}, \bibinfo{pages}{439}.

\bibitem[{\citenamefont{Jaksch and Zoller}(2003)}]{Jaksch:2003}
\bibinfo{author}{\bibnamefont{Jaksch}, \bibfnamefont{D.}}, and
  \bibinfo{author}{\bibfnamefont{P.}~\bibnamefont{Zoller}},
  \bibinfo{year}{2003}, \bibinfo{journal}{New Journal of Physics}
  \textbf{\bibinfo{volume}{5}}, \bibinfo{pages}{56.1}.

\bibitem[{\citenamefont{Johne} \emph{et~al.}(2010)\citenamefont{Johne, Shelykh,
  Solnyshkov, and Malpuech}}]{Johne09arxiv-Polariton-DDT}
\bibinfo{author}{\bibnamefont{Johne}, \bibfnamefont{R.}},
  \bibinfo{author}{\bibfnamefont{I.~A.} \bibnamefont{Shelykh}},
  \bibinfo{author}{\bibfnamefont{D.~D.} \bibnamefont{Solnyshkov}}, and
  \bibinfo{author}{\bibfnamefont{G.}~\bibnamefont{Malpuech}},
  \bibinfo{year}{2010}, \bibinfo{journal}{Phys. Rev. B}
  \textbf{\bibinfo{volume}{81}}, \bibinfo{pages}{125327}.

\bibitem[{\citenamefont{Juzeli\=unas and \"Ohberg}(2004)}]{Juzeliunas:2004}
\bibinfo{author}{\bibnamefont{Juzeli\=unas}, \bibfnamefont{G.}}, and
  \bibinfo{author}{\bibfnamefont{P.}~\bibnamefont{\"Ohberg}},
  \bibinfo{year}{2004}, \bibinfo{journal}{Phys. Rev. Lett.}
  \textbf{\bibinfo{volume}{93}}(\bibinfo{number}{3}), \bibinfo{pages}{033602}.

\bibitem[{\citenamefont{Juzeli\=unas}
  \emph{et~al.}(2005{\natexlab{a}})\citenamefont{Juzeli\=unas, \"Ohberg,
  Ruseckas, and Klein}}]{Juzeliunas:2005}
\bibinfo{author}{\bibnamefont{Juzeli\=unas}, \bibfnamefont{G.}},
  \bibinfo{author}{\bibfnamefont{P.}~\bibnamefont{\"Ohberg}},
  \bibinfo{author}{\bibfnamefont{J.}~\bibnamefont{Ruseckas}}, and
  \bibinfo{author}{\bibfnamefont{A.}~\bibnamefont{Klein}},
  \bibinfo{year}{2005}{\natexlab{a}}, \bibinfo{journal}{Phys. Rev. A}
  \textbf{\bibinfo{volume}{71}}(\bibinfo{number}{5}), \bibinfo{pages}{053614}.

\bibitem[{\citenamefont{Juzeli\=unas}
  \emph{et~al.}(2010)\citenamefont{Juzeli\=unas, Ruseckas, and
  Dalibard}}]{Juzeliunas:2010}
\bibinfo{author}{\bibnamefont{Juzeli\=unas}, \bibfnamefont{G.}},
  \bibinfo{author}{\bibfnamefont{J.}~\bibnamefont{Ruseckas}}, and
  \bibinfo{author}{\bibfnamefont{J.}~\bibnamefont{Dalibard}},
  \bibinfo{year}{2010}, \bibinfo{journal}{Phys. Rev. A}
  \textbf{\bibinfo{volume}{81}}, \bibinfo{pages}{053403}.

\bibitem[{\citenamefont{Juzeli\=unas}
  \emph{et~al.}(2008{\natexlab{a}})\citenamefont{Juzeli\=unas, Ruseckas, Jacob,
  Santos, and \"Ohberg}}]{Juz08PRL}
\bibinfo{author}{\bibnamefont{Juzeli\=unas}, \bibfnamefont{G.}},
  \bibinfo{author}{\bibfnamefont{J.}~\bibnamefont{Ruseckas}},
  \bibinfo{author}{\bibfnamefont{A.}~\bibnamefont{Jacob}},
  \bibinfo{author}{\bibfnamefont{L.}~\bibnamefont{Santos}}, and
  \bibinfo{author}{\bibfnamefont{P.}~\bibnamefont{\"Ohberg}},
  \bibinfo{year}{2008}{\natexlab{a}}, \bibinfo{journal}{Phys. Phys. Lett.}
  \textbf{\bibinfo{volume}{100}}, \bibinfo{pages}{200405}.

\bibitem[{\citenamefont{Juzeli\=unas}
  \emph{et~al.}(2008{\natexlab{b}})\citenamefont{Juzeli\=unas, Ruseckas,
  Lindberg, Santos, and \"Ohberg}}]{Juz08PRA}
\bibinfo{author}{\bibnamefont{Juzeli\=unas}, \bibfnamefont{G.}},
  \bibinfo{author}{\bibfnamefont{J.}~\bibnamefont{Ruseckas}},
  \bibinfo{author}{\bibfnamefont{M.}~\bibnamefont{Lindberg}},
  \bibinfo{author}{\bibfnamefont{L.}~\bibnamefont{Santos}}, and
  \bibinfo{author}{\bibfnamefont{P.}~\bibnamefont{\"Ohberg}},
  \bibinfo{year}{2008}{\natexlab{b}}, \bibinfo{journal}{Phys. Rev. A}
  \textbf{\bibinfo{volume}{77}}, \bibinfo{pages}{011802(R)}.

\bibitem[{\citenamefont{Juzeli\=unas}
  \emph{et~al.}(2005{\natexlab{b}})\citenamefont{Juzeli\=unas, Ruseckas, and
  \"Ohberg}}]{Juzeliunas:2005b}
\bibinfo{author}{\bibnamefont{Juzeli\=unas}, \bibfnamefont{G.}},
  \bibinfo{author}{\bibfnamefont{J.}~\bibnamefont{Ruseckas}}, and
  \bibinfo{author}{\bibfnamefont{P.}~\bibnamefont{\"Ohberg}},
  \bibinfo{year}{2005}{\natexlab{b}}, \bibinfo{journal}{J. Phys. B}
  \textbf{\bibinfo{volume}{38}}, \bibinfo{pages}{4171}.

\bibitem[{\citenamefont{Juzeli\=unas}
  \emph{et~al.}(2006)\citenamefont{Juzeli\=unas, Ruseckas, \"Ohberg, and
  Fleischhauer}}]{Juzeliunas:2006}
\bibinfo{author}{\bibnamefont{Juzeli\=unas}, \bibfnamefont{G.}},
  \bibinfo{author}{\bibfnamefont{J.}~\bibnamefont{Ruseckas}},
  \bibinfo{author}{\bibfnamefont{P.}~\bibnamefont{\"Ohberg}}, and
  \bibinfo{author}{\bibfnamefont{M.}~\bibnamefont{Fleischhauer}},
  \bibinfo{year}{2006}, \bibinfo{journal}{Phys. Rev. A}
  \textbf{\bibinfo{volume}{73}}(\bibinfo{number}{2}), \bibinfo{pages}{025602}.

\bibitem[{\citenamefont{Katsnelson}
  \emph{et~al.}(2006)\citenamefont{Katsnelson, Novoselov, and
  Geim}}]{Katsnelson06NP}
\bibinfo{author}{\bibnamefont{Katsnelson}, \bibfnamefont{M.~I.}},
  \bibinfo{author}{\bibfnamefont{K.~S.} \bibnamefont{Novoselov}}, and
  \bibinfo{author}{\bibfnamefont{A.~K.} \bibnamefont{Geim}},
  \bibinfo{year}{2006}, \bibinfo{journal}{Nat. Phys.}
  \textbf{\bibinfo{volume}{2}}, \bibinfo{pages}{620}.

\bibitem[{\citenamefont{Klein and Jaksch}(2009)}]{Klein:2009a}
\bibinfo{author}{\bibnamefont{Klein}, \bibfnamefont{A.}}, and
  \bibinfo{author}{\bibfnamefont{D.}~\bibnamefont{Jaksch}},
  \bibinfo{year}{2009}, \bibinfo{journal}{Europhys. Lett.}
  \textbf{\bibinfo{volume}{85}}(\bibinfo{number}{1}), \bibinfo{pages}{13001}.

\bibitem[{\citenamefont{Kohmoto}(1989)}]{Kohmoto:1989a}
\bibinfo{author}{\bibnamefont{Kohmoto}, \bibfnamefont{M.}},
  \bibinfo{year}{1989}, \bibinfo{journal}{Phys. Rev. B}
  \textbf{\bibinfo{volume}{39}}(\bibinfo{number}{16}), \bibinfo{pages}{11943}.

\bibitem[{\citenamefont{Kolovsky}(2011)}]{Kolovsky:2011EPL}
\bibinfo{author}{\bibnamefont{Kolovsky}, \bibfnamefont{A.~R.}},
  \bibinfo{year}{2011}, \bibinfo{journal}{Europhys. Lett.}
  \textbf{\bibinfo{volume}{93}}, \bibinfo{pages}{20003}.

\bibitem[{\citenamefont{Koo} \emph{et~al.}(2009)\citenamefont{Koo, Kwon, Eom,
  Chang, Han, and Johnson}}]{Koo:2009Science}
\bibinfo{author}{\bibnamefont{Koo}, \bibfnamefont{H.~C.}},
  \bibinfo{author}{\bibfnamefont{J.~H.} \bibnamefont{Kwon}},
  \bibinfo{author}{\bibfnamefont{J.}~\bibnamefont{Eom}},
  \bibinfo{author}{\bibfnamefont{J.}~\bibnamefont{Chang}},
  \bibinfo{author}{\bibfnamefont{S.~H.} \bibnamefont{Han}}, and
  \bibinfo{author}{\bibfnamefont{M.}~\bibnamefont{Johnson}},
  \bibinfo{year}{2009}, \bibinfo{journal}{Science}
  \textbf{\bibinfo{volume}{325}}, \bibinfo{pages}{1515}.

\bibitem[{\citenamefont{Kr{\'a}l} \emph{et~al.}(2007)\citenamefont{Kr{\'a}l,
  Thanopulos, and Shapiro}}]{Shapiro07RMP}
\bibinfo{author}{\bibnamefont{Kr{\'a}l}, \bibfnamefont{P.}},
  \bibinfo{author}{\bibfnamefont{I.}~\bibnamefont{Thanopulos}}, and
  \bibinfo{author}{\bibfnamefont{M.}~\bibnamefont{Shapiro}},
  \bibinfo{year}{2007}, \bibinfo{journal}{Rev. Mod. Phys}
  \textbf{\bibinfo{volume}{79}}, \bibinfo{pages}{53}.

\bibitem[{\citenamefont{Lamata} \emph{et~al.}(2007)\citenamefont{Lamata, Leon,
  Schatz, and Solano}}]{Lamata2007PRL}
\bibinfo{author}{\bibnamefont{Lamata}, \bibfnamefont{L.}},
  \bibinfo{author}{\bibfnamefont{T.}~\bibnamefont{Leon}},
  \bibinfo{author}{\bibnamefont{Schatz}}, and
  \bibinfo{author}{\bibfnamefont{E.}~\bibnamefont{Solano}},
  \bibinfo{year}{2007}, \bibinfo{journal}{Phys. Lett. Lett.Rev.}
  \textbf{\bibinfo{volume}{98}}, \bibinfo{pages}{253005}.

\bibitem[{\citenamefont{Larson} \emph{et~al.}(2010)\citenamefont{Larson,
  Martikainen, and Collin}}]{Larson2010PRA}
\bibinfo{author}{\bibnamefont{Larson}, \bibfnamefont{J.}},
  \bibinfo{author}{\bibfnamefont{J.-P.} \bibnamefont{Martikainen}}, and
  \bibinfo{author}{\bibfnamefont{A.}~\bibnamefont{Collin}},
  \bibinfo{year}{2010}, \bibinfo{journal}{Phys. Rev. A}
  \textbf{\bibinfo{volume}{82}}, \bibinfo{pages}{043620}.

\bibitem[{\citenamefont{Larson and Sj\"oqvist}(2009)}]{Larson09PRA}
\bibinfo{author}{\bibnamefont{Larson}, \bibfnamefont{J.}}, and
  \bibinfo{author}{\bibfnamefont{E.}~\bibnamefont{Sj\"oqvist}},
  \bibinfo{year}{2009}, \bibinfo{journal}{Phys. Rev. A}
  \textbf{\bibinfo{volume}{79}}, \bibinfo{pages}{043627}.

\bibitem[{\citenamefont{Lewenstein}
  \emph{et~al.}(2007)\citenamefont{Lewenstein, Sanpera, Ahufinger, Damski, De,
  and Sen}}]{Lewenstein:2007}
\bibinfo{author}{\bibnamefont{Lewenstein}, \bibfnamefont{M.}},
  \bibinfo{author}{\bibfnamefont{A.}~\bibnamefont{Sanpera}},
  \bibinfo{author}{\bibfnamefont{V.}~\bibnamefont{Ahufinger}},
  \bibinfo{author}{\bibfnamefont{B.}~\bibnamefont{Damski}},
  \bibinfo{author}{\bibfnamefont{A.~S.} \bibnamefont{De}}, and
  \bibinfo{author}{\bibfnamefont{U.}~\bibnamefont{Sen}}, \bibinfo{year}{2007},
  \bibinfo{journal}{Adv. Phys.}
  \textbf{\bibinfo{volume}{56}}(\bibinfo{number}{2}), \bibinfo{pages}{243}.

\bibitem[{\citenamefont{Li} \emph{et~al.}(2007)\citenamefont{Li, Bruder, and
  Sun}}]{Li-Brud-Sun07PRL}
\bibinfo{author}{\bibnamefont{Li}, \bibfnamefont{Y.}},
  \bibinfo{author}{\bibnamefont{Bruder}}, and
  \bibinfo{author}{\bibfnamefont{C.~P.} \bibnamefont{Sun}},
  \bibinfo{year}{2007}, \bibinfo{journal}{Phys. Lett. Lett.}
  \textbf{\bibinfo{volume}{99}}, \bibinfo{pages}{130403}.

\bibitem[{\citenamefont{Lignier} \emph{et~al.}(2007)\citenamefont{Lignier,
  Sias, Ciampini, Singh, Zenesini, Morsch, and Arimondo}}]{Lignier:2007}
\bibinfo{author}{\bibnamefont{Lignier}, \bibfnamefont{H.}},
  \bibinfo{author}{\bibfnamefont{C.}~\bibnamefont{Sias}},
  \bibinfo{author}{\bibfnamefont{D.}~\bibnamefont{Ciampini}},
  \bibinfo{author}{\bibfnamefont{Y.}~\bibnamefont{Singh}},
  \bibinfo{author}{\bibfnamefont{A.}~\bibnamefont{Zenesini}},
  \bibinfo{author}{\bibfnamefont{O.}~\bibnamefont{Morsch}}, and
  \bibinfo{author}{\bibfnamefont{E.}~\bibnamefont{Arimondo}},
  \bibinfo{year}{2007}, \bibinfo{journal}{Phys. Rev. Lett.}
  \textbf{\bibinfo{volume}{99}}(\bibinfo{number}{22}), \bibinfo{pages}{220403}.

\bibitem[{\citenamefont{Lim} \emph{et~al.}(2008)\citenamefont{Lim, Smith, and
  Hemmerich}}]{Lim:2008a}
\bibinfo{author}{\bibnamefont{Lim}, \bibfnamefont{L.-K.}},
  \bibinfo{author}{\bibfnamefont{C.~M.} \bibnamefont{Smith}}, and
  \bibinfo{author}{\bibfnamefont{A.}~\bibnamefont{Hemmerich}},
  \bibinfo{year}{2008}, \bibinfo{journal}{Phys. Rev. Lett.}
  \textbf{\bibinfo{volume}{100}}(\bibinfo{number}{13}), \bibinfo{eid}{130402}
  (pages~\bibinfo{numpages}{4}).

\bibitem[{\citenamefont{Lin} \emph{et~al.}(2011)\citenamefont{Lin,
  Jimenez-Garcia, and Spielman}}]{Lin:2011}
\bibinfo{author}{\bibnamefont{Lin}, \bibfnamefont{Y.}},
  \bibinfo{author}{\bibfnamefont{K.}~\bibnamefont{Jimenez-Garcia}}, and
  \bibinfo{author}{\bibfnamefont{I.}~\bibnamefont{Spielman}},
  \bibinfo{year}{2011}, \bibinfo{journal}{Nature}
  \textbf{\bibinfo{volume}{471}}, \bibinfo{pages}{83}.

\bibitem[{\citenamefont{Lin}
  \emph{et~al.}(2009{\natexlab{a}})\citenamefont{Lin, Compton,
  Jim\'{e}nez-Garc\'{\i}a, Porto, and Spielman}}]{Lin2009b}
\bibinfo{author}{\bibnamefont{Lin}, \bibfnamefont{Y.-J.}},
  \bibinfo{author}{\bibfnamefont{R.~L.} \bibnamefont{Compton}},
  \bibinfo{author}{\bibfnamefont{K.}~\bibnamefont{Jim\'{e}nez-Garc\'{\i}a}},
  \bibinfo{author}{\bibfnamefont{J.~V.} \bibnamefont{Porto}}, and
  \bibinfo{author}{\bibfnamefont{I.~B.} \bibnamefont{Spielman}},
  \bibinfo{year}{2009}{\natexlab{a}}, \bibinfo{journal}{Nature}
  \textbf{\bibinfo{volume}{462}}, \bibinfo{pages}{628}.

\bibitem[{\citenamefont{Lin}
  \emph{et~al.}(2009{\natexlab{b}})\citenamefont{Lin, Compton, Perry, Phillips,
  Porto, and Spielman}}]{Lin:2009}
\bibinfo{author}{\bibnamefont{Lin}, \bibfnamefont{Y.~J.}},
  \bibinfo{author}{\bibfnamefont{R.~L.} \bibnamefont{Compton}},
  \bibinfo{author}{\bibfnamefont{A.~R.} \bibnamefont{Perry}},
  \bibinfo{author}{\bibfnamefont{W.~D.} \bibnamefont{Phillips}},
  \bibinfo{author}{\bibfnamefont{J.~V.} \bibnamefont{Porto}}, and
  \bibinfo{author}{\bibfnamefont{I.~B.} \bibnamefont{Spielman}},
  \bibinfo{year}{2009}{\natexlab{b}}, \bibinfo{journal}{Phys. Rev. Lett.}
  \textbf{\bibinfo{volume}{102}}, \bibinfo{pages}{130401}.

\bibitem[{\citenamefont{Littlejohn and Weigert}(1993)}]{Littlejohn:1993a}
\bibinfo{author}{\bibnamefont{Littlejohn}, \bibfnamefont{R.~G.}}, and
  \bibinfo{author}{\bibfnamefont{S.}~\bibnamefont{Weigert}},
  \bibinfo{year}{1993}, \bibinfo{journal}{Phys. Rev. A}
  \textbf{\bibinfo{volume}{48}}(\bibinfo{number}{2}), \bibinfo{pages}{924}.

\bibitem[{\citenamefont{Lukin}(2003)}]{Lukin:2003}
\bibinfo{author}{\bibnamefont{Lukin}, \bibfnamefont{M.~D.}},
  \bibinfo{year}{2003}, \bibinfo{journal}{Rev. Mod. Phys.}
  \textbf{\bibinfo{volume}{75}}(\bibinfo{number}{2}), \bibinfo{pages}{457}.

\bibitem[{\citenamefont{Luttinger}(1951)}]{Luttinger:1951}
\bibinfo{author}{\bibnamefont{Luttinger}, \bibfnamefont{J.~M.}},
  \bibinfo{year}{1951}, \bibinfo{journal}{Phys. Rev.}
  \textbf{\bibinfo{volume}{84}}(\bibinfo{number}{4}), \bibinfo{pages}{814}.

\bibitem[{\citenamefont{Mandel} \emph{et~al.}(2003)\citenamefont{Mandel,
  Greiner, Widera, Rom, H{\"a}nsch, and Bloch}}]{Mandel:2003a}
\bibinfo{author}{\bibnamefont{Mandel}, \bibfnamefont{O.}},
  \bibinfo{author}{\bibfnamefont{M.}~\bibnamefont{Greiner}},
  \bibinfo{author}{\bibfnamefont{A.}~\bibnamefont{Widera}},
  \bibinfo{author}{\bibfnamefont{T.}~\bibnamefont{Rom}},
  \bibinfo{author}{\bibfnamefont{T.~W.} \bibnamefont{H{\"a}nsch}}, and
  \bibinfo{author}{\bibfnamefont{I.}~\bibnamefont{Bloch}},
  \bibinfo{year}{2003}, \bibinfo{journal}{Phys. Rev. Lett.}
  \textbf{\bibinfo{volume}{91}}, \bibinfo{pages}{010407}.

\bibitem[{\citenamefont{Mead}(1992)}]{Mead:1992}
\bibinfo{author}{\bibnamefont{Mead}, \bibfnamefont{C.~A.}},
  \bibinfo{year}{1992}, \bibinfo{journal}{Rev. Mod. Phys.}
  \textbf{\bibinfo{volume}{84}}, \bibinfo{pages}{51}.

\bibitem[{\citenamefont{Mead and Truhlar}(1979)}]{Mead:1979}
\bibinfo{author}{\bibnamefont{Mead}, \bibfnamefont{C.~A.}}, and
  \bibinfo{author}{\bibfnamefont{D.~G.} \bibnamefont{Truhlar}},
  \bibinfo{year}{1979}, \bibinfo{journal}{J. Chem. Phys.}
  \textbf{\bibinfo{volume}{70}}(\bibinfo{number}{5}), \bibinfo{pages}{2284}.

\bibitem[{\citenamefont{Merkl} \emph{et~al.}(2008)\citenamefont{Merkl, Zimmer,
  Juzeli\=unas, and \"Ohberg}}]{Merkl08EPL}
\bibinfo{author}{\bibnamefont{Merkl}, \bibfnamefont{M.}},
  \bibinfo{author}{\bibfnamefont{F.~E.} \bibnamefont{Zimmer}},
  \bibinfo{author}{\bibfnamefont{G.}~\bibnamefont{Juzeli\=unas}}, and
  \bibinfo{author}{\bibfnamefont{P.}~\bibnamefont{\"Ohberg}},
  \bibinfo{year}{2008}, \bibinfo{journal}{Europhys. Lett.}
  \textbf{\bibinfo{volume}{83}}, \bibinfo{pages}{54002}.

\bibitem[{\citenamefont{Messiah}(1961{\natexlab{a}})}]{Messiah_T_sym}
\bibinfo{author}{\bibnamefont{Messiah}, \bibfnamefont{A.}},
  \bibinfo{year}{1961}{\natexlab{a}}, \emph{\bibinfo{title}{Quantum
  {M}echanics, Chapter XV, \S~19}}, volume~\bibinfo{volume}{II}
  (\bibinfo{publisher}{North-Holland Publishing Company},
  \bibinfo{address}{Amsterdam}).

\bibitem[{\citenamefont{Messiah}(1961{\natexlab{b}})}]{Messiah_adiab}
\bibinfo{author}{\bibnamefont{Messiah}, \bibfnamefont{A.}},
  \bibinfo{year}{1961}{\natexlab{b}}, \emph{\bibinfo{title}{Quantum
  {M}echanics, Chapter XVII, \S~13}}, volume~\bibinfo{volume}{II}
  (\bibinfo{publisher}{North-Holland Publishing Company},
  \bibinfo{address}{Amsterdam}).

\bibitem[{\citenamefont{M\"oller and Cooper}(2009)}]{Moller:2009}
\bibinfo{author}{\bibnamefont{M\"oller}, \bibfnamefont{G.}}, and
  \bibinfo{author}{\bibfnamefont{N.~R.} \bibnamefont{Cooper}},
  \bibinfo{year}{2009}, \bibinfo{journal}{Phys. Rev. Lett.}
  \textbf{\bibinfo{volume}{103}}(\bibinfo{number}{10}),
  \bibinfo{pages}{105303}.

\bibitem[{\citenamefont{M\"oller and Cooper}(2010)}]{Moller:2010}
\bibinfo{author}{\bibnamefont{M\"oller}, \bibfnamefont{G.}}, and
  \bibinfo{author}{\bibfnamefont{N.~R.} \bibnamefont{Cooper}},
  \bibinfo{year}{2010}, \bibinfo{journal}{Phys. Rev. A}
  \textbf{\bibinfo{volume}{82}}(\bibinfo{number}{6}), \bibinfo{pages}{063625}.

\bibitem[{\citenamefont{Moody} \emph{et~al.}(1986)\citenamefont{Moody, Shapere,
  and Wilczek}}]{Moody86PRL}
\bibinfo{author}{\bibnamefont{Moody}, \bibfnamefont{J.}},
  \bibinfo{author}{\bibfnamefont{A.}~\bibnamefont{Shapere}}, and
  \bibinfo{author}{\bibfnamefont{F.}~\bibnamefont{Wilczek}},
  \bibinfo{year}{1986}, \bibinfo{journal}{Phys. Rev. Lett.}
  \textbf{\bibinfo{volume}{56}}(\bibinfo{number}{9}), \bibinfo{pages}{893}.

\bibitem[{\citenamefont{Moody} \emph{et~al.}(1989)\citenamefont{Moody, Shapere,
  and Wilczek}}]{Moody:1989}
\bibinfo{author}{\bibnamefont{Moody}, \bibfnamefont{J.}},
  \bibinfo{author}{\bibfnamefont{A.}~\bibnamefont{Shapere}}, and
  \bibinfo{author}{\bibfnamefont{F.}~\bibnamefont{Wilczek}},
  \bibinfo{year}{1989}, in \emph{\bibinfo{booktitle}{Geometric {P}hases in
  {P}hysics}}, edited by
  \bibinfo{editor}{\bibfnamefont{A.}~\bibnamefont{Shapere}} and
  \bibinfo{editor}{\bibfnamefont{F.}~\bibnamefont{Wilczek}}
  (\bibinfo{publisher}{World Scientific}, \bibinfo{address}{Singapore}), pp.
  \bibinfo{pages}{160--183}.

\bibitem[{\citenamefont{Mueller}(2004)}]{Mueller:2004}
\bibinfo{author}{\bibnamefont{Mueller}, \bibfnamefont{E.~J.}},
  \bibinfo{year}{2004}, \bibinfo{journal}{Phys. Rev. A}
  \textbf{\bibinfo{volume}{70}}(\bibinfo{number}{4}), \bibinfo{pages}{041603}.

\bibitem[{\citenamefont{Nenciu}(1991)}]{Nenciu:1991}
\bibinfo{author}{\bibnamefont{Nenciu}, \bibfnamefont{G.}},
  \bibinfo{year}{1991}, \bibinfo{journal}{Rev. Mod. Phys.}
  \textbf{\bibinfo{volume}{63}}(\bibinfo{number}{1}), \bibinfo{pages}{91}.

\bibitem[{\citenamefont{Neto} \emph{et~al.}(2009)\citenamefont{Neto, Guinea,
  Peres, Novoselov, and Geim}}]{Castro-Neto:2009RMP}
\bibinfo{author}{\bibnamefont{Neto}, \bibfnamefont{A.~H.~C.}},
  \bibinfo{author}{\bibfnamefont{F.}~\bibnamefont{Guinea}},
  \bibinfo{author}{\bibfnamefont{N.~M.~R.} \bibnamefont{Peres}},
  \bibinfo{author}{\bibfnamefont{K.~S.} \bibnamefont{Novoselov}}, and
  \bibinfo{author}{\bibfnamefont{A.~K.} \bibnamefont{Geim}},
  \bibinfo{year}{2009}, \bibinfo{journal}{Rev. Mod. Phys.}
  \textbf{\bibinfo{volume}{81}}, \bibinfo{pages}{109}.

\bibitem[{\citenamefont{Osterloh} \emph{et~al.}(2005)\citenamefont{Osterloh,
  Baig, Santos, Zoller, and Lewenstein}}]{Osterloh:2005}
\bibinfo{author}{\bibnamefont{Osterloh}, \bibfnamefont{K.}},
  \bibinfo{author}{\bibfnamefont{M.}~\bibnamefont{Baig}},
  \bibinfo{author}{\bibfnamefont{L.}~\bibnamefont{Santos}},
  \bibinfo{author}{\bibfnamefont{P.}~\bibnamefont{Zoller}}, and
  \bibinfo{author}{\bibfnamefont{M.}~\bibnamefont{Lewenstein}},
  \bibinfo{year}{2005}, \bibinfo{journal}{Phys. Rev. Lett.}
  \textbf{\bibinfo{volume}{95}}(\bibinfo{number}{1}), \bibinfo{pages}{010403}.

\bibitem[{\citenamefont{Pietil\"{a} and M\"{o}tt\"{o}nen}(2009)}]{Pietila09PRL}
\bibinfo{author}{\bibnamefont{Pietil\"{a}}, \bibfnamefont{V.}}, and
  \bibinfo{author}{\bibfnamefont{M.}~\bibnamefont{M\"{o}tt\"{o}nen}},
  \bibinfo{year}{2009}, \bibinfo{journal}{Phys. Lett. Lett.}
  \textbf{\bibinfo{volume}{102}}, \bibinfo{pages}{080403}.

\bibitem[{\citenamefont{Polak and Kope\ifmmode~\acute{c}\else
  \'{c}\fi{}}(2009)}]{Polak:2009}
\bibinfo{author}{\bibnamefont{Polak}, \bibfnamefont{T.~P.}}, and
  \bibinfo{author}{\bibfnamefont{T.~K.}
  \bibnamefont{Kope\ifmmode~\acute{c}\else \'{c}\fi{}}}, \bibinfo{year}{2009},
  \bibinfo{journal}{Phys. Rev. A}
  \textbf{\bibinfo{volume}{79}}(\bibinfo{number}{6}), \bibinfo{pages}{063629}.

\bibitem[{\citenamefont{Rashba}(1960)}]{Rashba60}
\bibinfo{author}{\bibnamefont{Rashba}, \bibfnamefont{E.~I.}},
  \bibinfo{year}{1960}, \bibinfo{journal}{Sov. Phys. Sol. St.}
  \textbf{\bibinfo{volume}{2}}, \bibinfo{pages}{1224}.

\bibitem[{\citenamefont{Ruostekoski}(2009)}]{Ruostekoski:2009}
\bibinfo{author}{\bibnamefont{Ruostekoski}, \bibfnamefont{J.}},
  \bibinfo{year}{2009}, \bibinfo{journal}{Phys. Rev. Lett.}
  \textbf{\bibinfo{volume}{103}}, \bibinfo{pages}{080406}.

\bibitem[{\citenamefont{Ruostekoski}
  \emph{et~al.}(2002)\citenamefont{Ruostekoski, Dunne, and
  Javanainen}}]{Ruostekoski:2002}
\bibinfo{author}{\bibnamefont{Ruostekoski}, \bibfnamefont{J.}},
  \bibinfo{author}{\bibfnamefont{G.~V.} \bibnamefont{Dunne}}, and
  \bibinfo{author}{\bibfnamefont{J.}~\bibnamefont{Javanainen}},
  \bibinfo{year}{2002}, \bibinfo{journal}{Phys. Rev. Lett.}
  \textbf{\bibinfo{volume}{88}}, \bibinfo{pages}{180401}.

\bibitem[{\citenamefont{Ruseckas} \emph{et~al.}(2005)\citenamefont{Ruseckas,
  Juzeli\=unas, \"Ohberg, and Fleischhauer}}]{Ruseckas:2005}
\bibinfo{author}{\bibnamefont{Ruseckas}, \bibfnamefont{J.}},
  \bibinfo{author}{\bibfnamefont{G.}~\bibnamefont{Juzeli\=unas}},
  \bibinfo{author}{\bibfnamefont{P.}~\bibnamefont{\"Ohberg}}, and
  \bibinfo{author}{\bibfnamefont{M.}~\bibnamefont{Fleischhauer}},
  \bibinfo{year}{2005}, \bibinfo{journal}{Phys. Rev. Lett.}
  \textbf{\bibinfo{volume}{95}}(\bibinfo{number}{1}), \bibinfo{pages}{010404}.

\bibitem[{\citenamefont{Rusin and Zawadzki}(2009)}]{Rusin2009PRA}
\bibinfo{author}{\bibnamefont{Rusin}, \bibfnamefont{T.~M.}}, and
  \bibinfo{author}{\bibfnamefont{W.}~\bibnamefont{Zawadzki}},
  \bibinfo{year}{2009}, \bibinfo{journal}{Phys. Rev. A}
  \textbf{\bibinfo{volume}{80}}, \bibinfo{pages}{045416}.

\bibitem[{\citenamefont{Rusin and Zawadzki}(2010)}]{Rusin2010PRD}
\bibinfo{author}{\bibnamefont{Rusin}, \bibfnamefont{T.~M.}}, and
  \bibinfo{author}{\bibfnamefont{W.}~\bibnamefont{Zawadzki}},
  \bibinfo{year}{2010}, \bibinfo{journal}{Phy. Rev. D}
  \textbf{\bibinfo{volume}{82}}, \bibinfo{pages}{125031}.

\bibitem[{\citenamefont{Sarma} \emph{et~al.}(2006)\citenamefont{Sarma,
  Freedman, and Nayak}}]{Sarma:2006}
\bibinfo{author}{\bibnamefont{Sarma}, \bibfnamefont{S.~D.}},
  \bibinfo{author}{\bibfnamefont{M.}~\bibnamefont{Freedman}}, and
  \bibinfo{author}{\bibfnamefont{C.}~\bibnamefont{Nayak}},
  \bibinfo{year}{2006}, \bibinfo{journal}{Physics Today} , \bibinfo{pages}{32}.

\bibitem[{\citenamefont{Satija} \emph{et~al.}(2008)\citenamefont{Satija, Dakin,
  Vaishnav, and Clark}}]{Satija:2008}
\bibinfo{author}{\bibnamefont{Satija}, \bibfnamefont{I.~I.}},
  \bibinfo{author}{\bibfnamefont{D.~C.} \bibnamefont{Dakin}},
  \bibinfo{author}{\bibfnamefont{J.~Y.} \bibnamefont{Vaishnav}}, and
  \bibinfo{author}{\bibfnamefont{C.~W.} \bibnamefont{Clark}},
  \bibinfo{year}{2008}, \bibinfo{journal}{Phys. Rev. A}
  \textbf{\bibinfo{volume}{77}}(\bibinfo{number}{4}), \bibinfo{pages}{043410}.

\bibitem[{\citenamefont{Schliemann}
  \emph{et~al.}(2003)\citenamefont{Schliemann, Egues, and
  Loss}}]{Schliemann03PRL-DDT-Balanced}
\bibinfo{author}{\bibnamefont{Schliemann}, \bibfnamefont{J.}},
  \bibinfo{author}{\bibfnamefont{J.~C.} \bibnamefont{Egues}}, and
  \bibinfo{author}{\bibfnamefont{D.}~\bibnamefont{Loss}}, \bibinfo{year}{2003},
  \bibinfo{journal}{Phys. Rev. Lett.} \textbf{\bibinfo{volume}{90}},
  \bibinfo{pages}{146801}.

\bibitem[{\citenamefont{Schliemann}
  \emph{et~al.}(2006)\citenamefont{Schliemann, Loss, and
  Westervelt}}]{Schlieman06PRB}
\bibinfo{author}{\bibnamefont{Schliemann}, \bibfnamefont{J.}},
  \bibinfo{author}{\bibfnamefont{D.}~\bibnamefont{Loss}}, and
  \bibinfo{author}{\bibfnamefont{R.~M.} \bibnamefont{Westervelt}},
  \bibinfo{year}{2006}, \bibinfo{journal}{Phys. Rev. B}
  \textbf{\bibinfo{volume}{73}}, \bibinfo{pages}{085323}.

\bibitem[{\citenamefont{Smith}(1969)}]{Smith69PR}
\bibinfo{author}{\bibnamefont{Smith}, \bibfnamefont{F.~T.}},
  \bibinfo{year}{1969}, \bibinfo{journal}{Phys. Rev.}
  \textbf{\bibinfo{volume}{179}}, \bibinfo{pages}{111}.

\bibitem[{\citenamefont{Song and Foreman}(2009)}]{Song2009PRA}
\bibinfo{author}{\bibnamefont{Song}, \bibfnamefont{J.-J.}}, and
  \bibinfo{author}{\bibfnamefont{B.~A.} \bibnamefont{Foreman}},
  \bibinfo{year}{2009}, \bibinfo{journal}{Phys. Rev. A}
  \textbf{\bibinfo{volume}{80}}, \bibinfo{pages}{045602}.

\bibitem[{\citenamefont{S\o{}rensen}
  \emph{et~al.}(2005)\citenamefont{S\o{}rensen, Demler, and
  Lukin}}]{Sorensen:2005}
\bibinfo{author}{\bibnamefont{S\o{}rensen}, \bibfnamefont{A.~S.}},
  \bibinfo{author}{\bibfnamefont{E.}~\bibnamefont{Demler}}, and
  \bibinfo{author}{\bibfnamefont{M.~D.} \bibnamefont{Lukin}},
  \bibinfo{year}{2005}, \bibinfo{journal}{Phys. Rev. Lett.}
  \textbf{\bibinfo{volume}{94}}, \bibinfo{pages}{086803}.

\bibitem[{\citenamefont{Spielman}(2009)}]{Spielman:2009}
\bibinfo{author}{\bibnamefont{Spielman}, \bibfnamefont{I.~B.}},
  \bibinfo{year}{2009}, \bibinfo{journal}{Phys. Rev. A}
  \textbf{\bibinfo{volume}{79}}(\bibinfo{number}{6}), \bibinfo{pages}{063613}.

\bibitem[{\citenamefont{Stanescu} \emph{et~al.}(2008)\citenamefont{Stanescu,
  Anderson, and Galitski}}]{Stanescu08PRA}
\bibinfo{author}{\bibnamefont{Stanescu}, \bibfnamefont{T.}},
  \bibinfo{author}{\bibfnamefont{B.}~\bibnamefont{Anderson}}, and
  \bibinfo{author}{\bibfnamefont{V.}~\bibnamefont{Galitski}},
  \bibinfo{year}{2008}, \bibinfo{journal}{Phys. Rev. A}
  \textbf{\bibinfo{volume}{78}}, \bibinfo{pages}{023616}.

\bibitem[{\citenamefont{Stanescu} \emph{et~al.}(2010)\citenamefont{Stanescu,
  Galitski, and Sarma}}]{Stanescu:2010}
\bibinfo{author}{\bibnamefont{Stanescu}, \bibfnamefont{T.~D.}},
  \bibinfo{author}{\bibfnamefont{V.}~\bibnamefont{Galitski}}, and
  \bibinfo{author}{\bibfnamefont{S.~D.} \bibnamefont{Sarma}},
  \bibinfo{year}{2010}, \bibinfo{journal}{Phys. Rev. A}
  \textbf{\bibinfo{volume}{82}}, \bibinfo{pages}{013608}.

\bibitem[{\citenamefont{Stanescu} \emph{et~al.}(2007)\citenamefont{Stanescu,
  Zhang, and Galitski}}]{Stanescu07PRL}
\bibinfo{author}{\bibnamefont{Stanescu}, \bibfnamefont{T.~D.}},
  \bibinfo{author}{\bibfnamefont{C.}~\bibnamefont{Zhang}}, and
  \bibinfo{author}{\bibfnamefont{V.}~\bibnamefont{Galitski}},
  \bibinfo{year}{2007}, \bibinfo{journal}{Phys. Rev. Lett}
  \textbf{\bibinfo{volume}{99}}, \bibinfo{pages}{110403}.

\bibitem[{\citenamefont{Teodorescu and Winkler}(2009)}]{Winkler09PRB}
\bibinfo{author}{\bibnamefont{Teodorescu}, \bibfnamefont{V.}}, and
  \bibinfo{author}{\bibfnamefont{R.}~\bibnamefont{Winkler}},
  \bibinfo{year}{2009}, \bibinfo{journal}{Phys. Rev. B}
  \textbf{\bibinfo{volume}{80}}, \bibinfo{pages}{041311(R)}.

\bibitem[{\citenamefont{Thouless} \emph{et~al.}(1982)\citenamefont{Thouless,
  Kohmoto, Nightingale, and den Nijs}}]{Thouless:1982a}
\bibinfo{author}{\bibnamefont{Thouless}, \bibfnamefont{D.~J.}},
  \bibinfo{author}{\bibfnamefont{M.}~\bibnamefont{Kohmoto}},
  \bibinfo{author}{\bibfnamefont{M.~P.} \bibnamefont{Nightingale}}, and
  \bibinfo{author}{\bibfnamefont{M.}~\bibnamefont{den Nijs}},
  \bibinfo{year}{1982}, \bibinfo{journal}{Phys. Rev. Lett.}
  \textbf{\bibinfo{volume}{49}}(\bibinfo{number}{6}), \bibinfo{pages}{405}.

\bibitem[{\citenamefont{Tinkham}(1996)}]{tink96}
\bibinfo{author}{\bibnamefont{Tinkham}, \bibfnamefont{M.}},
  \bibinfo{year}{1996}, \emph{\bibinfo{title}{Introduction to
  {S}uperconductivity}} (\bibinfo{publisher}{McGraw-Hill}).

\bibitem[{\citenamefont{Tung} \emph{et~al.}(2006)\citenamefont{Tung,
  Schweikhard, and Cornell}}]{Tung:2006}
\bibinfo{author}{\bibnamefont{Tung}, \bibfnamefont{S.}},
  \bibinfo{author}{\bibfnamefont{V.}~\bibnamefont{Schweikhard}}, and
  \bibinfo{author}{\bibfnamefont{E.~A.} \bibnamefont{Cornell}},
  \bibinfo{year}{2006}, \bibinfo{journal}{Phys. Rev. Lett.}
  \textbf{\bibinfo{volume}{97}}(\bibinfo{number}{24}), \bibinfo{pages}{240402}.

\bibitem[{\citenamefont{Unanyan} \emph{et~al.}(1998)\citenamefont{Unanyan,
  Fleischhauer, Shore, and Bergmann}}]{Unanyan98OC}
\bibinfo{author}{\bibnamefont{Unanyan}, \bibfnamefont{R.~G.}},
  \bibinfo{author}{\bibfnamefont{M.}~\bibnamefont{Fleischhauer}},
  \bibinfo{author}{\bibfnamefont{B.~W.} \bibnamefont{Shore}}, and
  \bibinfo{author}{\bibfnamefont{K.}~\bibnamefont{Bergmann}},
  \bibinfo{year}{1998}, \bibinfo{journal}{Opt. Commun.}
  \textbf{\bibinfo{volume}{155}}, \bibinfo{pages}{144}.

\bibitem[{\citenamefont{Unanyan} \emph{et~al.}(1999)\citenamefont{Unanyan,
  Shore, and Bergmann}}]{Unanyan99PRA}
\bibinfo{author}{\bibnamefont{Unanyan}, \bibfnamefont{R.~G.}},
  \bibinfo{author}{\bibfnamefont{B.~W.} \bibnamefont{Shore}}, and
  \bibinfo{author}{\bibfnamefont{K.}~\bibnamefont{Bergmann}},
  \bibinfo{year}{1999}, \bibinfo{journal}{Phys. Rev. A}
  \textbf{\bibinfo{volume}{59}}, \bibinfo{pages}{2910}.

\bibitem[{\citenamefont{Vaishnav and Clark}(2008)}]{Vaishnav08PRL}
\bibinfo{author}{\bibnamefont{Vaishnav}, \bibfnamefont{J.~Y.}}, and
  \bibinfo{author}{\bibfnamefont{C.~W.} \bibnamefont{Clark}},
  \bibinfo{year}{2008}, \bibinfo{journal}{Phys. Phys. Lett.}
  \textbf{\bibinfo{volume}{100}}, \bibinfo{pages}{153002}.

\bibitem[{\citenamefont{Vaishnav} \emph{et~al.}(2008)\citenamefont{Vaishnav,
  Ruseckas, Clark, and Juzeli\=unas}}]{Vaisnav08PRL-DDT}
\bibinfo{author}{\bibnamefont{Vaishnav}, \bibfnamefont{J.~Y.}},
  \bibinfo{author}{\bibfnamefont{J.}~\bibnamefont{Ruseckas}},
  \bibinfo{author}{\bibfnamefont{C.~W.} \bibnamefont{Clark}}, and
  \bibinfo{author}{\bibfnamefont{G.}~\bibnamefont{Juzeli\=unas}},
  \bibinfo{year}{2008}, \bibinfo{journal}{Phys. Rev. Lett.}
  \textbf{\bibinfo{volume}{101}}, \bibinfo{pages}{265302}.

\bibitem[{\citenamefont{Visser and Nienhuis}(1998)}]{Visser:1998}
\bibinfo{author}{\bibnamefont{Visser}, \bibfnamefont{P.~M.}}, and
  \bibinfo{author}{\bibfnamefont{G.}~\bibnamefont{Nienhuis}},
  \bibinfo{year}{1998}, \bibinfo{journal}{Phys. Rev. A}
  \textbf{\bibinfo{volume}{57}}(\bibinfo{number}{6}), \bibinfo{pages}{4581}.

\bibitem[{\citenamefont{Vitanov} \emph{et~al.}(2001)\citenamefont{Vitanov,
  Fleischhauer, Shore, and Bergmann}}]{Vitanov01AAMOP}
\bibinfo{author}{\bibnamefont{Vitanov}, \bibfnamefont{N.~V.}},
  \bibinfo{author}{\bibfnamefont{M.}~\bibnamefont{Fleischhauer}},
  \bibinfo{author}{\bibfnamefont{B.~W.} \bibnamefont{Shore}}, and
  \bibinfo{author}{\bibfnamefont{K.}~\bibnamefont{Bergmann}},
  \bibinfo{year}{2001}, \bibinfo{journal}{Adv. At., Mol., Opt. Phys.}
  \textbf{\bibinfo{volume}{46}}, \bibinfo{pages}{55}.

\bibitem[{\citenamefont{Wang} \emph{et~al.}(2010)\citenamefont{Wang, Gao, Jian,
  and Zhai}}]{Wang:2010}
\bibinfo{author}{\bibnamefont{Wang}, \bibfnamefont{C.}},
  \bibinfo{author}{\bibfnamefont{C.}~\bibnamefont{Gao}},
  \bibinfo{author}{\bibfnamefont{C.-M.} \bibnamefont{Jian}}, and
  \bibinfo{author}{\bibfnamefont{H.}~\bibnamefont{Zhai}}, \bibinfo{year}{2010},
  \bibinfo{journal}{Phys. Lett. Lett.} \textbf{\bibinfo{volume}{105}},
  \bibinfo{pages}{160403}.

\bibitem[{\citenamefont{Wang and Gong}(2006)}]{Wang:2006a}
\bibinfo{author}{\bibnamefont{Wang}, \bibfnamefont{Y.-F.}}, and
  \bibinfo{author}{\bibfnamefont{C.-D.} \bibnamefont{Gong}},
  \bibinfo{year}{2006}, \bibinfo{journal}{Phys. Rev. B}
  \textbf{\bibinfo{volume}{74}}(\bibinfo{number}{19}), \bibinfo{eid}{193301}
  (pages~\bibinfo{numpages}{4}).

\bibitem[{\citenamefont{Weigert and Littlejohn}(1993)}]{Weigert:1993}
\bibinfo{author}{\bibnamefont{Weigert}, \bibfnamefont{S.}}, and
  \bibinfo{author}{\bibfnamefont{R.~G.} \bibnamefont{Littlejohn}},
  \bibinfo{year}{1993}, \bibinfo{journal}{Phys. Rev. A}
  \textbf{\bibinfo{volume}{47}}(\bibinfo{number}{5}), \bibinfo{pages}{3506}.

\bibitem[{\citenamefont{Wilczek and Zee}(1984)}]{Wilczek:1984}
\bibinfo{author}{\bibnamefont{Wilczek}, \bibfnamefont{F.}}, and
  \bibinfo{author}{\bibfnamefont{A.}~\bibnamefont{Zee}}, \bibinfo{year}{1984},
  \bibinfo{journal}{Phys. Rev. Lett.}
  \textbf{\bibinfo{volume}{52}}(\bibinfo{number}{24}), \bibinfo{pages}{2111}.

\bibitem[{\citenamefont{Williams} \emph{et~al.}(2010)\citenamefont{Williams,
  Al-Assam, and Foot}}]{Williams:2010}
\bibinfo{author}{\bibnamefont{Williams}, \bibfnamefont{R.~A.}},
  \bibinfo{author}{\bibfnamefont{S.}~\bibnamefont{Al-Assam}}, and
  \bibinfo{author}{\bibfnamefont{C.~J.} \bibnamefont{Foot}},
  \bibinfo{year}{2010}, \bibinfo{journal}{Phys. Rev. Lett.}
  \textbf{\bibinfo{volume}{104}}(\bibinfo{number}{5}), \bibinfo{pages}{050404}.

\bibitem[{\citenamefont{Winkler}(2003)}]{Winkler03Review}
\bibinfo{author}{\bibnamefont{Winkler}, \bibfnamefont{R.}},
  \bibinfo{year}{2003}, \emph{\bibinfo{title}{Spin--Orbit Coupling Effects in
  Two-Dimensional Electron and Hole Systems}} (\bibinfo{publisher}{Springer},
  \bibinfo{address}{Berlin}).

\bibitem[{\citenamefont{Wu}(2008)}]{Wu:2008}
\bibinfo{author}{\bibnamefont{Wu}, \bibfnamefont{C.}}, \bibinfo{year}{2008},
  \bibinfo{journal}{Phys. Rev. Lett.}
  \textbf{\bibinfo{volume}{101}}(\bibinfo{number}{18}),
  \bibinfo{pages}{186807}.

\bibitem[{\citenamefont{Xiao} \emph{et~al.}(2010)\citenamefont{Xiao, Chang, and
  Q.Niu}}]{Xiao10RMP}
\bibinfo{author}{\bibnamefont{Xiao}, \bibfnamefont{D.}},
  \bibinfo{author}{\bibfnamefont{M.-C.} \bibnamefont{Chang}}, and
  \bibinfo{author}{\bibnamefont{Q.Niu}}, \bibinfo{year}{2010},
  \bibinfo{journal}{Rev. Mod. Phys} \textbf{\bibinfo{volume}{82}},
  \bibinfo{pages}{1959}.

\bibitem[{\citenamefont{Ye} \emph{et~al.}(2008)\citenamefont{Ye, Kimble, and
  Katori}}]{Ye:2008}
\bibinfo{author}{\bibnamefont{Ye}, \bibfnamefont{J.}},
  \bibinfo{author}{\bibfnamefont{H.~J.} \bibnamefont{Kimble}}, and
  \bibinfo{author}{\bibfnamefont{H.}~\bibnamefont{Katori}},
  \bibinfo{year}{2008}, \bibinfo{journal}{Science}
  \textbf{\bibinfo{volume}{320}}(\bibinfo{number}{5884}),
  \bibinfo{pages}{1734}.

\bibitem[{\citenamefont{Yi} \emph{et~al.}(2008)\citenamefont{Yi, Daley,
  Pupillo, and Zoller}}]{Yi:2008}
\bibinfo{author}{\bibnamefont{Yi}, \bibfnamefont{W.}},
  \bibinfo{author}{\bibfnamefont{A.}~\bibnamefont{Daley}},
  \bibinfo{author}{\bibfnamefont{G.}~\bibnamefont{Pupillo}}, and
  \bibinfo{author}{\bibfnamefont{P.}~\bibnamefont{Zoller}},
  \bibinfo{year}{2008}, \bibinfo{journal}{New Journal of Physics}
  \textbf{\bibinfo{volume}{10}}(\bibinfo{number}{7}), \bibinfo{pages}{073015}.

\bibitem[{\citenamefont{Yip}(2010)}]{Yip:2010}
\bibinfo{author}{\bibnamefont{Yip}, \bibfnamefont{S.-K.}},
  \bibinfo{year}{2010}, \bibinfo{journal}{arXiv:1008.2263} .

\bibitem[{\citenamefont{Zee}(1988)}]{Zee88PRA}
\bibinfo{author}{\bibnamefont{Zee}, \bibfnamefont{A.}}, \bibinfo{year}{1988},
  \bibinfo{journal}{Phys. Rev. A}
  \textbf{\bibinfo{volume}{38}}(\bibinfo{number}{1}), \bibinfo{pages}{1}.

\bibitem[{\citenamefont{Zhang} \emph{et~al.}(2005)\citenamefont{Zhang, Li, and
  Sun}}]{Zhang:2005}
\bibinfo{author}{\bibnamefont{Zhang}, \bibfnamefont{P.}},
  \bibinfo{author}{\bibfnamefont{Y.}~\bibnamefont{Li}}, and
  \bibinfo{author}{\bibfnamefont{C.~P.} \bibnamefont{Sun}},
  \bibinfo{year}{2005}, \bibinfo{journal}{Eur.Phys.J. D}
  \textbf{\bibinfo{volume}{36}}, \bibinfo{pages}{229}.

\bibitem[{\citenamefont{Zhang} \emph{et~al.}(2010)\citenamefont{Zhang, Gong,
  and Oh}}]{Oh09-tripod}
\bibinfo{author}{\bibnamefont{Zhang}, \bibfnamefont{Q.}},
  \bibinfo{author}{\bibfnamefont{J.}~\bibnamefont{Gong}}, and
  \bibinfo{author}{\bibfnamefont{C.}~\bibnamefont{Oh}}, \bibinfo{year}{2010},
  \bibinfo{journal}{Ann. Phys.} \textbf{\bibinfo{volume}{325}},
  \bibinfo{pages}{1219}.

\bibitem[{\citenamefont{Zhu} \emph{et~al.}(2006)\citenamefont{Zhu, Fu, Wu,
  Zhang, and Duan}}]{Zhu:2006PRA}
\bibinfo{author}{\bibnamefont{Zhu}, \bibfnamefont{S.-L.}},
  \bibinfo{author}{\bibfnamefont{H.}~\bibnamefont{Fu}},
  \bibinfo{author}{\bibfnamefont{C.-J.} \bibnamefont{Wu}},
  \bibinfo{author}{\bibfnamefont{S.-C.} \bibnamefont{Zhang}}, and
  \bibinfo{author}{\bibfnamefont{L.-M.} \bibnamefont{Duan}},
  \bibinfo{year}{2006}, \bibinfo{journal}{Phys. Rev. Lett.}
  \textbf{\bibinfo{volume}{97}}, \bibinfo{pages}{240401}.

\bibitem[{\citenamefont{Zutic} \emph{et~al.}(2004)\citenamefont{Zutic, Fabian,
  and Das~Sarma}}]{Zutic04RMP}
\bibinfo{author}{\bibnamefont{Zutic}, \bibfnamefont{I.}},
  \bibinfo{author}{\bibfnamefont{J.}~\bibnamefont{Fabian}}, and
  \bibinfo{author}{\bibfnamefont{S.}~\bibnamefont{Das~Sarma}},
  \bibinfo{year}{2004}, \bibinfo{journal}{Rev. Mod. Phys.}
  \textbf{\bibinfo{volume}{76}}(\bibinfo{number}{2}), \bibinfo{pages}{323}.

\bibitem[{\citenamefont{Zygelman}(1990)}]{Zygelman:1990}
\bibinfo{author}{\bibnamefont{Zygelman}, \bibfnamefont{B.}},
  \bibinfo{year}{1990}, \bibinfo{journal}{Phys. Rev. Lett.}
  \textbf{\bibinfo{volume}{64}}, \bibinfo{pages}{256}.

\end{thebibliography}

\end{document}